\documentclass[12pt,doublespace]{JHEP3}

\usepackage{amsmath,amssymb,amsfonts,xspace} 
\usepackage{epsfig}

\DeclareMathOperator{\Li}{Li}

\renewcommand{\Im}{\imag}

\newcommand{\be}{\begin{equation}}
\newcommand{\ee}{\end{equation}}
\newcommand{\beq}{\begin{eqnarray}}
\newcommand{\eeq}{\end{eqnarray}}
\newcommand{\ben}{\begin{eqnarray}\displaystyle}
\newcommand{\een}{\end{eqnarray}}
\newcommand{\bea}[2]{\be\label{#2}\begin{array}{#1}}
\newcommand{\eea}{\end{array}\ee}

\def\Tr{\,{\rm Tr}\, }
\def\det{\,{\rm det}\, }

\def\sign{{\rm sign}}

\def\Im{\,{\rm Im}\, }

\def\({\left(}
\def\){\right)}
\def\[{\left[}
\def\]{\right]}
\def\p{\partial}

\def\11{1\!\! 1}

\newcommand{\bZ}{\mathbb{Z}}

\newcommand{\de}{\mathrm{d}}

\newcommand{\I}{\mathrm{i}}
\newcommand{\cA}{\mathcal{A}}
\newcommand{\cB}{\mathcal{B}}
\newcommand{\cL}{\mathcal{L}}

\newcommand{\cH}{\mathcal{H}}
\newcommand{\cP}{\mathcal{P}}

\newcommand{\cK}{\mathcal{K}}
\newcommand{\cM}{\mathcal{M}}

\newcommand{\cN}{\mathcal{N}}

\newcommand{\cX}{\mathcal{X}}

\newcommand{\IR}{\mathbb{R}}

\newcommand{\IZ}{\mathbb{Z}}

\newcommand{\hh}{g}
\newcommand{\ggx}{g}

\newcommand{\argu}[2]{(#1\gamma_1+#2\gamma_2)}
\newcommand{\argy}[3]{(#1\gamma_1+#2\gamma_2,#3)}
\newcommand{\kscom}[1]{\kappa(#1)}

\def\varpi{t}

\newcommand{\nn}{\nonumber}
\newcommand{\kahler}{{K\"ahler}\xspace}

\def\bse{\begin{subequations}}
\def\ese{\end{subequations}}

\def\qli2{{\bf E}}

\newcommand\bOm{\bar\Omega}
\newcommand\tOm{\widetilde\Omega}

\def\ea#1\ea{\begin{align}#1\end{align}}

\usepackage{graphicx}
\usepackage[arrow,curve,matrix,arc]{xy}

\newcommand{\DTb}{\bar{\rm DT}}

\numberwithin{equation}{section}
\title{Wall Crossing from Boltzmann Black Hole Halos}

\author{Jan Manschot$^{1}$, Boris Pioline$^{2}$, Ashoke Sen$^{3}$
\\
$^1$ {\it Institut de Physique Th\'eorique, CEA Saclay, CNRS-URA 2306,\\
91191 Gif sur Yvette, France}
\\
$^2$ {\it Laboratoire de Physique Th\'eorique et Hautes
Energies, CNRS UMR 7589, \\
Universit\'e Pierre et Marie Curie,
4 place Jussieu, 75252 Paris cedex 05, France} \\
$^3$ Harish-Chandra Research Institute,
Chhatnag
Road, Jhusi, Allahabad 211019, India
\\

\vspace*{2mm} {\tt e-mail: \email{
jan.manschot@cea.fr, pioline@lpthe.jussieu.fr,
sen@hri.res.in}
} \vspace*{-3mm}

}

\abstract{A key question in the study of $\cN=2$ supersymmetric string or field 
theories is to understand the decay of BPS bound states across 
walls of marginal stability in the space of parameters or vacua. 
By representing the potentially unstable bound states
as multi-centered black hole solutions in $\cN=2$ supergravity, we provide two 
fully general and explicit
formul\ae\  for the change in the (refined) index across the wall. 
The first, ``Higgs
branch"  formula relies on Reineke's results for invariants of quivers
without oriented loops, specialized to the Abelian case. The second, ``Coulomb branch"  formula results from
evaluating the symplectic volume of the classical phase space of multi-centered solutions
by localization. We provide extensive evidence that these new formul\ae\  agree 
with each other and with the mathematical  results of 
Kontsevich and Soibelman (KS) and Joyce and Song (JS). 
The main physical insight behind our results is that the 
Bose-Fermi statistics of individual black holes participating 
in the bound state can be traded for Maxwell-Boltzmann
statistics, provided the (integer) index $\Omega(\gamma)$ of the internal degrees
of freedom carried by each black hole 
is replaced by an effective (rational) index  $\bar\Omega(\gamma)=
\sum_{m|\gamma} \Omega(\gamma/m)/m^2$. 
A similar map also exists for the refined index.
This observation provides a 
physical rationale for the appearance of the 
rational Donaldson-Thomas invariant 
$\bar\Omega(\gamma)$ in the works of KS and JS.
The simplicity of the wall-crossing
formula for 
rational invariants allows us to  generalize
the ``semi-primitive wall-crossing formula" 
to arbitrary decays of the type 
$\gamma\to M\gamma_1+N\gamma_2$ with $M=2,3$.
}

\begin{document}

\section{Introduction and summary} \label{sintro}

In quantum field theories and string theory vacua with
extended supersymmetry, the spectrum of BPS states
can sometimes be determined exactly in a weakly coupled
region of the space of parameters (or vacua). In extrapolating
 the BPS spectrum to another point in parameter
 space, one must be wary of two issues: BPS states may pair up and disappear, and single particle
 states may decay into the continuum of multi-particle states.
 The first issue can be evaded by considering a suitable index 
 $\Omega(\gamma;t^a)$, where $\gamma$ is the vector of 
 electric and magnetic charges carried by the state and $t^a$ 
 parametrizes the value of the couplings (or moduli), designed such 
 that contributions from long multiplets cancel. The index $\Omega(\gamma;t^a)$ is then 
a piecewise constant function of the parameters $t^a$. To deal with the
second problem, it is important to understand how $\Omega(\gamma;t^a)$ changes 
across certain codimension-one subspaces of the
parameter space, known as  walls of marginal stability, where a 
single-particle BPS state becomes marginally
unstable against decay into two (or more) BPS 
states ~\cite{Cecotti:1992rm,Seiberg:1994rs,
Seiberg:1994aj,Bilal:1996sk, Douglas:2000gi}.

Initial progress in this direction for  four-dimensional string vacua 
came from supergravity, where BPS states are represented 
by (in general multi-centered) classical black hole solutions. 
Since the class of multi-centered solutions 
potentially unstable at a certain wall
of marginal stability exists  only on one side of the 
wall~\cite{Denef:2000nb,Denef:2002ru,Bates:2003vx},  
the discontinuity $\Delta\Omega(\gamma)$ in $\Omega(\gamma, t^a)$ is equal to 
the index of the multi-centered solutions with total charge $\gamma$, up to a sign
depending whether one enters or leaves the side on which these solutions  
exist~\cite{Denef:2007vg}.
Based on this physical picture, one easily finds that the jump 
of the index in the simplest case, where the 
only  configuration that may appear or disappear across
the wall is a two-centered solution with primitive charge vectors $\gamma_1$,
$\gamma_2$, is given by the  ``primitive wall-crossing formula" ~\cite{Denef:2007vg}:
\be 
\label{primwcf}   
\Delta \Omega(\gamma)= (-1)^{\gamma_{12}+1}\, 
|\gamma_{12}|\,  \Omega(\gamma_1)\, \Omega(\gamma_2)\ .
\ee
With some more effort one can also 
compute $\Delta\Omega(\gamma)$ in 
``semi-primitive" cases, where the relevant multi-centered solutions 
which appear or disappear across the wall of marginal stability are
halos of black holes with charges in multiple of $\gamma_2$, orbiting
around a core of charge $\gamma_1$~\cite{Denef:2007vg}.

While applying this method to the general ``non-primitive" case seemed out of 
reach up until now, the breakthrough came from the mathematical front, with the 
works of  Kontsevich and Soibelman (KS) \cite{ks,Kontsevich:2009xt}
and Joyce and Song (JS) \cite{MR2357325,Joyce:2008pc,Joyce:2009xv}. In these
works, general formul\ae\  were derived for the discontinuity of 
generalized Donaldson-Thomas (DT) invariants under changes of stability conditions. 
It is generally believed that generalized DT invariants are the appropriate
mathematical embodiment of the physical BPS invariants.
Although the KS and JS wall-crossing formul\ae\  look 
very different, there is by now much evidence that they
are equivalent\footnote{We have been informed by D. Joyce
of a general argument showing the equivalence of the KS and JS
wall-crossing formulae. The equality of the DT invariants as defined
by KS and JS seems, on the other hand, less firmly established.}.
Since these two formul\ae\  appeared, much efforts have been 
devoted towards interpreting, deriving and checking 
these wall-crossing formul\ae\ in various physical
settings \cite{Diaconescu:2007bf,Gaiotto:2008cd,Alexandrov:2008gh,
Jafferis:2008uf,Cecotti:2009uf,Cecotti:2010fi,Gaiotto:2010be,Andriyash:2010qv,
Manschot:2009ia}.  
Our goal in this paper is to 
rederive the wall-crossing formula using
multi-centered black hole
solutions in supergravity and extend the  
original black hole halo picture to the
general ``non-primitive" case.
We also carry out extensive comparisons between our
formul\ae\ and those of KS and JS.

One intriguing aspect of the KS and JS wall-crossing 
formul\ae\
is the appearance of two types of BPS-invariants, an
integer-valued invariant $\Omega$ (roughly $\Tr(-1)^F$) 
and a rational-valued invariant $\bar\Omega$. The two 
are related by the ``multi-cover formula"
\be \label{efirst}
\bar\Omega(\gamma)=\sum_{m|\gamma}m^{-2}\, 
\Omega(\gamma/m)\, ,
\ee
where the sum runs over all positive integers $m$ such that $\gamma/m$ lies 
in the charge lattice. 
We shall take \eqref{efirst} as 
the definition of $\bar\Omega$.
Similar divisor sums appear in various instances in quantum field theory,
e.g. in Schwinger's computation  of pair creation in an external electric field (see {\it e.g.}
Eqs. (4.118) and (4.119) in 
\cite{Itzykson:1980rh}), and in enumerative geometry, e.g.
the multi-cover formul\ae\  for Gromov-Witten invariants \cite{Aspinwall:1991ce}, 
which indeed are naturally understood from pair creation
arguments \cite{Gopakumar:1998ii,Gopakumar:1998jq}. The rational invariants
$\bar\Omega$ also arose in constructions 
of modular invariant black hole partition functions consistent with 
wall-crossing 
\cite{Manschot:2010xp,Manschot:2010nc,Nishinaka:2010fh}, and in studies of 
D-brane instanton corrections to the hypermultiplet moduli space 
metric \cite{Alexandrov:2008gh,Alexandrov:2010np,Alexandrov:2010ca}.

For the purposes of computing the jump in the BPS spectrum, 
the rational  invariants  turn out to be especially convenient. Indeed,
one consequence of the JS/KS wall-crossing
formul\ae\  is that the variation $\Delta\bar\Omega(\gamma)$ 
of the rational invariants across a wall of marginal stability, 
when expressed in terms of the rational invariants 
$\bar\Omega(\gamma')$ on one 
side of the wall, involves only ``charge conserving'' terms, i.e. sums of products 
of $\bar\Omega(\gamma_i)$ for different $\gamma_i$ such that 
$\sum_i \gamma_i=\gamma$.  In contrast, the variation of the integer invariants
$\Delta\Omega(\gamma)$, expressed  in terms of the integer invariants
$\Omega(\gamma)$,  
does not satisfy this property, and as a result, involves 
considerably more terms. Needless to say, physical charge is conserved 
no matter what invariant one chooses to consider.

Our main new insight, at the basis of the results presented below,
is the following physical explanation of this phenomenon. 
In computing $\Delta\Omega(\gamma)$ from the index
associated  with a $n$-centered black hole solution carrying total 
charge $\gamma=\sum_{i=1}^n \gamma_i$, each center must be treated as 
a point-like particle carrying $|\Omega(\gamma_i)|$ 
internal states. 
When some of the $\gamma_i$ coincide, the
corresponding centers must  in addition obey Bose or Fermi statistics, depending on the
sign of  $\Omega(\gamma_i)$. As a result of (anti-)symmetrizing the 
many-body wave-function, the total index associated with
such a configuration involves, in addition to the 
product $\prod_{i=1}^n\Omega(\gamma_i)$, terms of lower degree in 
$\Omega(\gamma_i)$ -- {\it e.g.} two identical bosons of degeneracy $\Omega$
will give a degeneracy of $\Omega(\Omega+1)/2$.
The terms of lower degree in $\Omega$ violate
the charge conservation property defined above.
However, due  to special properties of the interactions between 
centers (namely, the no-force condition between centers with mutually
local charges), we show that it is possible to map the 
problem of computing the index of multi-centered
black holes with individual centers
satisfying Bose/Fermi statistics to an
equivalent problem where the centers 
satisfy instead Maxwell-Boltzmann statistics. In this
Boltzmannian reformulation, each center carries 
an  effective (in general non integer) 
index $\bar\Omega(\gamma)$ related to 
$\Omega(\gamma)$ via \eqref{efirst}, 
and charge conservation is manifest. 
This provides a 
physical rationale of the charge conservation property
of the  wall-crossing formula written in terms of 
the rational invariants $\bar\Omega$.

The same argument generalizes for the refined `index' $\Omega_{\rm ref}(\gamma,y)$, 
defined roughly as $\Tr (-1)^F \, y^{2J_3}$, 
which keeps track of the angular momentum 
of the BPS states.
However  this refined `index' is only protected (i.e. immune to contributions 
of long multiplets)
in the presence of a $SU(2)_R$ 
symmetry \cite{Gaiotto:2010be}. 
Such a symmetry exists in $\cN=2$ 
supersymmetric field theory, but not in string theory or supergravity. As a result,
this  refined `index' will in general be different at weak and strong coupling (more generally
its value will depend on both vector multiplets and hypermultiplets).  
Nevertheless, one may still investigate the variation of $\Omega_{\rm ref}(\gamma,y)$
across lines of marginal stability in vector multiplet moduli space. In fact, 
KS have provided a wall-crossing formula for motivic Donaldson-Thomas 
invariants, which are conjectured to be equal to the refined 
invariants $\Omega_{\rm ref}(\gamma,y)$ at weak coupling, where the derived
category description of D-branes is appropriate
\cite{Dimofte:2009bv,Dimofte:2009tm}. Similarly, one may ask about the 
wall-crossing formula in the strong coupling region where the supergravity
picture is appropriate. 
As for the standard index, we find that the 
variation $\Delta\Omega_{\rm ref}^\pm(\gamma,y)$ can be computed by treating centers as Boltzmannian particles carrying internal states with effective refined index 
\be \label{esecond}
\bar \Omega_{\rm ref}(\gamma,y) 
\equiv \sum_{m|\gamma} 
\frac{y - y^{-1}}{m\, (y^m - y^{-m})}\, 
\Omega_{\rm ref}(\gamma/m, y^m) 
\, .
\ee
In this formulation, charge conservation is again manifest.
At $y=1$, $\Omega_{\rm ref}(\gamma,y)$ reduces to $\Omega(\gamma)$ and 
\eqref{esecond} to \eqref{efirst}.

While the arguments  
above rely on representing BPS states as multi-centered 
solutions in supergravity, it is clear that it extends to the case of $\cN=2$ 
supersymmetric gauge theories which can be obtained as rigid limits of 
supergravity theories \cite{Denef:1998sv}. In general, we expect that BPS
solitons in the Coulomb phase can be represented as classical multi-centered
solutions of the Abelian gauge theory at low energy, albeit singular ones.
For the purposes of computing the wall-crossing, the singularity is irrelevant,
and the problem can still be reduced to the quantum mechanics of point-like
particles interacting by Coulomb law and scalar interactions. In particular, solitons
with mutually local charges ($\langle\gamma_1,\gamma_2\rangle =0$) do not interact, 
and the above Bose-Fermi/Boltzmann equivalence carries over.

Finally, it is worth pointing out that a similar phenomenon occurs 
for non-primitive wall-crossing in
$\cN=4$ supersymmetric string 
theories~\cite{Banerjee:2008pu}.
In this case only two-centered configurations contribute, 
and the only non-trivial effect comes from symmetrization~\cite{Sen:2008ht,Dabholkar:2008zy}. 
The variation of the index is thus given by the  primitive wall-crossing formula
\eqref{primwcf}, provided  $\Omega(\gamma)$ is replaced by with $\tilde\Omega(\gamma)=
\sum_{m\vert\gamma} \Omega(\gamma/m)$ in this formula. Note that in contrast 
to the effective index \eqref{efirst} relevant for $\cN=2$ BPS states, the
effective  index $\tilde\Omega(\gamma)$ relevant for $\cN=4$ dyons 
does not include any factor of $1/m^2$ in its definition.
This difference can be traced to
the presence of extra fermion zero modes carried by 
a quarter BPS dyon in
$\cN=4$ supersymmetric theories. Trace over these
fermion zero modes for a system of identical particles
produces an extra factor of $m^2$ in \eqref{efirst}
compared to that for half BPS dyons in $\cN=2$
supersymmetric theories.

We shall now summarise our main results. Consider a wall
of marginal stability on which the central charges 
$Z_{\gamma_1}$ and $Z_{\gamma_2}$ of
two charge vectors $\gamma_1$ and $\gamma_2$
align. Assume further that, possibly after a change of basis in the lattice 
spanned by $\gamma_1$ and $\gamma_2$,  BPS states 
carrying charge $M\gamma_1+N\gamma_2$ exist
only for $(M\ge 0, N\ge 0)$ and 
$(M\le 0, N\le 0)$.
Then on one side of the wall, which we call the
chamber $c^-$, we have 
$\langle\gamma_1,\gamma_2\rangle \,
{\rm Im}(Z_{\gamma_1}
\bar Z_{\gamma_2})>0$, and 
there are
multi-centered bound states with individual 
centers carrying
charges of the form $m_i\gamma_1+n_i\gamma_2$
with different integers $m_i,n_i\geq 0$.
Here $\langle\gamma_1,\gamma_2\rangle$ is the
symplectic inner product between $\gamma_1$ and
$\gamma_2$. On the other
side of the wall, called the chamber $c^+$,
there are no bound states of this form. Let us
denote by $\Omega^\pm(\alpha)$ the index 
$\Tr'(-1)^{2J_3}$  on the
two sides of the wall for a charge vector $\alpha=M\gamma_1+N\gamma_2$ with $M,N\ge 0$.
($\Tr'$ denotes the trace
after removing the fermion zero modes
associated with broken supersymmetries.)
Then the physical reasoning outlined above shows that the wall-crossing formula,
expressed in terms of the rational  
invariants \eqref{efirst}, must take the form
\be \label{efirst2}
\bar\Omega^-(\gamma) -\bar\Omega^+(\gamma) =
\sum_{n\geq 2 }\, 
  \sum_{\substack{ \{\alpha_1,\dots, \alpha_n\} \\
\gamma= \alpha_1+\dots +\alpha_n}}\, 
\frac{g(\{\alpha_i\})}{|{\rm Aut}(\{\alpha_i\})|}
 \prod\nolimits_{i=1}^n \bOm^+(\alpha_i) \ ,
 \ee
 where the sum runs over all possible unordered
 decompositions
 of $\alpha$ into vectors $\alpha_1,\dots, \alpha_n$,
 each of which is a linear combination of 
 $\gamma_1$ and $\gamma_2$ with non-negative
 integer coefficients. Here,  
 $|{\rm Aut}(\{\alpha_i\})|$ is the symmetry factor appropriate
 for Maxwell-Boltzmann statistics, namely 
 the order of the subgroup of the 
 permutation group of $n$ elements which preserves 
 the ordered set $(\alpha_1,\dots, \alpha_n)$, for a fixed 
 (arbitrary) choice of 
 ordering.\footnote{Thus if the set  
 $\{\alpha_i\}$ consists of $m_1$ copies of $\beta_1$,
 $m_2$ copies of $\beta_2$ etc. then
 $|{\rm Aut}(\{\alpha_i\})|=\prod_k m_k!.$} 
 Of course, one could 
 instead decide to absorb this symmetry factor 
 in the normalization of $g(\{\alpha_i\})$.
 The point of the normalization chosen in \eqref{efirst2} is that 
 $g(\alpha_1,\dots, \alpha_n)$ can now be identified as 
the index associated with an $n$-centered black hole
configuration in
supergravity, with the individual centers carrying
charges $\alpha_1,\alpha_2,\dots, \alpha_n$ and 
treated as {\it distinguishable} particles, and furthermore {\it carrying
no internal degeneracy}. Clearly, the same considerations
imply that the wall-crossing formula for  refined invariants
takes an analog form
\be \label{efirst3}
\bar\Omega_{\rm ref}^-(\gamma,y) -\bar\Omega^+_{\rm ref}
(\gamma,y) =
\sum_{n\geq 2 }\, 
  \sum_{\substack{ \{\alpha_1,\dots, \alpha_n\} \\
\gamma= \alpha_1+\dots +\alpha_n}}\, 
\frac{g_{\rm ref}(\{\alpha_i\},y)}{|{\rm Aut}(\{\alpha_i\})|}
 \prod\nolimits_{i=1}^n \bOm_{\rm ref}^+(\alpha_i,y) \ ,
 \ee
 where $g_{\rm ref}(\{\alpha_i\},y)$ computes the 
 refined `index' of the same $n$-centered black hole
configuration, and reduces to $g(\{\alpha_i\})$ at $y=1$. 
In order to complete the wall-crossing formula we need
to specify the factor $g_{\rm ref}(\{\alpha_i\},y)$ (or
its $y=1$ limit  $g(\{\alpha_i\})$ ). While these factors
can be extracted from the KS and JS formulae,  we shall present
two novel ways for computing them, which we call
the ``Higgs branch" and the ``Coulomb branch" formulae. We have
checked in many cases the equivalence of these prescriptions
with the KS and JS formulae, although we have not yet been
able to prove the equivalence rigorously.

The ``Higgs branch" formula is based on Denef's observation \cite{Denef:2002ru} 
that the spectrum of supersymmetric bound states of 
multi-centered black holes can be computed in the framework of quiver quantum 
mechanics. This description is appropriate at weak coupling, the arrows of the quiver 
describing the open strings stretched between two D-branes. 
Due to the fact that the charges carried by the various centers lie on a two-dimensional
sublattice of the full charge lattice, the relevant quiver  turns out to have no oriented loops. 
A formula for the motivic invariants of such quivers was given 
by Reineke in \cite{MR1974891}. Furthermore,
since the constituents of the bound states are to be treated
as distinguishable particles without internal multiplicity, the relevant 
quiver carries dimension one vector spaces at each node
(equivalently, corresponds to a $U(1)^n$ gauge theory).
Reineke's formula simplifies in this case, leading
to 
\begin{equation} 
\label{higgsf}
 g_{\rm ref}(\alpha_1,\dots ,\alpha_n, y)=(-y)^{-1+n
 -\sum_{i<j} \alpha_{ij}} \,
 (y^2-1)^{1-n} \, 
 \sum_{\rm partitions}(-1)^{s-1} y^{2\sum_{a\leq b}\sum_{j<i }
 \alpha_{ji}\, 
 m^{(a)}_i \, m^{(b)}_j }\, .
\end{equation}
Here we have denoted by
$\alpha_{ij}=\langle \alpha_i, \alpha_j\rangle$ 
 the symplectic inner product between the vectors
$\alpha_i$ and $\alpha_j$, and have ordered
the $\alpha_i$'s such that $\langle \alpha_i,\alpha_j\rangle
> 0$ for $i< j$ (assuming that none of the vectors $\alpha_i$ coincide).
The sum runs over all ordered partitions of 
$(\alpha_1+\cdots +\alpha_n)$ into $s$ vectors
$\beta^{(a)}$ ($1\le a\le s$, $1\le s\le n$)
such that
\begin{enumerate}
\item $\sum_a \beta^{(a)} = \alpha_1+\cdots + \alpha_n$
\item $\beta^{(a)} = \sum_i m^{(a)}_i \alpha_i$ with
$m^{(a)}_i=0$ or $1$ for each $a,i$.
\item 
$\left\langle \sum_{a=1}^b \, \beta^{(a)}, 
\alpha_1+\cdots + \alpha_n\right\rangle > 0 \quad
\forall \quad b \ \hbox{with} \ 1\le b\le s-1
$
\end{enumerate}
When some of the $\alpha_i$'s coincide,  the value of $g_{\rm ref}(\alpha_1,\dots ,\alpha_n, y)$ 
can still be obtained from \eqref{higgsf} by taking the limit $\alpha_i\to\alpha_j$:
even though the $\alpha_i$'s are supposed to be valued
in the two dimensional lattice spanned by $\gamma_1$
and $\gamma_2$, eq. \eqref{shiggs} defines a continuous function 
of the  $\alpha_i$'s and this limit is well-defined. We have checked 
agreement with the KS and JS formulae for distinct $\alpha_i$ with  $n\leq 5$,
and in many cases where some of the $\alpha_i$ coincide. While it is not 
surprising that the Reineke formula is consistent with the JS formula (since the latter
applies to moduli space of quiver representations), it is remarkable that Abelian 
quivers (i.e. quivers carrying a dimension-one vector space at each node) encode the 
complete information about wall-crossing. In 
Appendix  \S\ref{sUNquiver} we show
that the index of certain non-Abelian quivers without oriented loops
can be reduced to the Abelian case using 
the same black hole halo picture.

Our second way of computing $g_{\rm ref}(\{\alpha_i\},y)$ instead follows from 
quantizing the moduli space of multi-centered BPS solutions, 
as first proposed in \cite{deBoer:2008zn}. This description is most appropriate at strong
coupling, when the classical supergravity description is valid.  Using the fact that
the moduli space $\cM_n$ of $n$-centered solutions carries a natural symplectic structure
and a Hamiltonian action of $SU(2)$, we evaluate the integral of $y^{2J_3}$ over this 
classical phase space by localization. The fixed points of the action of $J_3$ on $\cM_n$
are collinear configurations where all centers lie on the $z$-axis, with relative distances
determined by 
\be \label{denef1d}
\sum_{j=1\atop j\ne i}^n \frac{\alpha_{\sigma(i)\sigma(j)}}
{z_{\sigma(j)} - z_{\sigma(i)}} \, 
\sign(j-i)
= \Lambda\sum_{j=1\atop j\ne i}^n 
\alpha_{\sigma(i)\sigma(j)}\,  ,
\ee
where
$\Lambda$ is a positive real constant which 
can be removed by
rescaling the $z_i$'s, and
$\sigma$ is the permutation which determines the order of the centers along the
axis, $z_{\sigma(i)}<z_{\sigma(j)}$ if $i<j$. In this way 
we arrive at  the ``Coulomb branch formula"
\be \label{coulombf}
g_{\rm ref}(\{\alpha_i\},y)
= (-1)^{\sum_{i<j} \alpha_{ij} +n-1}
(y - y^{-1})^{1-n} \, \sum_{\rm permutations \, \sigma} \,  
s(\sigma)\, y^{
\sum_{i<j} \alpha_{\sigma(i)\sigma(j)}
}\, ,
\ee
where the sum runs over the subset of the permutation group in $n$ elements
for which the equations \eqref{denef1d} 
admit a solution. The factor $s(\sigma)$
originates from the determinant of the action of $J_3$ on the tangent space 
at the fixed points, and evaluates to the sign
\be
s(\sigma) = (-1)^{\#\{i; \sigma(i+1)<\sigma(i)\}}\ .
\ee
While eq. \eqref{coulombf} is still implicit, since it requires solving 
the equations \eqref{denef1d} (or rather, characterizing the permutations $\sigma$
such that  \eqref{denef1d}  admits a solution), 
it provides us with an economic 
way of determining $g_{\rm ref}(\{\alpha_i\},y)$, 
since each permutation has a unique $y$ dependence
and hence there are no cancellations between 
different permutations. In contrast both 
the JS formula 
and ``Higgs branch" formula \eqref{higgsf}
involves extensive
cancellations between different terms. We shall in fact
see in \S\ref{scomp} that motivated by the
Coulomb branch formula one can find an algorithm to
identify the uncancelled contributions in the Higgs branch
formula without having to evaluate all the terms 
given in \eqref{higgsf}.

The use of the rational invariants $\bar\Omega$ also
allows us to use the KS formula to 
derive explicit formulas for the
change in the index in some special cases. This
includes sectors carrying charges of the form
$2\gamma_1+N\gamma_2$ and 
$3\gamma_1+N\gamma_2$ for primitive vectors
$\gamma_1$, $\gamma_2$ and arbitrary integer
$N$. This generalizes the semi-primitive 
wall-crossing formula of \cite{Denef:2007vg} which describes
the change in the index in the sector with
charge $\gamma_1+N\gamma_2$, and some earlier
results on higher-rank ADHM and DT invariants \cite{Stoppa:2009,Toda:2009,Chuang:2010wx}.

The rest of the paper is organised as follows. In
\S\ref{sbolt}  we describe how the problem of computing
the index of multi-centered black holes can be mapped
to an equivalent problem with the individual centers
obeying Maxwell-Boltzmann statistics. We use this to derive some
general properties of the wall-crossing formula 
{\it e.g.} charge conservation, and also
reproduce the primitive and semi-primitive wall
crossing formula. 
We also generalize the results to
the case of refined 
index. 
In \S\ref{squiver} we compute the index 
associated to $n$-centered black hole configurations in two
different ways, 
first by mapping the problem to a quiver
quantum mechanics and second by quantizing  the classical phase space of 
multi-centered solutions. This leads to the Coulomb 
and Higgs branch formulae described above. 
In \S\ref{sec_KS} we review the 
KS wall-crossing formula, and recast it 
in terms of the rational invariants $\bar\Omega$, 
 making the charge conservation property manifest.
We verify that the motivic KS formula agrees
with the results of \S\ref{sbolt} and \S\ref{squiver} 
in many cases, 
and obtain higher order generalizations of the 
semi-primitive wall-crossing formula. 
In \S\ref{sjoyce} we review the wall-crossing formula
due to Joyce and Song and compare it to the
KS, Higgs branch and Coulomb formulae. 
We find agreement in all cases that we consider.
In Appendix \ref{sswall} we illustrate the  general wall
crossing formul\ae\ in some special cases.
In Appendix \ref{secD0D6}
we apply  the results of
\S\ref{sec_KS} to analyze the spectrum of  D0-D6 bound
states on a
Calabi-Yau 3-fold.
In Appendix \ref{egensw} we check our results  
against the spectrum of BPS states
in $\cN=2$ $SU(2)$ gauge theories.
In Appendix \ref{sUNquiver} we show 
how our Boltzmann gas
picture allows one to  express
the Poincar\'e polynomial associated with quiver quantum
mechanics with
$U(N)$ factors in terms of the Poincar\'e polynomial
of Abelian quivers. This lends further support
to the validity of the Boltzmann gas picture of
multi-centered black holes.

\section{Boltzmannian view of the wall-crossing}
\label{sbolt}

\subsection{BPS states in $\cN=2$ supergravity}

We consider $\cN=2$ supergravity in 4 dimensions, coupled to $n_v$ vector multiplets. 
Let $\cM$ be the vector multiplet
moduli space, parametrized by complex scalar fields $t^a$, 
$a=1,\dots ,n_v$, and $\Gamma$ be the lattice of 
electromagnetic charges. $\Gamma$ is a lattice of rank dimension $2n_v+2$, equipped
with an integer symplectic pairing. We choose a Lagrangian decomposition 
$\Gamma=\Gamma_m \oplus \Gamma_e$, such that the symplectic pairing 
is given by 
\be
\langle \gamma, \gamma' \rangle = q_\Lambda p'^{\Lambda} - q'_\Lambda p_\Lambda \in \IZ
\ee
where $\gamma=(p^\Lambda,q_\Lambda)$, $\gamma'=(p'^\Lambda,q'_\Lambda)$. 
The mass of a BPS state with charge $\gamma$ is equal to  the absolute value of 
the central charge $Z_\gamma$, defined by 
\be
Z_\gamma = e^{\cK/2} (q_\Lambda X^\Lambda-p^\Lambda
F_\Lambda)\ ,
\ee
where $\cK$, $X^\Lambda$ and $F_\Lambda$ are
appropriate functions of the moduli fields $\{t^a\}$.
Let $\cH_\gamma(t^a)$ be the Hilbert space of states with charge $\gamma$ in the vacuum
where the scalars asymptote to $t^a$ at spatial infinity. 
The index 
\be 
\Omega(\gamma;t^a) = -\frac12 \Tr_{\cH_\gamma(t^a)} (-1)^{2J_3} (2J_3)^2 
\ee
defines an integer $\Omega(\gamma;t^a)$, which counts the number of BPS states 
with sign (the overall coefficient ensures that a half-hypermultiplet contributes one unit 
to $\Omega(\gamma;t^a)$). 
Alternatively we could define the index as
\be \label{ealt}
\Omega(\gamma ; t^a) = \Tr'_{\cH_\gamma(t^a)} (-1)^{2J_3} 
\ee
where $\Tr'$ denotes trace over BPS states, computed after
removing the contribution from the fermion zero modes
associated with the supersymmetries broken by the
state.
Mathematically, the BPS invariants $\{\Omega(\gamma;t^a), \gamma\in\Gamma\}$ 
are the generalized Donaldson-Thomas (DT) invariants for the derived 
category of coherent sheaves (in type IIA on a Calabi-Yau threefold $\cX$) or
the Fukaya category (in type IIB on a Calabi-Yau 
threefold $\cX$).

\subsection{Wall crossing: Preliminaries} \label{spril}

The BPS invariants $\Omega(\gamma;t^a)$ are locally constant functions of $t^a$, but may jump 
on codimension one subspaces of $\cM$ (line of marginal stability), where $\gamma$ can be
written as the sum $\gamma= M \gamma_1+ N\gamma_2$, where $M,N$ are two non-zero integers and $\gamma_1$
and $\gamma_2$ are two primitive (non-zero) vectors such that the phases of $Z_{\gamma_1}$ and $Z_{\gamma_2}$ are aligned. 
We denote the hyperplane 
where the phases of  $\gamma_1,\gamma_2$ are aligned by $\cP(\gamma_1,\gamma_2)$. 
Since the dependence of $\Omega(\gamma;t^a)$ on
$\{t^a\}$  is mild, we shall drop $t^a$ from the
argument of $\Omega$ and use superscripts $\pm$ to
indicate which side of $\cP(\gamma_1,\gamma_2)$
we are computing the index.

Clearly, $\cP(\gamma_1,\gamma_2)$ 
depends only on the two-plane spanned by $\gamma_1$
and $\gamma_2$ inside $\Gamma$. 
For a given choice of $\gamma$ and of this two plane, 
$(M,N)$ and $(\gamma_1,\gamma_2)$ are uniquely defined up to a common 
action of $SL(2,\IZ)$.
We shall now make a special choice of 
$(\gamma_1,\gamma_2)$ 
such that there are no BPS states carrying charges
of the form $M\gamma_1-N\gamma_2$ 
or $-M\gamma_1+N\gamma_2$ with 
$M,N>0$ \cite{Andriyash:2010qv}.
For this reason it will be convenient to introduce
the notation
\be \label{ebpsset}
\tilde\Gamma: \quad \{ M\gamma_1 + N\gamma_2,
\quad M,N\ge 0, \quad (M,N)\ne (0,0)\}\, ,
\ee
and focus on BPS states carrying charge in
$\tilde\Gamma$.
For definiteness we shall choose
$\langle \gamma_1,\gamma_2\rangle<0$.

We shall be considering the jump in the index 
$\Omega(M\gamma_1+N\gamma_2)$
across the wall $\cP(\gamma_1,\gamma_2)$ for
non-negative integers $M,N$.
We shall denote by $Z_\gamma$ the central charge
associated with the charge $\gamma$, and
choose $\Omega^+$ to describe the index
in the chamber in which $\arg(Z_{\gamma_1})>\arg(Z_{\gamma_2})$.
 In this case $\arg(Z_{M\gamma_1+N\gamma_2})$
is greater (less) than
$\arg(Z_{M'\gamma_1+N'\gamma_2})$
if $M/N$ is
greater (less) than $M'/N'$. We shall denote this
chamber by $c_+$.
For $\Omega^-$ the ordering of 
$\arg (Z_{M\gamma_1+N\gamma_2})$ is
reversed, and the corresponding chamber will be called
$c_-$.
Using the fact that $\langle\gamma_1,\gamma_2\rangle<0$
we now get in the chamber $c_+$
\be \label{etest}
\langle M\gamma_1+N\gamma_2,
M'\gamma_1+N'\gamma_2\rangle
\, \Im \left(Z_{M\gamma_1+N\gamma_2}
\bar Z_{M'\gamma_1+N'\gamma_2}\right) < 0\, .
\ee
This is the condition under which two-centered bound states of
black holes carrying charges $M\gamma_1+N\gamma_2$
and $M'\gamma_1+N'\gamma_2$ 
do not exist\cite{Denef:2000nb}.
Thus $\Omega^+$'s label the index in the chamber in which
there are no two centered black holes with each center
carrying
charge in $\tilde\Gamma$. Generalizing this argument 
(see \S\ref{scoulomb}) one can show
that in the chamber $c^+$ there are also no
multi-centered black holes carrying charges
of the form
$(M_i\gamma_1+N_i\gamma_2)$ for different $(\{M_i\},
\{N_i\})$. In contrast $\Omega^-$'s label the index in the chamber
where such bound states are present. 

Note that $\Omega^+$ can receive contribution both from
single and multi-centered black holes, but these 
multi-centered configurations consist of centers whose charges
lie outside the sublattice spanned by $\gamma_1$
and $\gamma_2$. 
Due to this 
the distances between the centers
remain finite as the wall 
$\cP(\gamma_1,\gamma_2)$ is approached.
In contrast the distance between the centers
carrying charges in $\tilde\Gamma$
-- appearing in a configuration 
contributing to $\Omega^-$ -- goes to infinity in this
limit.
Thus the configurations which contribute to $\Omega^+$,
even though not necessarily single centered black holes,
can be treated as 
a single unit near this wall of marginal stability. 
For this reason we shall refer to
$\Omega^+\argu{M}{N}$ as the index
associated with a {\it black hole molecule}
of charge $M\gamma_1+N\gamma_2$. 
Using this analogy, the full system, containing
multiple molecules loosely bound to each other near
the wall $\cP(\gamma_1,\gamma_2)$, may be thought 
of as a  {\it molecular cluster}. 
While the black hole molecule may itself be made of atoms
(i.e. single centered black holes), the nature of
these constituents is irrelevant for the problem at hand.

Our aim is to provide a wall-crossing formula 
which expresses $\Omega^-$ in terms of
$\Omega^+$.
In supergravity the difference $\Omega^--\Omega^+$ is
accounted for by the bound states of multiple 
black hole molecules carrying charges of the form $M_i\gamma_1
+N_i\gamma_2$, since they contribute to $\Omega^-$ but
not to $\Omega^+$. Our goal in the rest of this section and
the next section
will be to use this viewpoint to derive the wall-crossing
formula.

\subsection{Bose/Fermi statistics to 
Maxwell-Boltzmann statistics}
\label{sstatistics}

Let $\gamma_0$ be a primitive point on the charge lattice and let 
$d(s\gamma_0)$
be the number of bosonic states
minus the number of fermionic states
of a black hole molecule\footnote{We exclude from this counting 
the fermionic zero modes associated with broken
supersymmetry. A formal way of
doing this is to use helicity supertraces~\cite{Bachas:1996bp,Gregori:1997hi} instead of
the Witten index.} carrying charge
$s\gamma_0$ moving in some appropriate 
background. 
We shall consider a system carrying total charge 
$k\gamma_0$ consisting
of a gas of mutually non-interacting
black hole molecules  carrying charges $s\gamma_0$ for 
different integers $s$.
A typical configuration will contain $m_s$ black
hole molecules with charge $s\gamma_0$, subject to the constraint
\be \label{eas0}
\sum_s s m_s = k\, .
\ee
The net contribution to the index from such configurations is given by
\be \label{eas0.1}
N_k= \sum_{\{m_s\}\atop \sum_s sm_s=k}\prod_s 
\left[{1\over m_s!} {(d(s\gamma_0) + m_s - 1)!
\over (d(s\gamma_0)-1))!} \right]\, .
\ee
For bosons $d(s\gamma_0)>0$, and the above result follows from
the fact that $m$ identical bosons occupying $d$ states produce
a degeneracy of $d_B=d(d+1) \cdots (d+m-1)/m!$.  For fermions
$d(s\gamma_0)<0$, and the result follows from the fact that
$m$ fermions occupying $|d|$ states have total degeneracy
$d_F=(|d|) (|d|-1) \cdots (|d|-m+1) / m!$ and index 
$(-1)^m d_F = d(d+1) \cdots (d+m-1)/m!$.
It will be convenient to encode (\ref{eas0.1}) into a generating
function
\ben \label{eas0.2}
f(x) = \sum_k N_k x^k
&=& \sum_{\{m_s\}}\prod_s 
{1\over m_s!} {(d(s\gamma_0) + m_s - 1)!
\over (d(s\gamma_0)-1))!} x^{ s m_s}\nonumber \\
&=& \prod_s (1 - x^s)^{-d(s\gamma_0)}\, . 
\een

We shall now prove that exactly the same contribution to the
index is obtained if for each $\gamma$ we replace $d(\gamma)$
by 
\be \label{edefdg}
\bar d(\gamma)\equiv \sum_{m|\gamma} m^{-1}d(\gamma/m)
\ee
and treat the particles as obeying 
Maxwell-Boltzmann statistics rather than
Bose or Fermi statistics. For this we calculate the generating
function of the index of a gas of 
non-interacting Boltzmann black hole
molecules carrying
charges $s\gamma_0$ for different integers $s$. This is given by
\ben \label{eas0.3}
g(x)&=&\sum_{\{m_s\}} \prod_{s} {1\over m_s!} \, (\bar d(s\gamma_0))^{m_s} 
x^{s m_s} 
= \exp\left(\sum_s \bar d(s\gamma_0) x^s\right)
\nonumber \\
&=& \exp\left( \sum_s \sum_{m|s} d(s\gamma_0/m)
m^{-1} x^s \right)
= \exp\left(\sum_k \sum_m d(k\gamma_0) m^{-1} 
 x^{mk}\right)
\nonumber \\
&=& \exp\left(-\sum_k d(k\gamma_0) \ln (1-x^k)\right)
= \prod_k (1 - x^k)^{-d(k\gamma_0)}\, .
\een
The $m_s!$ in the denominator in the first line is the
symmetry factor required for resolving the Gibbs paradox.
Comparing \eqref{eas0.3} 
with (\ref{eas0.2}) we see that the generating functions
$f(x)$ and $g(x)$ are the same. Thus we are allowed to treat the
black hole molecules as Boltzmann particles as long as we use the 
effective index $\bar d(\gamma)$.

In general $d(s\gamma_0)$ receives contribution from the
intrinsic index $\Omega^+(s\gamma_0)$ of the black
hole molecules and from the orbital 
degeneracy describing
its motion in the background of other charges.
The contribution to the orbital part of a black hole molecule
of charge
$s\gamma_0$ is expected not to be affected by the presence
of the other black hole molecules carrying charges
$l\gamma_0$ for any integer $l$ since the symplectic product
$\langle k\gamma_0, l\gamma_0\rangle$ vanishes and as
a consequence the particles are mutually
noninteracting. In particular the repulsive electromagnetic
interactions cancel against the attractive gravitational and
scalar exchange interactions.\footnote{At short distance they 
may interact via the exchange of massive string modes and
also via dipole and higher order interactions due to
massless particle exchange, but we
do not expect these interactions to affect the analysis
of supersymmetric index.} 
On the other hand the
orbital degeneracy 
is expected to depend on the background produced
by other black hole molecules 
carrying charge not proportional to $\gamma_0$.
We shall not need the explicit form of this contribution which is
hard to compute in general  when there are multiple 
other black
hole molecules distributed in space, 
but use the fact that 
for a charged particle in a magnetic field the number of
states per unit area in the lowest Landau level
is proportional to the 
charge carried by the particle. 
To be more specific, we compare a configuration of
a molecule of charge $s\gamma_0$ moving in the background
of other molecules carrying arbitrary charges with a
configuration of $s$ closeby molecules each of charge
$\gamma_0$ moving in the same background. 
In this case the phase space volume element
for the molecule of charge
$s\gamma_0$ will be $s$ times the phase space volume 
element
for each molecule of charge $\gamma_0$.
Thus
for a fixed
background the orbital density of states for a black hole 
molecule
carrying charge
$s\gamma_0$,  being proportional to the phase space
volume element, will be $s$ times the orbital density of states of
a molecule of charge $\gamma_0$.
Thus we have $d(s\gamma_0) / d(\gamma_0) 
= s  \Omega^+(s\gamma_0) / \Omega^+ (\gamma_0)$, and
hence\footnote{Note that in this argument the 
sign of $d$, and hence the
statistics of the particle, is determined by the sign of
$\Omega^+$. Although the orbital angular momentum
contributes to the index of the final configuration, they arise
from the electromagnetic field, and hence do not affect the
statistics of the individual particles in the halo.
}
\be \label{eas2}
\bar d(s\gamma_0)/ d(\gamma_0) = 
\sum_{m|s} m^{-1} d(s\gamma_0/m) / d(\gamma_0)
=  s\, \sum_{m|s} m^{-2} \Omega^+(s\gamma_0/m) 
/ \Omega^+(\gamma_0)\, .
\ee
Comparing this with 
\be \label{eas3}
d(s\gamma_0) / d(\gamma_0) =  s \, \Omega^+(s\gamma_0)
/\Omega^+(\gamma_0)\, 
\ee
we see that replacing
$d(\gamma)$ by $\bar d(\gamma)$ is equivalent to replacing
$\Omega^+(\gamma)$ by 
\be \label{eas4}
{\bar\Omega^+(\gamma) = \sum_{m|\gamma} m^{-2} 
\Omega^+(\gamma/m)
\, .}
\ee
We shall see in \S\ref{sec_KS} 
that the fractional DT invariants $\bar\Omega(\gamma)$
arise naturally in the 
KS wall-crossing formula.

We end this section with a word of caution. 
For a generic interacting
system of bosons or fermions
the effect of statistics and interaction cannot always be
disentangled to map the problem to an equivalent problem
with Boltzmann particles. Consider for example the case
of an interacting system of two identical particles
for which at a certain energy eigenvalue $E$ we have $n_s$
symmetric and $n_a$ anti-symmetric wave-functions.
In this case we 
shall not get any simple map from the bosonic or fermionic
system to  a system of Boltzmann particles. Only if the
identical particles are non-interacting so that multi-particle
wave-functions can be constructed from (anti-)symmetric
products of single particle wave functions, we have a
simple map from a Bose/Fermi gas to a Boltzmann gas.

\subsection{General wall-crossing formula
and charge conservation} \label{scharge}

The analysis of \S\ref{sstatistics} leads to the following
prescription for computing wall-crossing from supergravity
black holes. Suppose in the  chamber $c_-$
we have a configuration
of multi-centered black hole
molecules, consisting of $m_{r,s}$ centers
of charge $(r\gamma_1+s\gamma_2)$ for different values
of $(r,s)$. 
These molecules interact via long range electromagnetic,
gravitational and other massless particle exchange
interactions.
We now consider a quantum mechanics of this system
regarding the different centers as {\it distinguishable particles}, 
each with
unit degeneracy, and
denote by $g(\{m_{r,s}\})$ trace of $(-1)^{2J_3}$
in this
quantum mechanics. 
Then
the wall-crossing formula will be given by
\ben \label{egenwall}
&&\Delta\bar\Omega\argu{M}{N}
\equiv
\bar\Omega^-\argu{M}{N}
- \bar\Omega^+\argu{M}{N}\nonumber \\
&=&
{\sum_{\{m_{r,s}\}\atop
\sum_{r,s} r m_{r,s}=M, \sum_{r,s} s m_{r,s}=N} \, 
\hh(\{m_{r,s}\}) \prod_{r,s} \left[
{1\over (m_{r,s})!} \, \left(\bar\Omega^+(r\gamma_1 + s\gamma_2)
\right)^{m_{r,s}}
\right]}\, .\nonumber \\
\een
For $\gcd(M,N)=1$ we have 
$\bar\Omega^\pm\argu{M}{N}
= \Omega^\pm\argu{M}{N}$.
Eq. (\ref{egenwall}) then follows from the fact that
the left hand side represents the change in the index 
and the right hand side represents the total contribution from the
bound states of black hole molecules
which exist in the chamber $c_-$ but
not in the chamber $c_+$.
For $\gcd(M,N)>1$ the 
indices $\Omega^\pm\argu{M}{N}$ are
somewhat ill defined since the total index in the
sector of charge $(M\gamma_1+N\gamma_2)$ can receive
contribution also from unbound multiparticle states
carrying charges $(M\gamma_1+N\gamma_2)/s$
for integers $s\, |(M, N)$. 
Thus the unambiguous quantity is the effective 
index which enters the formula for the index of a bigger
system of which the system with charge $(M\gamma_1
+N\gamma_2)$ may form a subsystem
\cite{Andriyash:2010qv}.
This is what we call $\bar\Omega^\pm$  and this
is the quantity whose
jump is computed by the right hand side of (\ref{egenwall}).

A slightly different way of expressing \eqref{egenwall}
is:
\be \label{esab2}
\bar\Omega^-(\gamma) -\bar\Omega^+(\gamma) =
\sum_{n\geq 2 }\, 
  \sum_{\substack{ \{\alpha_1,\dots, \alpha_n\in
  \tilde\Gamma\} \\
\gamma= \alpha_1+\dots +\alpha_n}}\, 
\frac{g(\{\alpha_i\})}{|{\rm Aut}(\{\alpha_i\})|}
 \prod\nolimits_{i=1}^n \bOm^+(\alpha_i) \ ,
 \ee
 where the sum runs over all possible unordered
 decompositions
 of $\alpha$ into vectors $\alpha_1,\dots, \alpha_n\in
 \tilde\Gamma$. The integer $|{\rm Aut}(\{\alpha_i\})|$ is defined as
 follows. If all the $\alpha_i$'s are distinct then
 $|{\rm Aut}(\{\alpha_i\})|=1$. If on the other hand
 the set $(\alpha_1,\dots,\alpha_n)$ consists
 of $m_1$ copies of a vector $\beta_1$, $m_2$ 
 copies of a vector $\beta_2$ etc. then 
 $|{\rm Aut}(\{\alpha_i\})| =\prod_a m_a!$.
Finally $g(\alpha_1,\dots, \alpha_n)$ represents
the index associated with an $n$-centered black hole
configuration in
supergravity, with the individual centers carrying
charges $\alpha_1,\alpha_2,\dots, \alpha_n$ and 
treated as {\it distinguishable particles carrying
no internal degeneracy}. Note that by 
an abuse of notation we
have used for the argument of $g$ two
different representations of the index of multi-centered
black holes -- one where the arguments are
charges carried by individual centers and the other
where the arguments are integers specifying how
many of the centers  carry
a given charge vector $r\gamma_1+s\gamma_2$.

An immediate consequence of (\ref{esab2})
is `charge conservation' -- the sum of the charges
appearing in the arguments of $\bar\Omega^+$ on the right
hand side of the equation is equal to the argument of
$\Delta\bar\Omega$ on the left hand side of this equation.
In contrast if we had written the wall-crossing formula
using the indices $\Omega^+$ on the right hand side then
there is no manifest charge conservation. This is a 
consequence of the fact that the use of $\bar\Omega$ allows
us to use 
Maxwell-Boltzmann statistics for computing the
contribution to the index due to multiple black hole
molecules.
In contrast if we had used Bose or Fermi statistics
then manifest charge conservation is spoiled
by the symmetrization effect since the degeneracy of
$k$ identical particles carrying index $\Omega^+$ not only
contains a term proportional to $(\Omega^+)^k$ but also
other terms containing lower powers of $\Omega^+$.

\subsection{Semi-primitive wall-crossing from Boltzmann
gas  of black hole molecules} \label{sgas}

In this section we shall derive  
the semi-primitive
wall-crossing formula by counting the index of a
gas of black hole molecules carrying charges $s\gamma_2$
for different integers $s$, forming a halo
around another black hole molecule
of primitive charge $\gamma_1$. We denote by 
$k\gamma_2$ the
total charge carried by the black hole gas. As noted in
section \S\ref{sstatistics}, for this calculation we can regard
the gas as one obeying 
Maxwell-Boltzmann statistics as long as we
replace the index $\Omega^+(s\gamma_2)$ of a black hole
molecule
carrying charge $s\gamma_2$ by 
$\bar\Omega^+(s\gamma_2)$.
The orbital motion of a black hole molecule of charge
$s\gamma_2$ around a molecule of charge $\gamma_1$
produces states carrying angular momentum 
$(|\langle \gamma_1, s\gamma_2\rangle| -1)/2$, and gives 
a contribution $(-1)^{\langle \gamma_1, 
s\gamma_2\rangle +1}
|\langle \gamma_1, s\gamma_2\rangle|$ to the 
index\cite{Denef:2007vg}.
Taking into account this additional factor  
we get the total contribution to the
index from a single black hole molecule
of charge $s\gamma_2$ to be
\be \label{eas1.1}
(-1)^{\langle \gamma_1, s\gamma_2\rangle +1}\,
|\langle \gamma_1, s\gamma_2\rangle|\,
\bar\Omega^+ (s\gamma_2)\, .
\ee
Since we have already chosen 
$\langle \gamma_1, \gamma_2\rangle$ to be negative we
can drop the absolute value sign and the $+1$ from the exponent.
Thus if the halo consists of $m_s$ black hole
molecules of charge
$s\gamma_2$ then the net contribution to the index is
\be \label{eas1.2}
\prod_s \left[{1\over m_s!} \, \left((-1)^{\langle \gamma_1, s\gamma_2\rangle}
\langle \gamma_1, s\gamma_2\rangle\, 
\bar\Omega^+ (s\gamma_2)\right)^{m_s}\right]\, .
\ee
Summing over all possible values of $m_s$ subject to the
condition $\sum_s s m_s = N$, 
and multiplying this by the index $\Omega^+(\gamma_1)$
of the black hole molecule of charge $\gamma_1$
we get a total
contribution $\Omega^+(\gamma_1)\, \Omega_{\rm halo}(\gamma_1,N)$, where
\be \label{egk1}
\Omega_{\rm halo}(\gamma_1,N) \equiv 
\sum_{\{m_s\}\atop \sum_s s m_s = N}
\prod_s \left[ {1\over m_s!} \, \left((-1)^{\langle \gamma_1, s\gamma_2\rangle}
\langle \gamma_1, s\gamma_2\rangle\, 
\bar\Omega^+ (s\gamma_2)\right)^{m_s}\right] \, .
\ee
This is the jump in the index due to a bound state
of a black hole molecule of charge $\gamma_1$ and a halo of
black hole molecules carrying charges $s\gamma_2$ for 
different integers $s$.
In order to calculate the total change in the index
in the sector of charge $\gamma_1 + N\gamma_2$
across the wall of marginal stability
$\cP(\gamma_1,\gamma_2)$, we need to sum over
all possible bound states containing a core of charge
$\gamma_1+l\gamma_2$ and a halo of total charge
$(N-l)\gamma_2$. Thus we have to sum over several
terms of the form (\ref{egk1}) with $\gamma_1$ replaced
by $\gamma_1+ l\gamma_2$ and $N$ replaced by $(N-l)$ for
different integers $l$. 
This gives
\be \label{eomp1}
\Delta
\bar\Omega\argu{}{N}
= \sum_{l=0}^{N-1} 
\bar\Omega^+\argu{}{l} \, \Omega_{\rm halo}(\gamma_1,N-l)
\ee
where we have used the primitivity of
$\gamma_1 + \ell\gamma_2$ to replace
$\Omega^\pm(\gamma_1+\ell\gamma_2)$ by
$\bar \Omega^\pm(\gamma_1+\ell\gamma_2)$
for $0\le \ell\le N$.
This can be formalized in terms of the partition function 
\be \label{epart}
\bar Z^\pm(1,q) = 
\sum_{N=0}^\infty \bar\Omega^\pm\argu{}{N}\, q^N\, .
\ee
Then (\ref{eomp1}) can be written as
\be \label{epart2}
{\bar Z^-(1,q) = \bar Z^+(1,q)\, Z_{\rm halo}
(\gamma_1,q)\,
, }\ee
where
\be \label{edefhalo}
{Z_{\rm halo}(\gamma_1,q)=
\sum_{N\geq 0} \Omega_{\rm halo}(\gamma_1,N)\, q^N = 
\exp\left(\sum_{s=1}^{\infty} q^s (-1)^{\langle \gamma_1, 
s\gamma_2\rangle} \langle \gamma_1, s\gamma_2\rangle
\bar\Omega^+ (s\gamma_2)
\right)\, .}
\ee
We shall see later that this agrees with  
the KS and JS wall-crossing formula restricted
to the semi-primitive case 
(eqs.\eqref{Z1pm}, \eqref{Z1pmh} and \eqref{jsjs3}).
For $N=1$ using $\gamma_{12}<0$, 
we recover the primitive wall-crossing formula \eqref{primwcf}.

To recover the semi-primitive wall-crossing formula of
\cite{Denef:2007vg} from \eqref{epart2}, \eqref{edefhalo},
we proceed as follows.
First of all we note that the relation \eqref{eas4}
can be inverted as
\be
 \Omega(\gamma)   = 
 \sum_{d\vert\gamma} \frac{1}{d^2}\, \mu(d)  \, 
 \bOm(\gamma/d)\ ,
\ee
where $\mu(d)$ is the M\"obius function (i.e. 1 if $d$ is a product of an even number of distinct primes, $-1$ if $d$ is a product of an odd number of primes, or $0$ otherwise). 
Using 
the identity $\prod_{d=1}^{\infty}(1-q^d)^{\mu(d)/d}
=e^{-q}$ 
we can now express \eqref{edefhalo} as
\be\label{spwcf}
Z_{\rm halo}(\gamma_1,q)=  \prod_{k>0} 
\left( 1- (-1)^{k \gamma_{12}} q^k\right)^{ k \, 
|\gamma_{12}|\, 
\, \Omega^+(k\gamma_2)}\ .
\ee
Eqs.\eqref{epart2}, \eqref{spwcf} give precisely the
semi-primitive wall-crossing formula of
\cite{Denef:2007vg}.

\subsection{Refined `index' in supergravity} \label{smotivic}

Kontsevich and Soibelman also analyzed the wall 
crossing formula for the motivic DT invariants, which are quantized versions  
of the numerical DT
invariants $\Omega(\gamma;y)$. They enumerate the Betti numbers of the
moduli space of BPS-states for given charge $\gamma$ 
in the weak string coupling regime, whereas 
$\Omega(\gamma)$ 
equals
the Euler characteristics of this moduli space, 
up to a sign. Physically, the motivic DT invariants keep track of the angular 
momentum quantum numbers carried by the black hole
at weak string coupling where the system may be
represented as a collection of D6-D4-D2-D0 branes
wrapped on a Calabi-Yau 
3-fold\cite{Dimofte:2009bv,Dimofte:2009tm}. 
A simple way to do this is to introduce 
an extra factor
of $y^{2J_3}$ inside the trace in \eqref{ealt} 
besides the $(-1)^{2J_3}$ factor
that is already present in this definition of the
index.
Thus at a given point in moduli space, 
the refined `index' \footnote{We shall use the words
motivic and refined interchangeably.} 
$\Omega_{\rm ref}({\gamma,y})$ is 
defined by \cite{ks,Dimofte:2009bv,Dimofte:2009tm}
\be
\label{refdeg}
\Omega_{\rm ref}(\gamma,y) = 
\Tr'_{\cH(\gamma)} (-y)^{2J_3} 
\equiv \sum_{n\in \IZ} (-y)^n \, 
\Omega_{{\rm ref},n}(\gamma) \, ,
\ee
where $\Tr'$ denotes the trace over BPS states computed
after removing the contribution from the fermion
zero modes associated with the supersymmetries
broken by the black hole. Alternatively we could compute
the ordinary
trace over all the BPS states and then divide the result
by $(2-y-y^{-1})$ which represents the contribution from the
fermion zero modes.
The usual generalized DT invariants are obtained by setting $y=1$,
\be
\Omega(\gamma)\equiv \sum_{n\in\IZ} (-1)^n \Omega_{{\rm ref},n}(\gamma) 
= \Omega_{\rm ref}(\gamma,y=1)\ .
\ee

In string theory \eqref{refdeg} is not
an index since it is not protected against quantum
corrections as we switch on the string coupling.
In supersymmetric gauge theories one can define
an alternative version of this index
as \cite{Gaiotto:2010be}, 
\be
\label{refdegalt}
\Omega'_{\rm ref}(\gamma,y) = 
\Tr_{H(\gamma)} (2J_3) (-1)^{2J_3}(-y)^{2I_3+2J_3} 
\equiv \sum_{n\in \IZ} (-y)^n \, \Omega'_{
{\rm ref},n}(\gamma) 
\ee
where $I_3$ is the third component of the
$SU(2)_R$ symmetry. This is
protected against quantum corrections. We shall however
proceed with the definition \eqref{refdeg} since our
main interest is in string theory. 
Even though there is no general argument that protects this `index'
from changing as we vary the string coupling, and hence
the DT invariants and the black hole degeneracies may not
be equal,  we may nevertheless expect that the structure of
the wall-crossing formula at fixed coupling will remain the
same.  
Thus
we can regard the motivic KS formula as giving the
change in $\Omega_{\rm ref}(\gamma,y)$ across a
wall of marginal stability at fixed value of the string
coupling.
With this in mind we shall analyze the jump in the
motivic `index' in supergravity and then compare this
with the KS formula.
Our supergravity 
analysis, compared to that in \S\ref{sstatistics},
will be somewhat heuristic; however the final result
of this analysis will turn out to be consistent with
the KS motivic wall-crossing formula.

We begin by introducing some notations. We have
already denoted
by $\Omega_{\rm ref}(\gamma,y)$ the refined `index' 
computed by introducing a weight factor of $y^{2J_3}$
into the trace in \eqref{ealt}. More generally we shall use
the subscript $~_{\rm ref}$
to denote various quantities in which the trace over different
angular momentum states
has been performed
with a weight factor of $y^{2J_3}$.
With this the analog
of \eqref{edefdg} takes the form
\be \label{edefc}
\bar { d}(\gamma, J_3) = \sum_{m|\gamma, 2J_3}
m^{-1} \,  d(\gamma/m, J_3/m)\, .
\ee
A word of caution is warranted here.
Since the full system is rotationally invariant,  the
states of this system
can be characterized by their angular momentum. 
However when we examine the 
motion of one subsystem in the background
of the other, the background generically breaks rotational
invariance and hence states can no longer 
be classified by their angular momentum unless
the background is generated by a point source (or
a spherically symmetric source).
We shall nevertheless
proceed as if each subsystem consisting of a set
of identical particles moved in the background
produced by a point source so that
an assignment of angular momentum quantum
numbers to such individual subsystems were possible.
Based on this assumption we shall
arrive at an expression for the motivic index of
the whole system in terms of
the index carried by the individual molecules.
This procedure can be justified {\it a posteriori} by
the fact that it allows for a
physical understanding of the motivic generalization 
of the KS wall-crossing formula.

After multiplying \eqref{edefc} 
by $y^{2J_3}$ and summing over $J_3$ we
get
\ben \label{edefbtg}
\bar d_{\rm ref}(\gamma,y) &=&
\sum_{J_3} \sum_{m|\gamma, 2J_3}
m^{-1}  d(\gamma/m, J_3/m) y^{2J_3}
= \sum_{m|\gamma} \sum_{J_3'}m^{-1}
  d(\gamma/m, J_3')
y^{2mJ_3'}\nonumber \\
&=&
\sum_{m|\gamma} m^{-1} d_{\rm ref}(\gamma/m, y^m)\, .
\een
Our next task is to find the generalization of \eqref{eas2}.
Let us denote by
$ d_{\rm orb}(\gamma, J_3)$  the degeneracy due
to orbital motion of a black hole molecule of
charge $\gamma$ in some fixed background.
Again we pretend that the background is spherically
symmetric so that it makes sense to assign definite
angular momentum quantum numbers to the orbital states
of individual subsystems.
Then we have
\be \label{exu1}
 d_{\rm ref}(\gamma,y) =  \Omega_{\rm ref}^+(\gamma, y)
 d_{\rm orb;ref} (\gamma, y)\, ,
\ee
where 
\be \label{exu2}
d_{\rm orb;ref} (\gamma, y) = \sum_{J_3} 
 d_{\rm orb}(\gamma, J_3) \, y^{2J_3}\, .
\ee
Eq. \eqref{edefbtg} and \eqref{exu1} now give
\be \label{exu3}
\bar d_{\rm ref}(\gamma,y) = \sum_{m|\gamma} 
m^{-1}\, \Omega_{\rm ref}^+(\gamma/m, y^m)\, 
d_{\rm orb;ref} (\gamma/m, y^m)\, .
\ee
We shall now try to express 
$d_{\rm orb;ref} (\gamma/m, y^m)$ in
terms of $d_{\rm orb;ref} (\gamma, y)$. For this 
(still pretending that we have a rotationally invariant
subsystem) we shall decompose the 
orbital spectrum 
into $SU(2)$
representations and denote by $b(\gamma,J)$ the 
coefficient of the character of the representation of 
spin $J$ .
Then
we have
\be \label{exu4}
d_{\rm orb;ref} (\gamma, y) = \sum_J b(\gamma, J)
(y^{2J} + y^{2J-2} +\cdots + y^{-2J})
= \sum_J b(\gamma, J) { y^{2J+1} - y^{-2J-1}\over 
y - y^{-1}}
\, .
\ee
We now use the fact that for any $m\in\bZ^+$, and
for spherically symmetric background,
we have
\be \label{exu5}
b(\gamma, J) = b(m\gamma, J'), \qquad 2J'+1 = m(2J+1)\, .
\ee
Effectively \eqref{exu5} 
follows from the fact that increasing the charge of the
molecule by a factor of $m$ changes the angular momentum
carried by the lowest Landau level such that the
degeneracy of the Landau level gets scaled by a factor
of $m$. 
Using this we get
\be \label{exu6}
d_{\rm orb;ref} (\gamma/m, y^m)
= \sum_{J'} b(\gamma, J') \, 
\frac{y^{(2J'+1)} - y^{-(2J'+1)}}
{y^m - y^{-m}}
= 
\frac{y - y^{-1}}{y^m - y^{-m}}\, 
d_{\rm orb;ref} (\gamma, y) \, .
\ee
Substituting this into \eqref{exu3} we arrive at
\be \label{exu7}
\bar d_{\rm ref}(\gamma,y) =
\bar\Omega_{\rm ref}^+(\gamma,y)
d_{\rm orb;ref} (\gamma, y)\, ,
\ee
where the ``rational motivic invariants" 
$\bar \Omega_{\rm ref}$ are defined by 
\be \label{bOmref}
{
\bar \Omega_{\rm ref}(\gamma,y) 
\equiv \sum_{m|\gamma} 
\frac{y - y^{-1}}{m\, (y^m - y^{-m})}\, 
\Omega_{\rm ref}(\gamma/m, y^m) 
\, .}
\ee
This shows that in computing the refined index of the full system
we can treat the particles as obeying Maxwell-Boltzmann statistics
provided we replace $\Omega_{\rm ref}$ by $\bar\Omega_{\rm ref}$. 
As in the case of the classical DT invariants, the use of these invariants
ensures that only charge preserving terms appear in any 
wall-crossing formula. 

The rational motivic invariants have appeared  earlier
in other contexts, for example in the construction
of modular invariant
partition functions in \cite{Manschot:2010nc}. 
We would also like to point out that their
structure is very similar to the free energy which arises in the computations by Gopakumar and Vafa
\cite{Gopakumar:1998jq}. The only difference is that in the
Euclidean setup of \cite{Gopakumar:1998jq}, 
the factor $(y^d-y^{-d})$
on the right-hand side of (\ref{bOmref}) is replaced by
$(y^d-y^{-d})^2$. As in this case, the generating function of
the rational invariants $\bOm_{\rm ref}$ 
leads to a product formula
\be
\sum_{\gamma} \frac{\bar \Omega_{\rm ref}(\gamma,y)}{y-y^{-1}}\,e^{-\gamma\cdot \phi}={\rm
  log} \left[ \prod_{n\in \mathbb{Z}, \ell\geq 0 \atop \gamma} \left(
    1-y^{1+n+2\ell}\,e^{-\gamma\cdot \phi}\right)^{(-1)^n\Omega_{{\rm ref},n}(\gamma)}\right]\ ,
\ee
where $\phi$ is a vector of chemical potentials conjugate to the charge vector $\gamma$.
Note that this product structure is lost in the limit $y\to 1$.

The analog of \eqref{egenwall} now takes the 
form\footnote{Since a single centered
BPS black hole is expected to carry zero angular 
momentum\cite{Sen:2009vz,Sen:2010mz,Dabholkar:2010rm},
one might naively expect $\bOm^+_{\rm ref}(\gamma,y)$
to be independent of $y$. However
as discussed in \S\ref{spril}, we allow for centers which consist 
of multi-centered black holes whose relative
separation remains finite as we approach the wall of
marginal stability. As a result $\Omega_{\rm ref}(\gamma,y)$
can be a non-trivial function of $y$.}
\be \label{exu9}
\begin{split}
&\Delta\bar\Omega_{\rm ref}\argy{M}{N}{y}
\equiv
\bar\Omega_{\rm ref}^-\argy{M}{N}{y}
- \bar\Omega_{\rm ref}^+\argy{M}{N}{y} \\
=&
{\sum_{\{m_{r,s}\}\atop
\sum_{r,s} r m_{r,s}=M, \sum_{r,s} s m_{r,s}=N} \, 
g_{\rm ref}(\{m_{r,s}\}, y) \prod_{r,s} \left[
{1\over (m_{r,s})!} \, \left(\bar\Omega_{\rm ref}^+(r\gamma_1 
+ s\gamma_2, y)
\right)^{m_{r,s}}
\right]}\, ,
\end{split}
\ee
where
$g_{\rm ref}(\{m_{r,s}\}, y)$ 
measures $\Tr(-y)^{2J_3}$ from orbital motion
of a set of distinguishable particles, containing
$m_{r,s}$ number of particles carrying charges
$r\gamma_1+s\gamma_2$.
Similarly the analog of 
\eqref{esab2} is
\be \label{esab3}
\bar\Omega^-(\gamma,y) -\bar\Omega^+(\gamma,y) =
\sum_{n\geq 2 }\, 
  \sum_{\substack{ \{\alpha_1,\dots, \alpha_n\in
  \tilde\Gamma\} \\
\gamma= \alpha_1+\dots +\alpha_n}}\, 
\frac{g_{\rm ref}(\{\alpha_i\}, y)}{|{\rm Aut}(\{\alpha_i\})|}
 \prod\nolimits_{i=1}^n \bOm^+_{\rm ref}(\alpha_i, y) \ ,
 \ee
consistently with the  charge conservation property of the 
motivic wall
crossing formula when
expressed in terms of the combinations
$\bar\Omega_{\rm ref}^\pm$. Note that even though
our derivation of \eqref{esab3} has been marred 
by unreasonable assumption of spherical symmetry
in the dynamics of various subsystems, each term
in \eqref{esab3} is defined unambiguously so that
it can be put to test against known results.

We can also easily derive the semi-primitive version of
the motivic wall-crossing formula by following the
logic of \S\ref{sgas}. It takes the form\footnote{Note
that in the semi-primitive case our heuristic derivation becomes
rigorous since identical particles carrying charge 
$s\gamma_2$ move in the spherically symmetric
background produced by the charge 
$\gamma_1+l\gamma_2$. 
Since \cite{Andriyash:2010qv} argues that
general wall-crossing formula can be derived from the
semi-primitive formula, we can use this to justify our
general claim \eqref{esab3}.
}
\be
 \label{eomp1ref}
 \begin{split}
&\Delta\Omega_{\rm ref}\argy{}{N}{y}= \\
& {\sum_{l=0}^{N-1} 
\bar\Omega_{\rm ref}^+\argy{}{l}{y}
\sum_{\{m_s\}\atop \sum_s s m_s = N-l}
\prod_s \left[ {1\over m_s!} \, \left(\left({(-y)^{\langle \gamma_1, s\gamma_2\rangle} - (-y)^{-\langle \gamma_1, 
s\gamma_2\rangle}\over y - y^{-1}}\right)
\bar\Omega_{\rm ref}^+ (s\gamma_2,y)\right)^{m_s}
\right] \, .}
\end{split}
\ee
We shall see later that this is in perfect agreement with the
prediction of KS motivic wall-crossing formula
\eqref{ezhalo}.

\section{Multi-black hole bound states
and quiver quantum mechanics} \label{squiver}

In order to have a complete wall-crossing formula we need
to find explicit expressions for the functions 
$g(\{\alpha_i\})$, $g_{\rm ref}(\{\alpha_i\},
y)$ 
appearing in eqs.\eqref{esab2} and \eqref{esab3}
respectively. This requires finding the spectrum
of supersymmetric bound states of multi-black
hole configurations in supergravity.
As argued by Denef\cite{Denef:2000nb}, 
the supersymmetric 
quantum mechanics of multi-centered BPS configurations
can be viewed as the ``Coulomb phase" of a quiver matrix model, valid at strong coupling. 
At weak coupling, the
wave function is instead supported on the Higgs branch. It should be possible to compute
the function $g(\{\alpha_i\})$, $g_{\rm ref}
(\{\alpha_i\},y)$ from either description. 
In this section we shall describe both these
approaches.
 As we shall see, the description on the Higgs branch,
 described in \S\ref{shiggs}, allows us to solve the problem
 completely.
On the other hand the description on the Coulomb branch,
described in \S\ref{scoulomb}, also gives a complete
algorithm for finding $g(\{\alpha_i\})$, but it is
more difficult to solve it explicitly. 
Furthermore with some guesswork we can also
arrive at a specific proposal for $g_{\rm ref}(\{\alpha_i\},y)$
from the analysis on the Coulomb branch.

Without any loss of generality we can arrange the
$\alpha_i$'s so that 
\be \label{esa2}
\alpha_{ij} \equiv 
\langle\alpha_i, \alpha_j\rangle
\ge 0 \quad \hbox{for} \quad i<j\, .
\ee
Now if we 
represent a vector $M\gamma_1+N\gamma_2$ 
in $\tilde\Gamma$
by the point $(M,N)$ in the
Cartesian coordinate system, then in this representation
a pair of vectors $(\alpha,\beta)$ will follow clockwise
(anti-clockwise) order if $\langle\alpha,\beta\rangle$ is
positive (negative). 
The condition \eqref{esa2} then implies that
the vectors $\alpha_1,\cdots \alpha_n$ are arranged
in a clockwise fashion. 
Throughout this section we shall work with this
particular ordering of the $\alpha_i$'s.

\subsection{Higgs branch analysis} \label{shiggs}

As has been argued by Denef \cite{Denef:2002ru},
the bound state spectrum of multi-centered black holes
can also be computed using quiver quantum mechanics.
For computing $g_{\rm ref}(\{\alpha_i\},y)$ 
we need to study the
bound states of $n$ distinguishable particles. In this
case the quiver takes a simple form with $n$-nodes
each carrying a $U(1)$ factor, and $\alpha_{ij}$ arrows
from the node $i$ to the node $j$ for $i<j$. In particular
since the arrows always go from lower to higher
node, there are no oriented loops.

Now for quivers without oriented loops,
Reineke \cite{MR1974891} 
has computed the generating function of the 
Betti numbers
of the space of semi-stable solutions to the D-flatness conditions.
Physically they determine the number of 
supersymmetric quantum
states carrying given angular momentum $J_3$.
The formula takes a simple form when all nodes carry
$U(1)$ factors and we shall state the formula for this
special case. According to this formula
$\Tr\left( (-y)^{2J_3}\right)$, which can be
identified with the function $g_{\rm ref}(\{\alpha_i\}, y)$,
 is given by
\begin{equation} 
\label{thebigformula}
 g_{\rm ref}(\alpha_1,\dots, \alpha_n, y)=(-y)^{-L} \,
 (y^2-1)^{1-n} \, 
 \sum_{\rm partitions}(-1)^{s-1} y^{2\sum_{a\leq b}\sum_{j<i }
 \alpha_{ji}\, 
 m^{(a)}_i \, m^{(b)}_j }\, ,
\end{equation}
where the
sum runs over all ordered partitions of 
$(\alpha_1+\cdots + \alpha_n)$ into $s$ vectors
$\beta^{(a)}$ ($1\le a\le s$, $1\le s\le n$)
such that
\begin{enumerate}
\item $\sum_a \beta^{(a)} = \alpha_1+\cdots + \alpha_n$.
\item $\beta^{(a)} = \sum_i m^{(a)}_i \alpha_i$ with
$m^{(a)}_i=0$ or $1$ for each $a,i$.
\item 
$\arg \left( \sum_{a=1}^b \, Z_{\beta^{(a)}}\right) >
\arg \left( Z_{\alpha_1+\cdots + \alpha_n}\right)$ for all
$b$ between 1 and $s-1$ in the chamber $c^-$.
\end{enumerate}
Using the fact that in the chamber $c^-$ 
$\arg(Z_\alpha) > \arg(Z_\beta)$ implies $\langle
\alpha,\beta\rangle>0$, we can express condition 3. as
\be \label{econdombeta}
\left\langle \sum_{a=1}^b \, \beta^{(a)}, 
\alpha_1+\cdots + \alpha_n\right\rangle > 0 \quad
\forall \quad b \quad \hbox{with} \quad 1\le b\le s-1\ .
\ee
In \eqref{thebigformula} 
$L$ is a constant given by 
\be \label{edefL}
L = (1-n) + \sum_{i<j} \alpha_{ij}\, ,
\ee
in such a way that \eqref{thebigformula} is 
invariant under $y\to y^{-1}$.
Physically $L$ represents the maximum $2J_3$
eigenvalue that the system can carry.

We shall now illustrate how this formula works by
computing $g_{\rm ref}(\{\alpha_i\},y)$ for $n=2,3$
and 4.
First consider the $n=2$ case with $\alpha_{12}>0$.
In this case we have two possible ordered
partitions satisfying the conditions stated above:
\be \label{etwopart}
\{ \alpha_1+\alpha_2\}, \quad \{\alpha_1, \alpha_2\}
\, .
\ee
The first term contributes $y^{2\alpha_{12}}$ and the 
second
term contributes $-1$ to the sum. In total,
\be \label{erei2}
g_{\rm ref}(\alpha_1,\alpha_2,y) = (-y)^{1-\alpha_{12}}\, 
(y^2-1)^{-1} \, 
(y^{2\alpha_{12}} - 1)
= (-1)^{\alpha_{12}+1}\,
{y^{\alpha_{12}} - y^{-\alpha_{12}}
\over y - y^{-1}}\, .
\ee
This correctly produces the spectrum of two centered
bound states. Taking the $y\to 1$ limit gives
\be \label{eyto1}
g(\alpha_1,\alpha_2)= (-1)^{\alpha_{12}+1}\, \alpha_{12}.
\ee

Let us now turn to the $n=3$ case. 
We take
$\alpha_1$, $\alpha_2$, $\alpha_3$ to be 
three distinct elements (not necessarily primitive)
of $\tilde\Gamma$ such that
$\alpha_{12}$, $\alpha_{13}$ and $\alpha_{23}$
are all positive. For definiteness we shall choose the
$\alpha_i$'s such that $\alpha_{12}>\alpha_{23}$. 
In the convention described above
\eqref{esa2}, the charges listed below follow
a clockwise order as we move from left to right:
\be \label{eorder}
\alpha_1,\quad (\alpha_1+\alpha_2, \quad \alpha_1 +\alpha_3), 
\quad  \alpha_1 +\alpha_2
+ \alpha_3,  \quad 
\alpha_2, \quad \alpha_2+\alpha_3, \quad \alpha_3\, .
\ee
The relative ordering of the vectors inside (~)
is not determined unambiguously but is unimportant. The 
condition \eqref{econdombeta} is now easy to implement:
for every $b$, $\sum_{a=1}^b \beta^{(b)}$ must be one
of the vectors to the left of $\alpha_1+\alpha_2+
\alpha_3$ in the list \eqref{eorder}.
In this case the possible ordered
partitions
of $\alpha_1+\alpha_2+\alpha_3$
satisfying \eqref{econdombeta} are:
\be \label{erei4}
\{\alpha_1+\alpha_2+\alpha_3\}, \quad
\{\alpha_1, \alpha_2+\alpha_3\}, \quad
\{\alpha_1+ \alpha_2, \alpha_3\}, \quad
\{\alpha_1+ \alpha_3,\alpha_2\}, \quad
\{\alpha_1, \alpha_2, \alpha_3\}, \quad
\{\alpha_1, \alpha_3, \alpha_2\}\, .
\ee
This gives, after a cancelation between the second and the
last contribution,
\be \label{erei5}
\begin{split}
g_{\rm ref}(\alpha_1,\alpha_2,\alpha_3, y)
=& (-1)^{\alpha_{12}+\alpha_{13}+\alpha_{23}}\,
(y - y^{-1})^{-2} \, \\ &
\left( y^{\alpha_{12}+\alpha_{13}+\alpha_{23}}
- y^{\alpha_{12}-\alpha_{23} -\alpha_{13}} 
- y^{\alpha_{13}
+\alpha_{23}-\alpha_{12}} 
+y^{-\alpha_{12}-\alpha_{13}-\alpha_{23})} 
\right)  \\
=& (-1)^{\alpha_{12}+\alpha_{13}+\alpha_{23}}\,
{1\over \sinh^2\nu}
\sinh(\nu\alpha_{12}) \sinh(\nu(\alpha_{13}+
\alpha_{23}))\ ,
\end{split}
\ee
where $\nu\equiv\ln y$. Taking the
$y\to 1$ limit we get
\be \label{eyto2}
g(\alpha_1,\alpha_2,\alpha_3)
= (-1)^{\alpha_{12}+\alpha_{13}+\alpha_{23}}\, 
\alpha_{12} \, (\alpha_{13}+
\alpha_{23})\, .
\ee
If instead $\alpha_{12}<\alpha_{23}$, a similar reasoning leads to 
\beq \label{eyto3}
g_{\rm ref}(\alpha_1,\alpha_2,\alpha_3, y)
&=&(-1)^{\alpha_{12}+\alpha_{13}+\alpha_{23}}\nn\,
{1\over \sinh^2\nu}
\sinh(\nu\alpha_{23}) \sinh(\nu(\alpha_{12}+
\alpha_{13}))\ ,\\
&\stackrel{y\to 1}{\to}& (-1)^{\alpha_{12}+\alpha_{13}+\alpha_{23}}\, 
\alpha_{23} \, (\alpha_{12}+
\alpha_{13})\, .
\eeq

Next we consider the case $n=4$. We choose 4
vectors $\alpha_1,\alpha_2,\alpha_3,\alpha_4$
such that in the convention 
described below
\eqref{esa2}, different linear combinations of the
$\alpha_i$'s follow the following clockwise order
as we move from left to right in the list:
\be                                                                                                 
\label{e4body}                                                                                      
\begin{split}                                                                                       
\alpha_{1}\, ,\,  (\alpha_{1} + \alpha_{2}\, ,\,  \alpha_{1} +                                      
\alpha_{3}\, ,\,  \alpha_{1} + \alpha_{2} + \alpha_{3})\, ,\,                                       
 \alpha_{2}\, ,\,                                                                                   
(\alpha_{2} + \alpha_{3}\, ,\, \alpha_{1} + \alpha_{2} + \alpha_{4} )\, ,\\                         
   \alpha_{1} + \alpha_{2} + \alpha_{3} + \alpha_{4}\, ,\,                                          
  \alpha_{3}\, ,\, \alpha_{1} + \alpha_{3} + \alpha_{4}\, ,\,                                                                                                                                     
( \alpha_{1} + \alpha_{4}\, ,\,                                    
\alpha_{2} + \alpha_{3} + \alpha_{4}\, ,\,             
\alpha_{2} + \alpha_{4}\, ,\,                                                 
 \alpha_{3} + \alpha_{4})\, ,\,  \alpha_{4}\ , 
 \end{split}                                                                                        
\ee             
where again, the order of terms between brackets is irrelevant. 
We list below the allowed partitions
consistent with the three conditions described above
and the corresponding contribution to the summand: 
\vspace*{2mm}
\be \label{e4bodylist}
\begin{array}{|l|r|}
\hline
\{\alpha_1+\alpha_2+\alpha_3+\alpha_4\}  
&
\ y^{2(\alpha_{12}+\alpha_{13}+\alpha_{14}
+\alpha_{23}+\alpha_{24}+\alpha_{34})} \nonumber \\
\{\alpha_1,\alpha_2+\alpha_3+\alpha_4\}  
&
-y^{2(\alpha_{23}+\alpha_{24}+\alpha_{34})} \nonumber \\
\{\alpha_2,\alpha_1+\alpha_3+\alpha_4\}  
&
-y^{2(\alpha_{12}+\alpha_{13}+\alpha_{14}
+\alpha_{34})} \nonumber \\
\{\alpha_1+\alpha_2,\alpha_3+\alpha_4\}  
&
-y^{2(\alpha_{12}+\alpha_{34})} \nonumber \\
\{\alpha_1+\alpha_3,\alpha_2+\alpha_4\}  
&
-y^{2(\alpha_{13}
+\alpha_{23}+\alpha_{24})} \nonumber \\
\{\alpha_1+\alpha_2+\alpha_3,\alpha_4\}  
&
-y^{2(\alpha_{12}+\alpha_{13}
+\alpha_{23})} \nonumber \\
\{\alpha_2+\alpha_3,\alpha_1+\alpha_4\}  
&
-y^{2(\alpha_{12}+\alpha_{13}+\alpha_{14}
+\alpha_{23})} \nonumber \\
\{\alpha_1+\alpha_2+\alpha_4,\alpha_3\}  
&
-y^{2(\alpha_{12} +\alpha_{14}
 +\alpha_{24}+\alpha_{34})} \nonumber \\
\{\alpha_1,\alpha_2,\alpha_3+\alpha_4\}  
&
y^{2\alpha_{34}} \nonumber \\
\{\alpha_1,\alpha_2+\alpha_3,\alpha_4\}  
&
y^{2 \alpha_{23} } \nonumber \\
\{\alpha_1+\alpha_3,\alpha_2,\alpha_4\}  
&
y^{2( \alpha_{13}+\alpha_{23} )} \nonumber \\
\{\alpha_1,\alpha_3,\alpha_2+\alpha_4\}  
&
y^{2( \alpha_{23}+\alpha_{24} )} \nonumber \\
\{\alpha_2,\alpha_3,\alpha_1+\alpha_4\}  
&
y^{2(\alpha_{12}+\alpha_{13}+\alpha_{14}
 )} \nonumber \\
 \hline
\end{array}
\quad 
\begin{array}{|l|r|}
 \hline
\{\alpha_1,\alpha_2+\alpha_4,\alpha_3\}  
&
y^{2( \alpha_{24}+\alpha_{34})} \nonumber \\
\{\alpha_2,\alpha_1,\alpha_3+\alpha_4\}  
&
y^{2(\alpha_{12}+\alpha_{34})} \nonumber \\
\{\alpha_2,\alpha_1+\alpha_3,\alpha_4\}  
&
y^{2(\alpha_{12}+\alpha_{13})} \nonumber \\
\{\alpha_2,\alpha_1+\alpha_4,\alpha_3\}  
&
\ y^{2(\alpha_{12} +\alpha_{14}
 +\alpha_{34})} \nonumber \\
\{\alpha_1+\alpha_2,\alpha_3,\alpha_4\}  
&
y^{2\alpha_{12} } \nonumber \\
\{\alpha_1+\alpha_2,\alpha_4,\alpha_3\}  
&
y^{2(\alpha_{12} +\alpha_{34})} \nonumber \\
\{\alpha_2+\alpha_3,\alpha_1,\alpha_4\}  
&
y^{2(\alpha_{12}+\alpha_{13} 
+\alpha_{23} )} \nonumber \\
\{\alpha_1,\alpha_2,\alpha_3,\alpha_4\}  
&
-1 \nonumber \\
\{\alpha_1,\alpha_2,\alpha_4,\alpha_3\}  
&
-y^{2\alpha_{34}} \nonumber \\
\{\alpha_1,\alpha_3,\alpha_2,\alpha_4\}  
&
-y^{2\alpha_{23}} \nonumber \\
\{\alpha_2,\alpha_1,\alpha_3,\alpha_4\}  
&
-y^{2\alpha_{12}} \nonumber \\
\{\alpha_2,\alpha_3,\alpha_1,\alpha_4\}  
&
-y^{2(\alpha_{12}+\alpha_{13})} \nonumber \\
\{\alpha_2,\alpha_1,\alpha_4,\alpha_3\}  
&
-y^{2(\alpha_{12}+\alpha_{34})} \nonumber \\
 \hline
\end{array}
\ee
Adding these terms and substituting into
eq.\eqref{thebigformula} we arrive at
\be \label{e4bodyfin}
\begin{split}
& g_{\rm ref}(\alpha_1,\alpha_2,\alpha_3, \alpha_4,y) \\
= & (-1)^{\alpha_{12}+\alpha_{13}+\alpha_{14}+\alpha_{23}+\alpha_{24}+\alpha_{34}+1} {1\over (y-y^{-1})^3}\, \\ 
& \times \Big(
y^{\alpha_{12}+\alpha_{13}+\alpha_{14}-\alpha_{23}-\alpha_{24}-\alpha_{34}}-y^{
   \alpha_{12}+\alpha_{13}+\alpha_{14}+\alpha_{23}-\alpha_{24}-\alpha_{34}}+y^{
   \alpha_{12}-\alpha_{13}+\alpha_{14}-\alpha_{23}-\alpha_{24}+\alpha_{34}} \\
   & -y^{
   \alpha_{12}+\alpha_{13}+\alpha_{14}-\alpha_{23}-\alpha_{24}+\alpha_{34}}-y^{
   \alpha_{12}-\alpha_{13}+\alpha_{14}-\alpha_{23}+\alpha_{24}+\alpha_{34}}+y^{
   \alpha_{12}+\alpha_{13}+\alpha_{14}+\alpha_{23}+\alpha_{24}+\alpha_{34}} \\ &
   -y^{
   -\alpha_{12}-\alpha_{13}-\alpha_{14}-\alpha_{23}-\alpha_{24}-\alpha_{34}}+y^
   {-\alpha_{12}+\alpha_{13}-\alpha_{14}+\alpha_{23}-\alpha_{24}-\alpha_{34}}+y
   ^{-\alpha_{12}-\alpha_{13}-\alpha_{14}+\alpha_{23}+\alpha_{24}-\alpha_{34}} \\ &
   -
   y^{-\alpha_{12}+\alpha_{13}-\alpha_{14}+\alpha_{23}+\alpha_{24}-\alpha_{34}}
   +y^{-\alpha_{12}-\alpha_{13}-\alpha_{14}-\alpha_{23}+\alpha_{24}+\alpha_{34}
   }-y^{-\alpha_{12}-\alpha_{13}-\alpha_{14}+\alpha_{23}+\alpha_{24}+\alpha_{34
   }} \Big) \\
   =& (-1)^{\alpha_{12}+\alpha_{13}+\alpha_{14}+\alpha_{23}+\alpha_{24}+\alpha_{34}+1} {1\over \sinh^3\nu}\, \\
&   \Big[\sinh (\nu \alpha_{13}) 
\sinh (\nu (-\alpha_{12}+\alpha_{23}
+\alpha_{24})) \sinh (\nu (\alpha_{14}+\alpha_{34})) \\ &
+\sinh (\nu \alpha_{14})
   \sinh (\nu \alpha_{23}) \sinh (\nu (-\alpha_{12}-\alpha_{13}
   +\alpha_{24}+\alpha_{34})) \\ &
   +\sinh (\nu \alpha_{12}) 
   \sinh (\nu (\alpha_{13}+\alpha_{23})) 
   \sinh (\nu (\alpha_{14}+\alpha_{24}+\alpha_{34}))\Big]\ .
\end{split}
\ee
Taking the $y\to 1$ limit we get
\ben \label{enonmot4}
g(\alpha_1,\alpha_2,\alpha_3, \alpha_4)
&=&(-1)^{1+\sum_{i<j}\alpha_{ij}} \, 
\times \nonumber\\
&&\left[ \alpha_{12}\, \alpha_{13}\, \alpha_{24}\, + 
 \alpha_{13}\, \alpha_{14}\, \alpha_{24}\, + 
 \alpha_{12}\, \alpha_{23}\, \alpha_{24}\, + 
 \alpha_{14}\, \alpha_{23}\, \alpha_{24}\, \right. \nonumber\\ 
&&\left. + \alpha_{12}\, \alpha_{23}\, \alpha_{34}\, + 
 \alpha_{13}\, \alpha_{23}\, \alpha_{34}\, + 
 \alpha_{14}\, \alpha_{23}\, \alpha_{34}\, + 
 \alpha_{13}\, \alpha_{24}\, \alpha_{34}\, \right]\, .
 \een
 We have also  derived the analog of \eqref{e4bodyfin} for
$n=5$ but we suppress the result for the sake of brevity.
Similar results can be obtained for different choices of orderings, while
cases where some of the final states coincide can be obtained by taking 
suitable limits. 
For example
consider the case $g_{\rm ref}(\alpha_1,\alpha_2,
\alpha_2,y)$ with $\alpha_{12}>0$. This can be
considered as a special case of \eqref{erei5} 
in the limit $\alpha_3\to\alpha_2$ and
gives 
\be \label{especial2}
g_{\rm ref} (\alpha_1,\alpha_2,\alpha_2)
= {1\over \sinh^2\nu} \, \sinh(\nu\alpha_{12})
\sinh(2\nu\alpha_{12})\, .
\ee
We shall later verify that the various explicit 
results given in this section are in
perfect agreement with both KS and JS
wall-crossing formul\ae. 

Thus, eq. \eqref{thebigformula} provides
 a complete algorithm for computing the coefficient 
$g_{\rm ref}(\{\alpha_i\},y)$ in \eqref{esab3} for any number of $\alpha_i$'s.
This result is based on the study of multi-centered bound states
in supergravity, even though we had to rely on mathematical results
about moduli spaces of quiver representations. The key point is that 
we only needed invariants of Abelian quivers to compute the 
dynamics of internal (or hair) degrees of freedom, 
while each center could still be regarded as a macroscopic 
solution (provided its charge is large enough).
In the next subsection, we relinquish the gauge theoretical crutch provided by Reineke's formula
and directly quantize the internal degrees of freedom of the bound state.

\subsection{Coulomb branch analysis} \label{scoulomb}

In this subsection we shall try to reproduce the results
of \S\ref{shiggs} by directly quantizing
a configuration of multi-centered black holes.
We begin by
reviewing some relevant
properties of these solutions.
Consider a supersymmetric 
solution describing $n$ black holes, 
with different centers carrying charges 
$\alpha_1,\dots ,\alpha_n\in\tilde\Gamma$ 
located at $\vec r_1,\dots , 
\vec r_n$. We shall  define
\be \label{esadef}
r_{ij} = |\vec r_i - \vec r_j|\, .
\ee
The equations governing the locations $\vec r_i$ 
are \cite{Denef:2000nb}
\be \label{esa30pre}
\sum_{j=1\atop j\ne i}^n \frac{\alpha_{ij}}{r_{ij}}
=  c_i\, ,
\ee
where
\be
c_i \equiv 2 \, {\rm Im}\, (e^{-\I\phi} Z_{\alpha_i})\ ,
\qquad \phi = \arg(Z_{\alpha_1+\cdots +\alpha_n})\, .
\ee
Here
$Z_\alpha$ denotes the central charge for the charge
$\alpha$, computed with the asymptotic values of the
moduli fields. The constants
$c_i$ depend on the moduli through $Z_\alpha$,
and satisfy $\sum_{i=1,\dots , n} c_i=0$. \footnote{In the original
analysis of \cite{Denef:2000nb} each center was
regarded as a single centered black hole. We shall
use a slightly more general interpretation in which
each center is allowed to be a black hole molecule
-- composite of two or more single centered
black holes with charges of each center lying
outside $\tilde \Gamma$. By working at
a point close to the wall $\cP(\gamma_1,\gamma_2)$ we
can ensure that the distance between these molecules,
denoted by $r_{ij}$ in eq.\eqref{esa30pre}, is much larger
than the internal size of each molecule, and hence
\eqref{esa30pre} is a valid description of the inter-molecular
distance for this configuration.}

First we shall show that in the chamber $c_+$ none of
these solutions exist. For this note that at the
wall of marginal stability $\cP(\gamma_1,\gamma_2)$
the central charges of
$\gamma_1$ and $\gamma_2$, and hence
of all the vectors $\alpha_i$, become aligned. As a result, near this
wall the real
parts of $e^{-\I\phi} Z_{\alpha_i}$ are all positive;
we shall denote these by $A_i$. On the other hand
it follows from \eqref{esa2} that for $i=1$ the left hand
side of \eqref{esa30pre} is positive and for $i=n$ it is
negative. Thus we must have  
\be \label{esa4}
e^{-i\phi} Z_{\alpha_1} = A_1 + i B_1, \quad
e^{-i\phi} Z_{\alpha_n} = A_n - i B_n, \qquad
A_1, B_1, A_n, B_n > 0\, .
\ee
This gives a necessary condition for the multi-centered
solution to exist,
\be \label{esa5}
\langle \alpha_1, \alpha_n\rangle \, 
\hbox{Im} \, (Z_{\alpha_1} \bar Z_{\alpha_n})
= \langle \alpha_1, \alpha_n\rangle \, (A_1B_n + A_n B_1)
> 0\, .
\ee
On the other hand the chamber $c_+$ has been defined
such that the right hand side of \eqref{esa5} is
negative (see \eqref{etest}). 
This shows that a multi-centered solution of the type analyzed above
cannot exist in the chamber $c_+$. 
Note that this also proves that 
scaling solutions \cite{Denef:2002ru,Bena:2006kb,Denef:2007vg}, 
whose existence does not depend on the
moduli, cannot exist for charges $\alpha_1,\dots, \alpha_n\in \tilde \Gamma$.\footnote{We can also directly see this as
follows. For the scaling solutions the right hand side
of \eqref{esa30pre} vanishes~\cite{Denef:2007vg}. Now
for $i=1$ all the terms in the left hand side of
\eqref{esa30pre} are manifestly positive due to the
choice \eqref{esa2}, making it impossible
to satisfy this equation.}

{}From now on we work in the chamber $c_-$.
For an $n$-centered configuration, the location of the
centers is specified by $3n$ coordinates
$\vec r_i$ ($1\le i\le n$). Removing the
trivial center of mass degrees of freedom we are left
with $3n-3$ coordinates.
For multi-centered BPS solutions,
the relative distances $r_{ij} \equiv |\vec r_i-\vec r_j|$
must satisfy \eqref{esa30pre} for $i=1,\dots , n$. These
equations
are linearly dependent, since the sum over $i$ is trivially
satisfied. 
This gives $(n-1)$ independent constraints.
The moduli space of multi-centered solutions is then 
a $(2n-2)$ dimensional space
$\cM_n(c_1,\dots, c_n)$. 
In the case of interest here, where all $\alpha_i$ lie in $\tilde\Gamma$,
$\cM_n$ is compact.

As shown in \cite{deBoer:2008zn}, 
for fixed values of $c_i$,
$\cM_n$ carries a symplectic form $\omega$ given by the restriction
of the two-form\footnote{Our normalisation differs from a factor of two
from the one used in \cite{deBoer:2008zn}. This ensures that $\omega/(2\pi)$
has integer periods.}
\be \label{edefomega}
\omega =\frac14 \sum_{i<j} \alpha_{ij}\, 
\epsilon^{abc}
\frac{\de  r^a_{ij} \wedge  \de r^b_{ij}  \, r^c_{ij} 
}{|r_{ij}|^3}\ ,
\ee 
from $\IR^{3n-3}$ to  the moduli space $\cM_n(c_1,\dots, c_n)$.  This symplectic
form is invariant under $SU(2)$ rotations. The moment map 
associated to an infinitesimal rotation $\delta \vec r = \vec \epsilon \wedge \vec r$
is just $\vec \epsilon \cdot \vec J$, where  
\be \label{ejexp}
\vec J= \frac12 \sum_{i<j} \alpha_{ij}\, 
\frac{\vec r_{ij}}{|r_{ij}|}
\ee 
is the angular momentum. Thus the spectrum of
supersymmetric bound states can in
principle be obtained from geometric quantization
of this phase space and the information on
angular momentum, required for computing
$g_{\rm ref}$, can be found by studying the $J_3$
eigenvalues of these bound states.

We now  review the
results of \cite{deBoer:2008zn} 
for the
bound state spectrum of
3-centered black holes in the chamber 
$c_-$.\footnote{Note that if the three charge vectors
$\alpha_i$'s do not lie in a plane, then we can compute
the bound state degeneracy using attractor flow
trees \cite{Denef:2000nb}, {\it e.g.} we could first approach a wall where
the system splits into a pair of molecules, one with charge
$\alpha_1+\alpha_2$ and another with charge $\alpha_3$
and so  the index will be given by
$(-1)^{\alpha_{13}+\alpha_{23}+1} (\alpha_{13}
+\alpha_{23}) \Omega(\alpha_1+\alpha_2)
\Omega(\alpha_3)$. Then we can approach 
another wall where
the system with charge $\alpha_1+\alpha_2$ breaks up
into a pair of molecules of charges $\alpha_1$ and $\alpha_2$
with index $(-1)^{\alpha_{12}+1} \alpha_{12} \Omega(\alpha_1)
\Omega(\alpha_2)$. But when the three $\alpha_i$'s are
in the same plane spanned by $\gamma_1$
and $\gamma_2$, they all move away from each other
at a similar rate when we approach the wall of
marginal stability $\cP(\gamma_1,\gamma_2)$ and we
need to solve the 3-body bound state problem 
explicitly. Similarly if we have $n$ centers with their
charges lying in a plane then we need to explicitly solve
the $n$-body bound state problem.
}
In this case we have 
 $\langle \alpha_i, \alpha_j\rangle \, 
\hbox{Im} \, (Z_{\alpha_i} \bar Z_{\alpha_j})>0$ and 
hence
the clockwise ordering of the $\alpha_i$'s will
imply clockwise ordering of the $Z_{\alpha_i}$'s.
Furthermore
we shall restrict the
$\alpha_i$'s to satisfy
\be \label{constraint}
\alpha_{12} > \alpha_{23}
\ee 
so that
the clockwise ordering of the $\alpha_i$'s and their various
linear combinations are given by eq.\eqref{eorder}.
Thus
the same is true for the corresponding $Z$'s.
We now explicitly write out the equations
\eqref{esa30pre} as follows:
\be \label{esa6}
{\alpha_{12}\over r_{12}} +
{\alpha_{13}\over r_{13}}  = c_1\, ,
\quad 
-{\alpha_{12}\over r_{12}} +
{\alpha_{23}\over r_{23}}  = c_2\, , \quad
{\alpha_{13}\over r_{13}} +
{\alpha_{23}\over r_{23}}  = - c_3\, ,
\quad c_i \equiv 2 \, {\rm Im}\, (e^{-\I\phi} Z_{\alpha_i})
\, .
\ee
Since $e^{\I\phi} = Z_{\alpha_1+\alpha_2+
\alpha_3} / |Z_{\alpha_1+\alpha_2+
\alpha_3}|$, we have $c_1+c_2 + c_3=0$. 
Furthermore since according to
\eqref{eorder} $Z_{\alpha_1}$ and
$Z_{\alpha_1+\alpha_3}$
precedes  $Z_{\alpha_1+\alpha_2+\alpha_3}$
in the clockwise ordering while
$Z_{\alpha_2}$ and $Z_{\alpha_3}$ follow
it, we have
$c_1\ge 0$, $c_1+c_3\ge 0$, $c_2\le 0$, $c_3\le
0$. Thus we
can parametrize the $c_i$'s as
\be \label{epar}
c_1 = a, \quad c_2 = -a+b, \quad c_3=-b,
\quad
a,b\ge 0, \quad b\le a\, .
\ee

We can express the general solution to
\eqref{esa6} as \cite{deBoer:2008zn}
\be \label{egensol}
r_{12} = {\alpha_{12}\over \lambda - b}, \quad
r_{23} = {\alpha_{23}\over \lambda - a}, \quad
r_{13} = {\alpha_{13}\over a+b-\lambda}\, ,
\ee
for some constant real parameter $\lambda$. The
range of $\lambda$ is restricted by the positivity of
each $r_{ij}$  and also the triangle 
inequality satisfied by the $r_{ij}$'s. The
positivity of the $r_{ij}$'s together
with \eqref{epar} give $b\le a\le \lambda\le a+b$.
To study the consequences of the triangle
inequality we express them as
\ben \label{etriangle}
{\alpha_{12}\over \lambda - b}
+ {\alpha_{23}\over \lambda - a}
-{\alpha_{13}\over a+b-\lambda} \ge 0, \nonumber \\
{\alpha_{23}\over \lambda - a} +
{\alpha_{13}\over a+b-\lambda} - 
{\alpha_{12}\over \lambda - b} \ge 0, \nonumber \\
{\alpha_{13}\over a+b-\lambda} + 
{\alpha_{12}\over \lambda - b} - {\alpha_{23}\over \lambda - a}
\ge 0\, .
\een
We need to find solutions to these inequalities
in the range $a\le \lambda\le a+b$.
We begin near $\lambda=a+\epsilon$
for some small $\epsilon$. At this point $r_{23}$
is large and the last of eqs.\eqref{etriangle} is
violated. As we increase $\lambda$, at some value
the last equality is saturated when
\be \label{elower}
r_{23} = r_{13} + r_{12}\, .
\ee
It is easy to see that at this point the other two 
inequalities hold and hence above this bound
the allowed range of $\lambda$ begins. 
This continues till one of the other inequalities
fail to hold. It is easy to see that second inequality
continues to hold for $\lambda\le a+b$ but the
first inequality is violated beyond some value of $\lambda$
close to $a+b$ when
\be \label{eupper}
r_{13}= r_{12} + r_{23}\, .
\ee

The allowed range of angular momentum is given by
the classical angular momentum carried by the system
in the two extremes. At \eqref{elower} the points
$(\vec r_2, \vec r_1, \vec r_3)$ lie along a line and
hence the angular momentum is given by
$(\alpha_{13}+\alpha_{23} -\alpha_{12})/2$. On the
other hand at \eqref{eupper} the points
$(\vec r_1, \vec r_2, \vec r_3)$ lie along a line
and we have total angular momentum
$(\alpha_{13}+\alpha_{23}+\alpha_{12})/2$. Thus
we have
\be \label{efinang}
J_- =  (\alpha_{13}+\alpha_{23} -\alpha_{12})/2,
\quad J_+ = (\alpha_{13}+\alpha_{23}+\alpha_{12})/2
\, .
\ee
As was shown in
\cite{deBoer:2008zn} in quantum theory the upper
limit $J_+$ is shifted to $J_+-1$ and states
of all angular momentum between $J_-$ and
$J_+-1$ occur exactly once.
This gives~\cite{deBoer:2008zn}
\ben \label{ehy}
\ggx_{\rm ref}(\alpha_1,\alpha_2,\alpha_3;y)
&=& (-1)^{\alpha_{13} + \alpha_{23} +\alpha_{12}}
\sum_{J=J_-}^{J_+-1}\, {y^{2J+1} - y^{-2J-1}\over
y - y^{-1}}\nonumber \\
&=& (-1)^{\alpha_{13} + \alpha_{23} +\alpha_{12}}
{1\over (y - y^{-1})^2}\, 
(y^{2J_+} - y^{2J_-} - y^{-2J_-}+ y^{-2 J_+})
\nonumber \\
& = &
(-1)^{\alpha_{13} + \alpha_{23} +\alpha_{12}}
 {1\over \sinh^2 \nu}\,
\sinh(\nu(\alpha_{13} + \alpha_{23})) \,
\sinh( \nu\alpha_{12})
\, ,
\een
in agreement with  \eqref{erei5}. 

We shall now generalize 
Eqs. \eqref{ehy}  to an arbitrary number $n$ of centers.
For this we shall first simplify \eqref{esa30pre}.
Since we are interested in the situation where the
$\alpha_i$'s lie in a two dimensional plane we 
have
\be \label{enew1}
2\, {\rm Im} \, \left(e^{-\I\phi} Z_{\alpha_i}\right) =
\Lambda \, \langle \alpha_i, \sum_j \alpha_j
\rangle = \Lambda\, \sum_{j\ne i} \alpha_{ij}\, ,
\ee
for some positive 
constant $\Lambda$. This constant can be
removed by a rescaling of the variables $\vec r_i$, 
but we shall keep it in our subsequent equations.
This allows us to express \eqref{esa30pre} as:
\be \label{esa30}
\sum_{j=1\atop j\ne i}^n \frac{\alpha_{ij}}{r_{ij}}
= \Lambda\sum_{j\ne i} \alpha_{ij}\, .
\ee
Our strategy will be to relate  $\Tr(-1)^{2J_3} y^{2J_3}$ to an 
integral over the classical phase space $\cM_n$ of 
solutions 
to eq. \eqref{esa30}.
Now,  $(-1)^{2J_3}$ is a rapidly varying function
on $\cM_n$ and does not have a smooth
classical limit. Our experience with the quantum theory
for $n=2$ and $n=3$ nevertheless 
suggests that it takes the same value over all the quantum
states and is given by $(-1)^{2J_{max}-n+1}$,
where $J_{max}=\sum_{i<j}\alpha_{ij}/2$
is the maximum allowed classical angular momentum.
On the other hand, for $y$ close to 1, $y^{2J_3}$ is a slowly varying function
over the classical phase space and one expects that for
large $|\alpha_{ij}|$, its quantum expectation value is well 
approximated by integrating $y^{2J_3}$ over the classical phase space.
Thus, we are led to consider
\be \label{ephaseint}
g_{\rm classical} (\{\alpha_i\},y) \equiv  
{(-1)^{\sum_{i<j}\alpha_{ij} -n+1}\over (2\pi)^{n-1} (n-1)!}\,  \int_{\cM_n} 
\, e^{2\nu\, J_3}\, \omega^{n-1}\, ,
\qquad \nu\equiv \ln y\, .
\ee
This formula should well approximate the refined index $g_{\rm ref}$ 
at large $|\alpha_{ij}|$ and $y$ close to 1, but could in principle
be corrected in the full quantum theory. Our experience with the two 
and three centered cases, as well as an explicit evaluation of 
the 4-centered case presented below,  suggests that 
at $y=1$ the classical
phase space integral \eqref{ephaseint}
in fact agrees with the exact quantum index $g(\{\alpha_i\})$. 
In addition, the same integral \eqref{ephaseint}, after a minor amendment 
to be described shortly, appears to agree with the exact refined index
$g_{\rm ref}(\{\alpha_i\})$ for all values of $y$.

Now using the localization theorem of \cite{Duistermaat:1982vw},
we can express \eqref{ephaseint} as a sum over
contributions from fixed points of the Hamiltonian vector field associated
to the moment map $J_3$, i.e. rotations along the $z$-axis.  
Fixed points 
are therefore multi-centered black hole
configurations in which all centers are aligned along
the $z$-axis, in an appropriate order consistent with
\eqref{esa30}. Furthermore, since all the relative distances
between the centers are fixed by \eqref{esa30}, the
fixed points are isolated. Thus, fixed points of $J_3$
are labelled by  permutations $\sigma$ of
$1,2,\dots ,n$ such that the 
centers are
arranged in a given order along the $z$-axis, satisfying
$z_{\sigma(i)}<z_{\sigma(j)}$ if $i<j$. In this case
the constraint \eqref{esa30} takes the form
\be \label{esa40}
\sum_{j=1\atop j\ne i}^n \frac{\alpha_{\sigma(i)\sigma(j)}}
{z_{\sigma(j)} - z_{\sigma(i)}} \, 
\sign(j-i)
= \Lambda \sum_{j=1\atop j\ne i}^n 
\alpha_{\sigma(i)\sigma(j)}\,   ,
\ee
which is equivalent to the extremization of the ``superpotential"
\be \label{epotential}
W=-\frac12 \sum_{i\neq j} {\rm sign}(j-i) \alpha_{\sigma(i)\sigma(j)} \log|z_{\sigma(j)}-z_{\sigma(i)}| 
-   \Lambda \sum_{i\neq j} \alpha_{ij}    z_i \ .
\ee 
At such a fixed point the 
third component of the
classical
angular momentum is given by 
\be \label{elocal6}
J_3 = {1\over 2} \sum_{i<j} \alpha_{\sigma(i)\sigma(j)} \, 
\, .
\ee
The localization formula of \cite{Duistermaat:1982vw} now gives
\be \label{elocal5}
g_{\rm classical}(\{\alpha_i\},y)
= (-1)^{\sum_{i<j} \alpha_{ij} +n-1}
(2\nu)^{1-n} {\sum_\sigma}' s(\sigma)\, 
y^{\sum_{i<j} \alpha_{\sigma(i)\sigma(j)}}
\, ,
\ee
where $\sum'_\sigma$ denotes sum over only those
permutations for which a solution
to \eqref{esa40} exists,
and $s(\sigma)$ is the sign of 
the Hessian of the matrix
representing the action of $2\nu\, J_3$
on the tangent space of $\cM_n$
at the fixed
point. To compute $s(\sigma)$ we make a convenient 
choice of coordinates on $\cM_n$.
Without any loss of generality we can choose
$\vec r_1$ to be at the origin.
At a fixed point of the action of $J_3$, all the other points are then
along the $z$-axis. 
We now note that to first order the relative
distances between the centers remain unchanged if
we displace each of the $\vec r_i$ for $i\ge 2$
in the $(x-y)$ plane. Thus these $(2n-2)$ coordinates
provide us with a convenient parametrization of the
moduli space of the solution
near this fixed point. Let us denote them 
by $(x_i, y_i)$ ($2\le i\le n$). 
The action of $J_3$ on these coordinates is
simply an independent rotation in the $(x_i, y_i)$ plane 
for each $i$. The
Hessian of $2\nu J_3$ is given by $(2\nu)^{n-1}$ 
up to a sign $s(\sigma)$.
To determine the sign  we note that
in the coordinate system $\{x_i,y_i\}$ introduced
above, $J_3$ and $\omega$ take the form:
\be \label{ej3omega}
J_3 = {1\over 2} \sum_{i<j} \alpha_{\sigma(i)\sigma(j)}
- {1\over 4} M_{ij} (x_i x_j + y_i y_j) + \cdots,
\qquad \omega = 
\frac12 M_{ij} dx_i\wedge dy_j +\cdots \, ,
\ee
where $\cdots$ denote higher order terms and
\be \label{edefmij}
M_{ii} = \alpha_{1i} {z_i\over |z_i|^3}
+ \sum_{k\neq i, k\geq 2} \alpha_{ik} \, {z_k - z_i\over 
|z_k - z_i|^3}\, ,
\qquad M_{ij} = -\alpha_{ij} {z_j - z_i\over |z_j-z_i|^3}
\quad \hbox{for $i\neq j$}, \quad i,j\ge 2
\, .
\ee
It is worth noting that the matrix $M_{ij}$ is also equal to the Hessian of 
the superpotential \eqref{epotential} with respect to the $n-1$ variables
$z_2,\dots z_n$, with $z_1$ being set to zero.
The sign $s(\sigma)$ of
the Hessian
associated with the
fixed point is thus given by
\be \label{esigneq}
s(\sigma) = \sign \det M\, .
\ee
Although the prescription \eqref{esigneq} appears to treat $z_1$ on a different footing than
the other $z_i$ due to the gauge condition $z_1=0$, one could just as well have 
computed $s(\sigma)$ using a symmetric gauge condition $\sum_{i=1\dots n} z_i=0$.
Indeed, the same sign $s(\sigma)$ can be obtained as (the opposite of) the sign  of the
determinant of the Hessian of the superpotential 
\be \label{ewhat}
\hat W=W
+ {\lambda\over n}\, \, \sum_{i=1}^n z_i
\ee
with respect to all $z_i$, $i=1\dots n$ and to the Lagrange multiplier $\lambda$. To see
this, note that the Hessian of $\hat W$ with respect to $(\lambda,z_1,z_{i=2\dots n})$ is
given by \be \label{ewhat2}
\hat M = 
\begin{pmatrix} 
 0 & {1/ n} & A\cr {1/ n} & {\p^2 W/ \p z_1^2} 
 & W_1\cr
A^T & W_1^T & M
\end{pmatrix}
\ee
where both $A$ and $W_1$ are $(n-1)$ dimensional 
row matrices, with $A=(1/n, \cdots 1/n)$ and $W_1
= (\p^2 W/ \p z_1 \p z_2, \cdots \p^2 W/ \p z_1 \p z_n)$.
By adding the third to $(n+1)$'th rows to the 
second row and
third to $(n+1)$-th columns to the second column
and using the fact that $\sum_{i=1}^n \p^2 W/\p z_i \p z_j=0$
due to translation invariance, we can bring \eqref{ewhat2}
to the form
\be \label{ewhat3}
\begin{pmatrix}
0 & 1 & A \cr 1 & 0 & {\bf 0}\cr A^T & {\bf 0}^T & M
\end{pmatrix}\, ,
\ee
where ${\bf 0}$ denotes an $(n-1)$-dimensional row
matrix with all entries 0.
{}From this we see that $\det \hat M=-\det M$. 

If there are more than one solution of \eqref{esa40} 
for a given permutation
$\sigma$ then the right hand side of
\eqref{esigneq} should be replaced by
a sum of $\sign \det M$  over
all solutions.
Numerical evidence indicates however that 
there is at most one fixed point for a given 
permutation. Moreover, it suggests 
that
$s(\sigma)$ can  be expressed in terms of the 
permutation $\sigma$ through
\be \label{esign}
s(\sigma) = (-1)^{\#\{i; \sigma(i+1)<\sigma(i)\}}\ .
\ee
This is easily proven for the 
special critical points of $W$ 
 with 
${\rm sign}(z_{\sigma(i)}-z_{\sigma(j)})=
{\rm sign}(i-j)$ and
${\rm sign}(z_{\sigma(i)}-z_{\sigma(j)})=
{\rm sign}(j-i)$.
These represent solutions of \eqref{esa40}
corresponding to
permutations $(12\dots n)$ and $(n\dots 21)$.
Since these 
correspond to the global maximum and global minimum of $J_3$, respectively, 
the matrix $M_{ij}$ is  positive definite or negative definite, respectively, leading to 
$s(12\dots n)=1, s(n\dots 21)=(-1)^{n-1}$. 
As we shall see in \S\ref{scomp}, the  result \eqref{esign}
is required for consistency with the Higgs branch derivation
presented in \S\ref{shiggs}. This also
suggests that if there are more than one
fixed points for a given permutation their contributions should cancel 
pairwise leaving
behind the contribution from 
0 or 1 fixed point.

We do not expect the classical formula \eqref{ephaseint}
and hence \eqref{elocal5}
to reproduce the full $y$ dependent
quantum answer for $\Tr(-y)^{2J_3}$ -- after all the
quantization of angular momentum is not visible classically.  This is
apparent from \eqref{elocal5}: while the terms inside
the sum involve integral powers of $y=e^\nu$ and hence
are compatible with charge quantization, the
overall factor $(2\nu)^{1-n}$ does not respect charge
quantization. Comparison with the 
exact results \eqref{thebigformula}, \eqref{ehy}
suggests a remedy\footnote{Note added in v2:
in our subsequent work \cite{MPS2}, we derive this
multiplicative renormalisation from the Atiyah-Bott
Lefschetz fixed point formula for the equivariant index of the
Dirac operator on $\cM_n$.}
: replace the factor of
$(2\nu)^{1-n}$ by $(2\sinh\nu)^{1-n} =
(y - y^{-1})^{1-n}$. In the $y\to 1$ limit this will
approach the classical result in accordance with the
earlier observation that in this limit the classical and
quantum results agree.
Thus our proposal for the full quantum 
version of \eqref{elocal5} is
\be \label{elocal55}
g_{\rm ref}(\{\alpha_i\},y)
= (-1)^{\sum_{i<j} \alpha_{ij} +n-1}
(y - y^{-1})^{1-n} {\sum_\sigma}' s(\sigma)\, 
y^{
\sum_{i<j} \alpha_{\sigma(i)\sigma(j)}
}\, .
\ee
This reduces the problem of computing the function 
$g_{\rm ref}$
 to identifying which of the permutations
$\sigma$ are consistent with \eqref{esa40}.
This is  a  tedious but straightforward
procedure. Below we give the results for $n=3$ and
$n=4$ for the same order of various linear combinations
of the $\alpha_i$'s as given in \eqref{eorder}
and \eqref{e4body}.

For $n=3$ the detailed analysis of the configuration
space was carried out in \cite{deBoer:2008zn} 
some relevant details
of which were reviewed earlier in this section. Two
of the four collinear configurations are given in
\eqref{elower} and two others are given
by \eqref{eupper}.\footnote{For each of 
\eqref{elower} and \eqref{eupper} we have two
configurations related by  $z\to -z$ symmetry.}
This gives 
the following order of the centers
along the $z$-axis and the value of $s(\sigma)$,
\be \label{e3perm}
\{1,2,3;+\}, \{2,1,3;-\}, \{3,1,2;-\}, \{3,2,1;+\}\, ,
\ee
leading to 
\ben \label{e3res}
g_{\rm ref}(\alpha_1,\alpha_2,\alpha_3,y)
&=& (-1)^{\alpha_{12}+\alpha_{23}+\alpha_{13}}\,
(y-y^{-1})^{-2} \nonumber \\ &&
\times \Big( y^{\alpha_{12}+\alpha_{13}+\alpha_{23}}
- y^{\alpha_{13}+\alpha_{23}-\alpha_{12}}
- y^{\alpha_{12}-\alpha_{23} -\alpha_{13}}
+y^{-\alpha_{12}-\alpha_{13}-\alpha_{23}} \Big)\, ,
\nonumber \\
\een
in agreement with the result \eqref{ehy} following
from exact quantization of the 3-centered system.

Next we test \eqref{elocal55} by working out the result
for $n=4$. 
There are 12 fixed points, whose
orders and the associated $s(\sigma)$
are given by
\ben \label{e4order}
\{1,2,3,4;+\},
\{3,1,2,4;-\},
\{1,3,4,2;-\},
\{1,4,2,3;-\},
\{3,1,4,2;+\},
\{2,3,4,1;-\},
 \nonumber \\
\{1,4,3,2;+\},
\{2,4,1,3;-\},
\{3,2,4,1;+\},
\{2,4,3,1;+\},
\{4,2,1,3;+\}, 
\{4,3,2,1;-\}\, . \nonumber \\
\een
Eq.\eqref{elocal55} now gives
 \be \label{elocal7}
\begin{split}
g_{\rm ref}&(\alpha_1,\alpha_2,\alpha_3, \alpha_4,y) 
=  (-1)^{\alpha_{12}+\alpha_{13}+\alpha_{14}+\alpha_{23}+\alpha_{24}+\alpha_{34}+1} 
(y-y^{-1})^{-3}\, \\ 
 \times& \Big(
y^{   \alpha_{12}+\alpha_{13}+\alpha_{14}+\alpha_{23}+\alpha_{24}+\alpha_{34}} 
-y^{   \alpha_{12}-\alpha_{13}+\alpha_{14}-\alpha_{23}+\alpha_{24}+\alpha_{34}}
 -y^{   \alpha_{12}+\alpha_{13}+\alpha_{14}-\alpha_{23}-\alpha_{24}+\alpha_{34}}
\\   &
 -y^{   \alpha_{12}+\alpha_{13}+\alpha_{14}+\alpha_{23}-\alpha_{24}-\alpha_{34}}
+y^{   \alpha_{12}-\alpha_{13}+\alpha_{14}-\alpha_{23}-\alpha_{24}+\alpha_{34}} 
-y^{-\alpha_{12}-\alpha_{13}-\alpha_{14}+\alpha_{23}+\alpha_{24}+\alpha_{34}} 
\\ &
+ y^{\alpha_{12}+\alpha_{13}+\alpha_{14}-\alpha_{23}-\alpha_{24}-\alpha_{34}}
 -   y^{-\alpha_{12}+\alpha_{13}-\alpha_{14}+\alpha_{23}+\alpha_{24}-\alpha_{34}}
+y^{-\alpha_{12}-\alpha_{13}-\alpha_{14}-\alpha_{23}+\alpha_{24}+\alpha_{34}  }
\\ &
   +y^{-\alpha_{12}-\alpha_{13}-\alpha_{14}+\alpha_{23}+\alpha_{24}-\alpha_{34}} 
  +y^{-\alpha_{12}+\alpha_{13}-\alpha_{14}+\alpha_{23}-\alpha_{24}-\alpha_{34}}
  -y^{   -\alpha_{12}-\alpha_{13}-\alpha_{14}-\alpha_{23}-\alpha_{24}-\alpha_{34}}
\Big)
 \end{split}
 \ee
in agreement with \eqref{e4bodyfin}.

\subsection{Comparison of the results of Higgs branch 
and Coulomb branch analysis} \label{scomp}

To compare the results of the Coulomb branch
analysis described in \S\ref{scoulomb} 
with the Higgs branch computation described in 
\S\ref{shiggs}, note that the power $y^{\sum_{i<j} \alpha_{\sigma(i)\sigma(j)}}$ 
in \eqref{elocal55} matches the power of $y$ in
\eqref{thebigformula}, provided the ordered decomposition $\beta$ is chosen as follows:
break the sequence $\{\sigma(i),i=1,\dots , n\}$ 
into increasing subsequences $\{\sigma(i_{a-1}+1),\dots,
\sigma(i_a)\}$, $0=i_0<i_1<\dots<i_s=n$, where $s$
is the number of such  increasing subsequences. The vectors 
$\beta^{(a)}$ are then 
\be \label{ehiggscoulomb}
\beta^{(s+1-a)} =  
\sum_{i=i_{a-1}+1}^{i=i_a} \alpha_{\sigma(i)}\ .
\ee
For example for the permutation $\sigma(123) = (132)$
there are two possible choices for the increasing
subsequences: $\{\{13\}, \{2\}\}$ and $\{\{1\}, \{3\}, \{2\}\}$,
and the corresponding partitions are 
$\{\alpha_2, \alpha_1+\alpha_3\}$ and $\{\alpha_2,
\alpha_3, \alpha_1\}$.
To see that the partition \eqref{ehiggscoulomb} 
generates the correct power of $y$ associated
with the permutation $\sigma$, we note that 
\eqref{ehiggscoulomb} implies that $m_i^{s+1-a}
= \sum_{j=i_{a-1}+1}^{i_a} \delta_{i, \sigma(j)}$ in 
\eqref{thebigformula}. Leaving aside the factor of
$(y-y^{-1})^{-n+1}$, the power of $y$ in
\eqref{thebigformula} is now given by
\be \label{etestx}
-\sum_{i<j} \alpha_{ij} + 2\sum_{a\geq b}
\sum_{k=i_{a-1}+1}^{i_a} \sum_{l=i_{b-1}+1\atop
\sigma(l)<\sigma(k)}^{i_b}
\alpha_{\sigma(l)\sigma(k)}
= -\sum_{i<j} \alpha_{ij} + 2
\sum_{l<k\atop \sigma(l)<\sigma(k)} \alpha_{\sigma(l)
\sigma(k)}\, ,
\ee
where in the second step we have used the fact that
if $k,l$ belong to different subsequences labeled by
$a$ and $b$  then $a>b$ will
imply $k>l$. On the other hand if they belong to the
same subsequence then, since the subsequence is
increasing, the condition $\sigma(l)<\sigma(k)$ will
imply $l<k$. The right hand side
of \eqref{etestx} gives precisely the
power of $y$ associated with the permutation $\sigma$
as given in \eqref{elocal55}.

The sign associated with the partition \eqref{ehiggscoulomb}
in 
\eqref{thebigformula} is 
$(-1)^{s-1}$.
If the increasing subsequences are maximal, \i.e.\ chosen
so that it is not possible to build bigger increasing
subsequences, then this sign
is in agreement with the
rule given in \eqref{esign}.
For example, for
the permutation $\sigma(1234)=3142$ the maximal
increasing subsequences are $\{\{3\}, \{14\}, \{2\}\}$. 
Thus it  corresponds to the 
ordered decomposition 
$\{\alpha_2,\alpha_1+\alpha_4, \alpha_3\}$, contributing
with a positive sign.

Now,   there are typically many increasing subsequences
associated with a given permutation,  obtained by breaking
up the maximal increasing subsequences into smaller
increasing subsequences, and contributing with
different signs. Thus, 
in order to determine if a given permutation contributes
and with what sign, we have to combine the contributions
from these different terms.
In the previous example the other possible increasing
subsequence is $\{\{3\},\{1\},\{4\},\{2\}\}$, but the
corresponding partition $\{\alpha_2,\alpha_4,\alpha_1,
\alpha_3\}$ does not satisfy the condition 
\eqref{econdombeta} and hence does not contribute. 
In general however the situation is more complicated.
Consider for example the permutation $\sigma(1234)
= (3412)$. The corresponding increasing subsequences
are $\{\{34\}, \{12\}\}$, $\{\{3\}, \{4\}, \{12\}\}$,
$\{\{34\}, \{1\}, \{2\}\}$, and $\{\{3\},\{4\}, \{1\}, \{2\}\}$,
associated to the partitions 
$\{\alpha_1+\alpha_2,\alpha_3+\alpha_4\}$, 
$\{\alpha_1+\alpha_2,\alpha_4, \alpha_3\}$,
$\{\alpha_2,\alpha_1,\alpha_3+\alpha_4\}$
and
$\{\alpha_2,\alpha_1,\alpha_4,\alpha_3\}$, respectively.
For the order given in \eqref{e4body} only the
first and the third partitions are allowed by the rules 
\eqref{econdombeta}. They contribute with opposite
sign making the contribution vanish. This explains
why the permutation (3412) is absent from the list
\eqref{e4order}.

It is easy to convince oneself that 
all possible partitions of the vectors $(\alpha_1,\dots,
\alpha_n)$,  whether or not they satisfy 
the condition \eqref{econdombeta},  
are in  one to one correspondence with the set of all 
increasing subsequences of all the
permutations of $(12\dots n)$  via 
the rule \eqref{ehiggscoulomb}.
So, the complete Higgs branch 
contribution can be generated 
by beginning with the maximal increasing subsequences
associated with a given permutation and combining them
with the contribution from other increasing subsequences
associated with the same permutation.
The following observations are now in order:
\begin{enumerate}
\item If a given partition is not allowed by the rule
\eqref{econdombeta} then all its subpartitions 
are also disallowed. For example in the example
of the previous paragraph, once we know that
$\{\alpha_1+\alpha_2, \alpha_4,\alpha_3\}$ is not
allowed, we can immediately conclude that
$\{\alpha_2, \alpha_1, \alpha_4,\alpha_3\}$ is also not
allowed. 
\item If there are two or more maximal
increasing subsequences of
length two or more, then each of the maximal
increasing subsequences can be independently 
broken up into  smaller increasing subsequences. 
The compatibility of a  partition of a particular maximal 
increasing sequence with the condition 
\eqref{econdombeta}
or not is independent on the partitioning
of  the other maximal
increasing subsequences. For example
for the partition $\{\alpha_1+\alpha_2,
\alpha_3+\alpha_4\}$, the compatibility of  the
splitting of $\alpha_3+\alpha_4$ into $\{\alpha_4,\alpha_3\}$
can be determined independently of
whether $\alpha_1+\alpha_2$ is kept as a single
element, or has been split into $\{\alpha_2,\alpha_1\}$.
For this reason we can associate, to each maximal 
increasing subsequence, a weight given by a sum
of $\pm 1$ for each of the allowed splittings of that
subsequence ($+1$ for splitting into odd number of
subsequences, including the original maximal
increasing subsequence,
 and $-1$ for splitting into even number
of subsequences).
The final weight is given by the product
of the weights computed from each maximal
increasing subsequence.

In the example above the weight factor associated with
$\alpha_3+\alpha_4$ is 1 since it cannot be split,
while the weight factor associated with 
$\alpha_1+\alpha_2$
vanishes since it allows a split 
$\{\alpha_2,\alpha_1\}$ with
opposite sign. As a result the net weight is 
$1\times 0=0$.

\item The problem of determining the contribution from
a given permutation now reduces to computing the
weight factor associated with each
maximal increasing subsequence of
that permutation.
This can be done as follows. We begin with a particular
maximal increasing subsequence
and first consider all possible
partitions of this subsequence into two
smaller increasing subsequences.
For this we need to 
simply insert a comma at one place that indicates how
we divide the original subsequence. 
Not all such subsequences may generate partitions
allowed by \eqref{econdombeta}; 
let us assume that there are
$k$ possible places where we are allowed to
insert the comma. 
This gives $k$ terms, each with weight $-1$.
Now consider the possible 
partitions of the same maximal increasing subsequence 
into three increasing subsequences. It follows
from the rule \eqref{econdombeta} that the allowed
partitions are obtained by inserting a pair of commas
into two of the same $k$ possible positions. 
Thus there are $k(k-1)/2$ possible terms, each with
weight $1$.
This generalizes to partitioning into
arbitrary number of increasing subsequences. 
Thus the net  weight
factor is
$1-k + {k\choose 2} - {k\choose 3}+
\cdots = (1-1)^k$. This shows that
the weight factor vanishes for $k\ge 1$, and is 1 for
$k=0$ \i.e.\ when it is not possible to subpartition an
increasing subsequence satisfying \eqref{econdombeta}.
\end{enumerate}
This leads to the following simple rule for deciding when
a given permutation contributes and the sign of the
contribution:
{\it A given permutation contributes if its
maximal increasing subsequences generate a
partition satisfying \eqref{econdombeta} via
\eqref{ehiggscoulomb}, and none of the other
(non-maximal) increasing subsequences generate
an allowed partition.
The sign of the contribution is
given by \eqref{esign}.
}

We should of course keep in mind that the algorithm
described above is not an independent result derived
from the
Coulomb branch, it is required to
ensure that the Coulomb and the
Higgs branch results agree. It will be interesting to find
an independent derivation of this from
the Coulomb branch analysis by directly examining
the condition for the existence of solutions to
\eqref{esa30}.

Before concluding this section we shall demonstrate how
the algorithm given above can be used to give a derivation
of the semi-primitive wall-crossing formula. We choose
$\alpha_n$ to be $\gamma_1$ and 
$\alpha_1,\cdots \alpha_{n-1}$ to consist of $m_1$ copies
of $\gamma_2$, $m_2$ copies of $2\gamma_2$ etc. in some
fixed order (which can be decided by adding some arbitrary
small vector to each of these charges which will be taken
to zero at the end). Thus we have $n=1+\sum_s m_s$ and
$\sum_{i<j}\alpha_{ij} = - \gamma_{12} \sum_s s m_s$.
Since $\langle \gamma_1+k\gamma_2,
\gamma_1+\sum_s s m_s \gamma_2\rangle<0$ for 
$k<\sum_s s m_s $, it follows from the
\eqref{econdombeta} that $\gamma_1$ must be part
of the last partition. 
Consider now a permutation of
$1,\cdots n$. In order that $\gamma_1$ is part of the
last partition, the first maximal 
increasing subsequence in this permutation must
contain the element $n$ as its last element.
This subsequence cannot be
partitioned into smaller increasing subsequences since
then $\gamma_1$ will not be part of the last partition.
Furthermore in order that the permutation gives a
non-vanishing contribution the rest of the maximal
increasing subsequences must each have length 1,
since any maximal increasing subsequence of length
2 or more can be partitioned into smaller increasing
subsequences without violating \eqref{econdombeta}
and the result will vanish. This implies that the 
 rest of the elements must be arranged in
decreasing order in the permutation $\sigma$. Thus the
only freedom in choosing the permutation is in deciding
which elements are part of the first increasing subsequence.
Let there be $k_s$ copies of $s\gamma_2$ in this set;
this can be chosen in a total of $\prod_s {m_s\choose k_s}$
ways. 
The total number of partitions associated with this
permutation is $1+\sum_s(m_s-k_s)$.
Thus the net contribution to $g_{\rm ref}$ is given by
\be \label{enetgref}
\begin{split}
& (-y)^{ \sum_s m_s + \gamma_{12}\sum_s  m_s s}
(y^2-1)^{ - \sum_s m_s}
\prod_s \sum_{k_s=0}^{m_s} 
\, {m_s\choose k_s} (-1)^{m_s-k_s}
y^{-2\gamma_{12}  k_s s}
\\ 
= & 
\prod_s \left[  {(-y)^{\langle \gamma_1, s\gamma_2\rangle} - (-y)^{-\langle \gamma_1, 
s\gamma_2\rangle}\over y - y^{-1}}
\right]^{m_s}
\, ,
\end{split}
\ee
in agreement with \eqref{eomp1ref}.

\section{Wall crossing from the 
Kontsevich-Soibelman formula\label{sec_KS}}

The first solution to the problem of determining 
$g(\{\alpha_i\})$  was
given
by Kontsevich and Soibelman \cite{ks}
and also independently by Joyce and 
Song~\cite{Joyce:2008pc,Joyce:2009xv}. 
In this section we shall review the results of \cite{ks}
and compare them with our results.
Subsection \ref{sks} states
the KS wall-crossing formula, Subsection \ref{schks} explains charge
conservation and the following subsections apply the KS formula to
determine the jump of $\Omega(\gamma)$ in various cases.
In subsections \ref{srefined} and \ref{ssemiref} we
describe generalization of the KS formula to the
motivic index.

\subsection{The KS formula} \label{sks}
To state the KS formula, we introduce the Lie algebra $\cA$ spanned by abstract generators $\{e_\gamma,
\gamma\in \Gamma\}$, satisfying the commutation rule
\be
\label{KSalg}
[ e_{\gamma} , e_{\gamma'} ] =  
\kscom{\langle \gamma,\gamma'\rangle}\, e_{\gamma+\gamma'},
\ee
where we defined
\be
\label{defkappa}
\kappa(x) = (-1)^x\, x\ .
\ee
At a given point in moduli space labeled by the parameters
$\{t^a\}$, we introduce the operator 
\be
\label{Uclas}
U_\gamma(t^a) = \exp\left( \Omega(\gamma;t^a) \, 
\sum_{d=1}^{\infty} \frac{e_{d\gamma}}{d^2}\right)
\ee 
in the Lie group generated by 
$\cA$.
The KS wall-crossing formula \cite{ks,Gaiotto:2008cd} states that the product
\be
\label{KSprod}
A_{\gamma_1,\gamma_2} = \prod_{\substack{\gamma=M\gamma_1+N\gamma_2,\\
M\geq 0, N \geq 0}} U_\gamma\ ,
\ee
ordered 
so that as we move from the left to the right the corresponding
$Z_\gamma$'s are ordered clockwise, \i.e.\ according to decreasing
values of $\arg(Z_\gamma)$, 
stays
constant across the hyperplane of marginal stability $\cP(\gamma_1,\gamma_2)$. As
$t^a$ crosses this locus, $\Omega(\gamma;t^a)$ jumps and the order of the factors
is reversed, but the operator $A_{\gamma_1,\gamma_2}$ stays constant. 
Thus, the KS formula may be stated as the equality
\be \label{ewallfinU}
\prod_{\substack{M\geq 0,N\geq 0, \\  M/N\downarrow}}
U^+_{M\gamma_1+N\gamma_2}
=
\prod_{\substack{M\geq 0,N\geq 0, \\  M/N\uparrow}}
U^-_{M\gamma_1+N\gamma_2}\, ,
\ee
where $M/N\downarrow$ 
means that the terms in the
product are arranged from left to right in the order of
decreasing values of $M/N$ while
$M/N\uparrow$ implies opposite ordering of
the factors, and $U^\pm_\gamma$ are defined as in 
(\ref{Uclas}) with $\Omega(\gamma; t^a)$
replaced by $\Omega^\pm$.

Noting that the operators $U_{k\gamma}$ for different 
$k\geq 1$ commute, one may combine 
 them into 
 \be
 \label{Vclas}
 V_\gamma\equiv \prod_{k=1}^{\infty} U_{k\gamma} = 
 \exp\left( \sum_{\ell=1}^{\infty} \bOm(\ell\gamma) \, e_{\ell\gamma} \right), \qquad \bar\Omega(\gamma)
 =\sum_{m|\gamma} m^{-2}\Omega(\gamma/m)\, ,
 \ee
 and rewrite \eqref{KSprod} into a product over primitive vectors only,
 \be
 \label{KSprodprim}
A_{\gamma_1,\gamma_2} = \prod_{\substack{\gamma=M\gamma_1+N\gamma_2,\\
M\geq 0, N\geq 0,\gcd(M,N)=1}} V_\gamma\ .
 \ee
Using the definition of $\Omega^\pm$ given in
\S\ref{spril} the wall-crossing formula takes the
form
\be \label{ewallfin}
\prod_{\substack{M\geq 0,N\geq 0, \\ \gcd(M,N)=1, M/N\downarrow}}
V^+_{M\gamma_1+N\gamma_2}
=
\prod_{\substack{M\geq 0,N\geq 0, \\ \gcd(M,N)=1, M/N\uparrow}}
V^-_{M\gamma_1+N\gamma_2}\, ,
\ee
where $V^\pm_\gamma$ are defined as in 
(\ref{Vclas}) with $\bar\Omega$
replaced by $\bar\Omega^\pm$.

The invariants $\bOm^-\argu{M}{N}$ 
on one side of the wall 
can be determined in terms of the  
invariants $\bOm^+\argu{M}{N}$
on the other side
by expressing both sides of (\ref{ewallfin})
into single exponentials using the Baker-Cambell-Hausdorff (BCH) formula, and then
comparing the coefficients of each $e_\gamma$ on either side. 
These equations can be
solved iteratively to determine 
$\bar\Omega^-\argu{M}{N}$ in terms of the
$\bar\Omega^+$'s, starting with the lowest 
values of $(M,N)$.
This is most conveniently done
by projecting the 
relation \eqref{ewallfin} to the finite-dimensional algebra 
\be \label{edeftrunc}
\cA_{M,N} = \cA \slash \{ \sum_{m>M \, {\rm and/or}\,
n> N} \IR  \cdot e_{m\gamma_1+n\gamma_2} \}\ ,
\ee
and using the Baker-Campbell-Hausdorff (BCH) formula
to commute the factors (see later). 
Since $\cA_{M,N}$ is a finite dimensional algebra
generated by $m \gamma_1+n\gamma_2$ for
$0\le m\le M$ and $0\le n\le N$, we have a
finite number of equations relating 
$\Omega^-\argu{m}{n}$ to 
$\Omega^+\argu{m}{n}$. 
For example a trivial consequence of (\ref{ewallfin}) 
is the relation
\be \label{etriv}
\Omega^-\argu{M}{N}=\Omega^+\argu{M}{N}
\ee
whenever $M=0$ or $N=0$. This follows from
the fact  the algebras $\cA_{M,0}$ and
$\cA_{0,N}$ are commutative. 

In order to derive the semi-primitive wall-crossing
formula and generalizations thereof, it is also practical to work
with the infinite dimensional algebra $\cA_{M,\infty}$, and consider
the generating functions
\be
\label{defZNq}
Z^\pm(M,q) = \sum_{N=0}^{\infty} \Omega^\pm\argu{M}{N}\,q^{N}\ ,\qquad
\bar Z^\pm(M,q) = \sum_{N=0}^{\infty} \bOm^\pm\argu{M}{N}\,q^{N}\ ,
\ee
for fixed value of $M$. These two objects are related by 
\be
\bar Z^\pm(M,q) = \sum_{d|M} \frac{1}{d^2} Z^\pm(M/d,q^d)\ ,\qquad
Z^\pm(M,q) = \sum_{d|M} \frac{\mu(d)}{d^2} \bar Z^\pm(M/d,q^d)\ .
\ee

\subsection{Charge conservation from KS formula}
\label{schks}

In this section we shall draw attention to
one specific feature of the wall-crossing
formula given in (\ref{ewallfin}), namely `charge
conservation'. It follows from the algebra (\ref{KSalg}),
and the definition of $V_\gamma$ given in (\ref{Vclas})
that after combining each side into a single
exponential, the coefficient of $e_\gamma$ consists of 
a sum of products of the form $\prod_i \bar\Omega(\gamma_i)$
with $\sum_i \gamma_i =\gamma$, up to 
an overall numerical constant.
Thus, any relation that follows from
(\ref{ewallfin}) has the property that the sum of the charges
in the argument of $\bar\Omega$'s has the same value for
all the terms on either side of \eqref{ewallfin}.  Thus when we solve this to find
$\bar\Omega^-(\gamma)$ in terms of $\bar\Omega^+(\gamma)$
and products of $\bar\Omega^+(\gamma_i)$'s, each term
in the expression will have the property that the charges
in the argument of $\bar\Omega^+$ in the product 
will add up to $\gamma$. This is precisely the
`charge conservation' rule that followed from the use of
Maxwell-Boltzmann statistics in \S\ref{scharge}. Note that the
wall-crossing formula written in terms of
$\Omega^\pm$ does not have any such manifest charge
conservation. We shall see examples of this `charge conservation' rule
in the explicit examples to be described below.

When several walls are crossed consecutively, the black hole molecules
and bound molecular clusters can be decomposed into smaller molecules
and eventually just single atoms. 
This is the attractor flow conjecure
\cite{Denef:2000nb}. Knowing the indices of the 
atoms, one can in
principle determine the contribution to the index of the total
molecule, or equivalently flow tree. These flow trees are naturally parametrized
by nested lists, {\it e.g.} $((\gamma_1,\gamma_2),\gamma_3)$, which
need to satisfy the 'charge conservation' rule. The structure of the nested lists
is identical to the commutation relations of the KS formula in
terms of $\bar \Omega(\gamma)$, and allows to determine easily the
contribution of a molecule to the index \cite{Manschot:2010xp}.

\subsection{Primitive wall-crossing}

In $\cA_{1,1}$  the 
BCH formula reduces to 
\be
\label{BCH2}
e^X\, e^Y = 
e^{X+Y +{1\over 2} [X,Y]}\ ,
\ee
since all multiple commutators involving
three or more generators vanish.
The
wall-crossing equation takes the form
\ben \label{erel1}
&& \exp(\bar\Omega^+(\gamma_1)e_{\gamma_1})
\exp(\bar\Omega^+(\gamma_1+\gamma_2)
e_{\gamma_1+\gamma_2})
\exp(\bar\Omega^+(\gamma_2)e_{\gamma_2})\nonumber \\
&&\qquad = 
\exp(\bar\Omega^-(\gamma_2)e_{\gamma_2})
\exp(\bar\Omega^-(\gamma_1+\gamma_2)
e_{\gamma_1+\gamma_2})
\exp(\bar\Omega^-(\gamma_1)e_{\gamma_1})\, .
\een
{}From this we find the primitive wall-crossing
relation
\be \label{eprimks}
\bOm^-(1,1) = \bOm^+(1,1) + 
(-1)^{{\gamma_{12}}} {\gamma_{12}} \bOm^+(0,1) \bOm^+(1,0)\, ,
\ee
where for simplicity we have denoted $\bar\Omega^\pm
(M\gamma_1 + N\gamma_2)$ by 
$\bOm^\pm(M,N)$ and $\langle\gamma_1,\gamma_2
\rangle$ by $\gamma_{12}$.

\subsection{Generic 3-body and 4-body contributions} \label{s3decay}

We shall now extract the generic $n$-body contribution from the 
KS wall-crossing formula. To explain what is meant by `generic',
let $n=3$, and 
$\alpha_1$, $\alpha_2$, $\alpha_3$ be 
three distinct (not necessarily primitive) elements 
of $\tilde\Gamma$ such that their central
charges $Z_{\alpha_i}$, $i=1,2,3$ are arranged
in clockwise order in $c_-$. Then
$\alpha_{ij}\equiv \langle\alpha_i, \alpha_j\rangle >0$
for $i<j$. For definiteness we shall choose the
$\alpha_i$'s such that $\alpha_{12}>\alpha_{23}$.
In this case,
in the convention described below
\eqref{esa2},   the different linear
combinations of the $\alpha_i$'s 
will follow clockwise order as we move from
left to right in the list
\eqref{eorder}. Furthermore, as we
move from the left to the right in this list, the central
charges will follow clockwise order in the chamber
$c^-$ and anti-clockwise order in the chamber $c^+$.
We can now ask
the following question: {\it what is the coefficient of
$\bOm^+(\alpha_1) \bOm^+(\alpha_2)
\bOm^+(\alpha_3)$ in the expression 
of  $\bOm^-(\alpha_1+\alpha_2+\alpha_3) -
\bOm^+(\alpha_1+\alpha_2+\alpha_3)$ in terms of
sum of products of $\bOm^+$'s ?} 
We refer to this coefficient as the generic 3-body contribution
to wall-crossing. 
In order that
the KS formula be consistent with the explicit
computation of bound state
degeneracies  of black hole molecules, this coefficient must agree
with the quantity $\ggx(\alpha_1,\alpha_2,\alpha_3)$
computed in \S\ref{squiver}.  

In order to carry out this computation, 
we can pretend that all  $\bOm^+(\alpha)$
other than those for $\alpha=\alpha_1,\alpha_2,\alpha_3$
vanish.
In this case the left hand side of
\eqref{ewallfin}  
takes a
simple form 
$V^+_{\alpha_3} \, 
V^+_{\alpha_2} \, 
V^+_{\alpha_1}$. 
Our task is to manipulate this product so that in the
final expression $V_\alpha$'s follow the ordering of the
$\alpha$'s given in \eqref{eorder}.
For this it is most expedient to expand
each $V^\pm_\alpha$ in \eqref{ewallfin}
in a Taylor series expansion in
$\bOm^\pm(\alpha)$, manipulate the left hand side
so that in each term the $e_\alpha$'s follow
the order \eqref{eorder}, and then identify the
coefficients of each term on two sides. 
Since our goal is to find the term
involving $\bOm^+(\alpha_1)
\bOm^+(\alpha_2)\bOm^+(\alpha_3)$, we  can
focus on the term 
$\bOm^+(\alpha_1)\bOm^+(\alpha_2)\bOm^+(\alpha_3)
e_{\alpha_3}
e_{\alpha_2} e_{\alpha_1}$ in
the Taylor series expansion on the left hand side.
Repeated use of
\eqref{KSalg} 
gives
\be \label{ereorder}
\begin{split}
e_{\alpha_3} e_{\alpha_2} e_{\alpha_1}
= & e_{\alpha_1} e_{\alpha_2} e_{\alpha_3}
+ \kappa(\langle\alpha_2, \alpha_1\rangle)
e_{\alpha_1+\alpha_2} e_{\alpha_3}
+ \kappa(\langle\alpha_3, \alpha_1\rangle)
e_{\alpha_1+\alpha_3} e_{\alpha_2}
\\ &
+ \kappa(\langle\alpha_3, \alpha_2\rangle)
e_{\alpha_1} e_{\alpha_2+\alpha_3}
+ \kappa(\langle\alpha_2, \alpha_1\rangle)
\kappa(\langle\alpha_3, \alpha_1+\alpha_2\rangle)
e_{\alpha_1+\alpha_2+\alpha_3}\, .
\end{split}
\ee
Note that the $e_\beta$'s on the right hand side follow
the order given in \eqref{eorder}.
Since in the Taylor series expansion
on the right hand side of \eqref{ewallfin} $e_{\alpha_1
+\alpha_2+\alpha_3}$ is multiplied by
$\bOm^-(\alpha_1+\alpha_2+\alpha_3)$,
by comparing the coefficient of
$e_{\alpha_1+\alpha_2+\alpha_3}$ in the 
left and right hand side of \eqref{ewallfin} 
we get, from \eqref{ereorder},
\be \label{efres}
\Delta\bOm(\alpha_1+\alpha_2+\alpha_3)
=(-1)^{\alpha_{12}+\alpha_{23}
 +\alpha_{13}}\, \alpha_{12} \left( \alpha_{13} 
+ \alpha_{23}\right) \bOm^+(\alpha_1)\,\bOm^+(\alpha_2)
\, \bOm^+(\alpha_3) +\cdots
\ee
where $\cdots$ represent terms other than the 
one containing the product 
$\bOm^+(\alpha_1)\,\bOm^+(\alpha_2)
\, \bOm^+(\alpha_3)$. 
Happily, this agrees with the result \eqref{eyto2} from
the black hole bound state analysis.

In a similar fashion, let us consider the generic
4-body contribution. We assume the 
same ordering of the different linear combinations
of the $\alpha_i$'s as given in \eqref{e4body}.
Then one finds for the jump across the wall
\begin{eqnarray} \label{e4jump3}
\Delta \bar \Omega(\alpha_1+\alpha_2+\alpha_3+\alpha_4)
&=&- \, \bOm^+(\alpha_1)\,\bOm^+(\alpha_2)
\, \bOm^+(\alpha_3) \, \bOm^+(\alpha_4)\times \nonumber \\
&&\left[ \kappa(\alpha_{12}) \kappa\left( \alpha_{13} + \alpha_{23}\right) \kappa\left(
    \alpha_{14} + \alpha_{24}+\alpha_{34}\right) \nonumber \right.\\
&&+\kappa(\alpha_{13}) 
\kappa\left( \alpha_{14} + \alpha_{34}\right)
\kappa\left(\alpha_{21} + \alpha_{23}+\alpha_{24}\right) \label{eD4body}\\ 
&&\left. +\kappa(\alpha_{23})\kappa(\alpha_{14})
\kappa\left(\left<\alpha_2
    +\alpha_3,\alpha_1+\alpha_4\right>\right)
    \right]+\dots \nonumber\\
&=&(-1)^{1+\sum_{i<j}\alpha_{ij}} \, \bOm^+(\alpha_1)\,\bOm^+(\alpha_2)
\, \bOm^+(\alpha_3) \, \bOm^+(\alpha_4)\times \nonumber\\
&&\left[ \alpha_{12}\, \alpha_{13}\, \alpha_{24}\, + 
 \alpha_{13}\, \alpha_{14}\, \alpha_{24}\, + 
 \alpha_{12}\, \alpha_{23}\, \alpha_{24}\, + 
 \alpha_{14}\, \alpha_{23}\, \alpha_{24}\, \right. \nonumber\\ 
&&\left. + \alpha_{12}\, \alpha_{23}\, \alpha_{34}\, + 
 \alpha_{13}\, \alpha_{23}\, \alpha_{34}\, + 
 \alpha_{14}\, \alpha_{23}\, \alpha_{34}\, + 
 \alpha_{13}\, \alpha_{24}\, \alpha_{34}\, \right]+\dots , \nonumber
\end{eqnarray}
where the dots represent other contributions. 
This again agrees with the result \eqref{enonmot4}
of the black hole bound state analysis.
For
completeness also give the result for the two body 
contribution to $\Delta\bOm$:
\be \label{etwobody}
 \Delta \bar \Omega(\alpha_1+\alpha_2)
 = (-1)^{\alpha_{12}+1} \alpha_{12}
 \, \bOm^+(\alpha_1)\,\bOm^+(\alpha_2)
+\cdots\, .
\ee

Finally note that this method can be easily generalized
to the case when some of the $\alpha_i$'s are equal.
For example if we are looking for a term 
$\prod_i \left(\bOm^+(\alpha_i)\right)^{m_i}$ in the
expression for $\bOm^-\left(\sum_i m_i\alpha_i\right)$, 
then
we must expand the left hand side of \eqref{ewallfin}
so that in $V_{\alpha_i}$ we keep the 
$\left(\bOm^+(\alpha_i) \, e_{\alpha_i}\right)^{m_i}
/ m_i!$
term,  then carry out the rearrangement described
above, and finally identify the coefficient of
$e_{\sum_i m_i \alpha_i}$ in the resulting expression.
Alternatively we could simply use the generic $n$-body
formula for non-identical particles, take the 
limit
when several of the $\alpha_i$'s approach 
each other and then include the symmetry factor
$1/\prod_i m_i!$ in accordance 
with \eqref{esab2}.
It is easy to see how this rule arises from the KS
formula, -- given a factor of $(e_{\beta})^{m}$
for any vector $\beta$
we can replace it by $e_{\beta^{(1)}} e_{\beta^{(2)}}
\cdots e_{\beta^{(m)}}$ for $m$ distinct vectors
$\beta^{(1)}, \dots, \beta^{(m)}$
with slightly different 
phases,\footnote{For this 
manipulation we
can ignore the fact that the charges are allowed to
take values on a lattice.} carry
out the abovementioned manipulations for
rearranging the vectors and then take the limit
when all the $\beta^{(i)}$'s approach each other to
recover the desired result.
 There is {\it a priori} an
ambiguity in this procedure since in the final
configuration the relative
ordering between two vectors which only differ by
the replacement of $\beta^{(i)}$ by $\beta^{(j)}$ for
some pair $(i,j)$ is arbitrary, but this does not affect
the final result since changing this relative order
picks up a commutator factor that vanishes as
$\beta^{(i)}\to \beta^{(j)}$. Thus the only effect
of having identical particles will be the Boltzmann
symmetry factor
$1/m!$.

\subsection{Semi-primitive wall-crossing formul\ae\  and 
generalizations \label{sec_semi}}

A general wall-crossing formula involves
computing $\Omega^-(m\gamma_1+n\gamma_2)$
in terms of $\Omega^+(k\gamma_1+\ell\gamma_2)$
for $k\le m$, $\ell\le n$.
We define
the order of 
the wall-crossing formula as the
smaller of $m$ and $n$. In this and the following
subsections we
give wall-crossing formul\ae\ for increasing order, starting
with order one in this section and ending at order three. 
For simplicity we shall give the result for the case when
$m$ is fixed to be 1, 2 or 3, but the result can be easily
generalized to the case when $n$ is 1, 2 or 3
(see \S\ref{soppo}).
In the D6-D0 example described
in appendix \ref{secD0D6}, order
corresponds to the number of D6-branes or the rank of the sheaf.

\subsubsection{Order one\label{sec_rank1}}

To extract the semi-primitive wall-crossing formula from the KS formula, 
we project \eqref{KSprodprim}, 
(\ref{ewallfin}) to the algebra $\cA_{1,\infty}$:
Thus we have
\be
A_{\gamma_1,\gamma_2}=V^+_{\gamma_1}\, V^+_{\gamma_1+\gamma_2}\, 
V^+_{\gamma_1+2\gamma_2}\cdots V^+_{\gamma_2}
= V^-_{\gamma_2} \cdots V^-_{\gamma_1+2\gamma_2}\,
V^-_{\gamma_1+\gamma_2}\, V^-_{\gamma_1}
\ .
\ee
Noting that $e_{\gamma_1+N\gamma_2}$ all commute in $\cA_{1,\infty}$,
this can be rewritten as 
\be
\label{AXY}
A_{\gamma_1,\gamma_2} = e^{X^+_1}\, e^Y
= e^Y e^{X_1^-}
\ee
where 
\be
\label{defXY}
X^\pm_1 = \sum_{N=0}^{\infty} 
\bOm^\pm\argu{}{N}
\, e_{\gamma_1+ N \gamma_2}\ ,\qquad
Y=\sum_{\ell=1}^{\infty}  \bOm^+(\ell \gamma_2)\, e_{\ell \gamma_2}
=\sum_{\ell=1}^{\infty}  \bOm^-(\ell \gamma_2)\, e_{\ell \gamma_2}\ .
\ee

It follows from (\ref{AXY}) that
\be \label{ex1+}
X_1^- = e^{-Y} X_1^+ e^Y\, .
\ee
To evaluate the right hand side of (\ref{ex1+})
we first observe that, for a single term in $Y$,
\be \label{emiss}
e^{-\bOm^+(\ell \gamma_2) e_{\ell\gamma_2} }
\, e_{\gamma_1+ N \gamma_2}\, e^{\bOm^+(\ell \gamma_2) e_{\ell\gamma_2} }
= \sum_{n_{\ell}=0}^{\infty}
\frac{1}{n_\ell !}
\left[ \kscom{ \ell\gamma_{12}}
 \, \bOm^+(\ell \gamma_2) \right]^{n_{\ell}}\, 
e_{\gamma_1+ (N+\ell n_\ell)  \gamma_2}\ .
\ee
Thus,
\be
e^{-Y}\, e_{\gamma_1+ N \gamma_2}\, e^{Y}
= 
\sum_{\{n_k\}}
\left( \prod_\ell
\frac{
\left[
\kscom{ \ell\gamma_{12}}\, \bOm^+(\ell \gamma_2) 
\right]^{n_{\ell}}}{ n_\ell !}\right) \, 
e_{\gamma_1+ (N+\sum_{\ell} \ell n_\ell)  \gamma_2}\ ,
\ee
and from eq, (\ref{ex1+}),
\be
\label{eYXY}
X_1^- = 
\sum_{N=0}^{\infty}
\bOm^+\argu{}{N}
\sum_{\{n_k\}}
\left( \prod_\ell
\frac{
\left[
\kscom{ \ell\gamma_{12}}\, \bOm^+(\ell \gamma_2) \right]^{n_{\ell}}}{ n_\ell !}\, 
\right) e_{\gamma_1+ (N+\sum_{\ell} \ell n_\ell)  \gamma_2}\, .
\ee
Using the relation between $X_1^-$ and 
$\bOm^-\argu{}{N}$
given in (\ref{defXY}) we now get
\be
\label{defOm1m}
\bOm^-\argu{}{N} = \sum_{N'=0}^{N}
\bOm^+\argu{}{N'}\, \Omega_{\rm halo}(\gamma_1,N-N')
\ee
where we defined 
\be
\label{Omhalo}
\Omega_{\rm halo}(\gamma_1,N) = \sum_{\substack{\{n_k\} \\\sum k n_k = N}}
\prod_\ell
\frac{
\left[
\kscom{ \ell\gamma_{12}}\,  \bOm^+(\ell \gamma_2) \right]^{n_{\ell}}}{ n_\ell !}\ .
\ee
In terms of the 
partition function $Z^\pm(1,q)=\bar Z^\pm(1,q)$ defined in \eqref{defZNq}, we  obtain
\be
\label{Z1pm}
Z^-(1,q) =Z^+(1,q)\, Z_{\rm halo}(\gamma_1, q)
\ee
where
\be
\label{Z1pmh}
Z_{\rm halo}(\gamma_1, q) =
\sum_{N=0}^{\infty}
\Omega_{\rm halo}(\gamma_1, N)\, q^N 
=\exp\left(
\sum_{\ell=1}^{\infty} (-1)^{\langle \gamma_1, \ell \gamma_2 \rangle}
\langle \gamma_1, \ell\gamma_2 \rangle\, \bOm^+(\ell\gamma_2)\, q^\ell \right)\ ,
\ee
reproducing \eqref{epart2}, (\ref{edefhalo}).

\subsubsection{Order two \label{sec_rank2}}

We now extend the semi-primitive wall-crossing formula to order 2, i.e. 
compute $Z^-(2,q)$. To this aim we project \eqref{KSprodprim},
(\ref{ewallfin}) to the algebra 
$\cA_{2,\infty}$, 
\be
A_{\gamma_1,\gamma_2} =
\left[\prod_{k=0}^{\infty} V^+_{\gamma_1+k\gamma_2}\, V^+_{2\gamma_1+(2k+1)\gamma_2}\right]\, 
\cdot
V^+_{\gamma_2}
=V^-_{\gamma_2}\cdot \left[\prod_{k=\infty}^{0} 
V^-_{2\gamma_1+(2k+1)\gamma_2}\, V^-_{\gamma_1+k\gamma_2}
\right]\, .
\ee
We can combine all factors of
$V^\pm_{\gamma_1+N\gamma_2}$ 
and $V^\pm_{2\gamma_1+N\gamma_2}$ on either side
into a single exponential,
by using the level two truncation \eqref{BCH2} of the BCH formula.
Thus, we now have
\be
\label{AX2X1Y}
A_{\gamma_1,\gamma_2} = e^{X_1^+ +X_2^+}\, e^Y
= e^Y \, e^{X_1^-+X_2^-}\, ,
\ee
where $X_1^\pm,Y$ are the same as in \eqref{defXY},
while
\be
\label{defX2p}
X_2^\pm= \sum_{N=0}^{\infty} \tOm^\pm_2(N)
\, e_{2\gamma_1+ N \gamma_2} 
\ee
\be
\label{defOm2p}
\tOm^\pm_2(N)\equiv\bOm^\pm(2\gamma_1+ N \gamma_2) 
\pm \frac14 \sum_{i+j=N}
\kscom{|j-i| \gamma_{12}} \,
\bOm^\pm\argu{}{i}\, \bOm^\pm\argu{}{j}\ .
\ee
Eq.(\ref{AX2X1Y}) now implies that
\be 
\label{ex2yrel}
X_1^-= e^{-Y} X_1^+ e^Y \, \qquad X_2^-= e^{-Y} X_2^+ e^Y \, . 
\ee
Using the analog of (\ref{defOm1m}) with $\gamma_1\to2\gamma_1$
and eq, (\ref{defX2p}) we get
\be
\label{defOm2m}
\tOm^-_2 (N)=
\sum_{N'=0}^{N}
\tOm^+_2(N')\, \Omega_{\rm halo}(2\gamma_1,N-N')\ ,
\ee
where $\Omega_{\rm halo}$ is defined in \eqref{Omhalo}. Combining 
this  with  \eqref{defOm1m}, \eqref{defOm2p}, we arrive at 
\be \label{eordtwo}
\begin{split}
\bOm^-\argu{2}{N} =& \sum_{0\leq N'\leq N}
\bOm^+\argu{2}{N'}\, \Omega_{\rm halo}(2\gamma_1,N-N')\\
&\hspace*{-2cm}  + \frac12
\sum_{0\leq i<j; i+j\leq N}
\kscom{(j-i) \gamma_{12}}\, 
\bOm^+\argu{}{i}\, \bOm^+\argu{}{j}\, \Omega_{\rm halo}(2\gamma_1,N-i-j)\\
&\hspace*{-2cm} - \frac12
\sum_{\substack{0\leq j<i, i+j=N\\ 0\leq i'\leq i, 0\leq j'\leq j}}
\kscom{(j-i) \gamma_{12}} \,
\bOm^+\argu{}{i'}\, \bOm^+\argu{}{j'}\ \, \Omega_{\rm halo}(\gamma_1,i-i')\,
\Omega_{\rm halo}(\gamma_1,j-j').
\end{split}
\ee
This result generalizes eq. (4.10) in \cite{Chuang:2010wx}. 

The partition functions \eqref{defZNq} for $M=2$ are most conveniently expressed 
in terms  of ``modified partition functions"
\be
\widetilde Z_2^\pm(q) = \sum_{N=0}^{\infty} \tOm^\pm_2(N)\, q^N\ ,
\ee
as 
\be
\bar Z^\pm(2,q) = \widetilde 
Z_2^\pm(q) \pm \frac12 \bar Z^\pm(1,q) \star 
\bar Z^\pm(1,q)\ ,
\ee
where the star product is defined by 
\be
\sum_{i=0}^{\infty} a_i q^i \, \star 
\sum_{i=0}^{\infty} b_j q^j =  
\sum_{0\le i<j} \kscom{(i-j)\gamma_{12}} \, a_i \, b_j\, q^{i+j}\ . 
\ee
The relation \eqref{defOm2m} then simplifies to  a simple wall-crossing identity for 
the modified partition functions,
\be
\label{Z2pm}
\widetilde Z_2^-(q) = \widetilde 
Z_2^+(q)\, Z_{\rm halo}(2\gamma_1,q)
\ee
where $Z_{\rm halo}(2\gamma_1,q)$ is given by the same formula as in 
\eqref{Z1pmh} with $\gamma_1\to 2\gamma_1$. 
Thus, the effective description is still in terms of a halo of 
non-interacting Boltzmannian particles around a core with effective 
degeneracy $\tOm_2^\pm(N)$.

\subsubsection{Order three\label{sec_rank3}}
The projection of \eqref{KSprodprim}, \eqref{ewallfin}
to the algebra 
$\cA_{3,\infty}$ reads 
\ben \label{ebeg}
A_{\gamma_1,\gamma_2} &=& \left[\prod_{k=0}^{\infty}
V^+_{\gamma_1+k\gamma_2}\, 
V^+_{3\gamma_1+(3k+1)\gamma_2}\, 
V^+_{2\gamma_1+(2k+1)\gamma_2}
V^+_{3\gamma_1+(3k+2)\gamma_2}\right] \cdot
V^+_{\gamma_2}\nonumber \\
&=& 
V^-_{\gamma_2}\cdot \left[\prod_{k=\infty}^{0}
V^-_{3\gamma_1+(3k+2)\gamma_2}
V^-_{2\gamma_1+(2k+1)\gamma_2}
V^-_{3\gamma_1+(3k+1)\gamma_2}\, 
V^-_{\gamma_1+k\gamma_2}\right]\, .
\een
We can combine factors according to the $\gamma_1$-charge 
by using the level three truncation of the BCH formula:
\begin{eqnarray}
\label{genBCH}
\log\left(e^{X_0}e^{X_1}\dots e^{X_N} \right)&=&\sum^N_{i=0} X_i
+\frac{1}{2}\sum_{0\leq i<j\leq N} \left[X_i,X_j\right] + \frac{1}{12} \sum_{i,j= 0}^N \left[X_i,\left[X_i,X_j\right]\right]\nonumber \\
&& + \frac{1}{6}\left( \sum_{0\leq i<j<k\leq N} + \sum_{0\leq k<j<i\leq N}\right) \left[X_i,\left[X_j,X_k\right]\right]+\dots.
\end{eqnarray} 
Thus, we now have
\be
\label{AX3X2X1Y}
A_{\gamma_1,\gamma_2} = 
e^{X_1^++X_2^++X_3^+}\, e^Y
=  e^Y\, e^{X_1^-+X_2^-+X_3^-}\, .
\ee
$X_2^\pm,X_1^\pm,Y$ are the same as in  
\eqref{defX2p},\eqref{defXY}, \eqref{AXY}
while
\be
\label{defX3p}
X_3^\pm= \sum_{N=0}^{\infty} \tOm^\pm_3
(3\gamma_1+ N \gamma_2)
\, e_{3\gamma_1+ N \gamma_2} 
\ee
where we defined 
\be
\label{defOm3p}
\begin{split}
\tOm^\pm_3(N)=&\bOm^\pm(3\gamma_1+ N \gamma_2) \\
&\pm \frac12  \sum_{i+j=N} \kscom{|j-2i|\,  \gamma_{12} }\,
\bOm^\pm(\gamma_1+i\gamma_2)\, \bOm^\pm(2\gamma_1+j \gamma_2)\\ &
+\frac1{12} \sum_{2i+j=N} 
\left[ \kscom{(j-i) \gamma_{12}} \right]^2 \, [\bOm^\pm(\gamma_1+i\gamma_2)]^2\,
\bOm^\pm(\gamma_1+j\gamma_2)\\&
+  \sum_{\substack{0\leq i<j<k\\i+j+k=N}}
\kappa(i,j,k)\, \bOm^\pm\argu{}{i}\, \bOm^\pm\argu{}{j} \, \bOm^\pm\argu{}{k}\ ,
\end{split}
\ee
\be
\label{defkijk}
\begin{split}
\kappa(i,j,k) \equiv
\frac16 \kscom{(i-j)\gamma_{12}}\, \kscom{(i+j-2k)\gamma_{12}} 
+  \frac1{6}\kscom{(k-j)\gamma_{12}}\,\kscom{(j+k-2i)\gamma_{12}}\ .
\end{split}
\ee
Note that \eqref{defkijk} may be written differently using   
the Jacobi identity
\be
-\kscom{(j-i)\gamma_{12}}\, \kscom{(i+j-2k)\gamma_{12}} 
-\kscom{(k-j)\gamma_{12}}\,\kscom{(j+k-2i)\gamma_{12}}\, 
+ \kscom{(k-i)\gamma_{12}}\,\kscom{(i+k-2j)\gamma_{12}} = 0\ ,
\ee
and that $\kappa(i,j,k)$ is symmetric under exchange $i\leftrightarrow k$.
As before, 
eq, \eqref{AX3X2X1Y}, besides leading to eqs.\eqref{ex1+}
and \eqref{ex2yrel}, gives
\be \label{ex3eq}
X_3^- = e^{-Y} X_3^+ e^Y\, .
\ee
Using the analog of \eqref{defOm1m} with $\gamma_1\to
3\gamma_1$ and eqs.\eqref{defX3p}
we get
\be
\label{defOm3m}
\tilde\Omega^-_3(N) = \sum_{0\leq N'\leq N}
\tilde\Omega^+_3(N')\, \Omega_{\rm halo}(3\gamma_1,N-N')
\ee
where $\Omega_{\rm halo}$ is defined in \eqref{Omhalo}.
Putting together \eqref{defOm3p} and \eqref{defOm3m}, we
arrive at  
\be
\label{Om3pm}
\begin{split}
&\bOm^-(3\gamma_1+N\gamma_2)=
\sum_{N'}  \bOm^+(3\gamma_1+ N' \gamma_2) \,  
\Omega_{\rm halo}(3\gamma_1,N-N')\\
&+ \frac12  \sum_{\substack{i\geq 0,j \geq 0\\ i+j\leq N}} \kscom{|j-2i|\,  \gamma_{12} }|\,
\bOm^+(\gamma_1+i\gamma_2)\, \bOm^+(2\gamma_1+j \gamma_2)\,
\Omega_{\rm halo}(3\gamma_1,N-i-j)\\ 
&
+\frac1{12} \sum_{\substack{i\geq 0,j \geq 0\\ 2i+j\leq N}} 
\left[ \kscom{(j-i) \gamma_{12}} \right]^2 \, [\bOm^+(\gamma_1+i\gamma_2)]^2\,
\bOm^+(\gamma_1+j\gamma_2)\, \Omega_{\rm halo}(3\gamma_1,N-2i-j)\
\\&
+   \sum_{\substack{0\leq i<j<k\\ i+j+k\leq N}}
\kappa(i,j,k)\, \bOm^+\argu{}{i}\, \bOm^+\argu{}{j} \, 
\bOm^+\argu{}{k}\, \, \Omega_{\rm halo}(3\gamma_1,N-i-j-k)\ ,\\
&+\frac12 \sum_{\substack{i+j=N\\ 0\leq i'\leq i,0\leq j'\leq j}}
 \kscom{|j-2i|  \gamma_{12}} \,
\bOm^+(\gamma_1+i'\gamma_2)\, \bOm^+(2\gamma_1+j' \gamma_2)\, 
\Omega_{\rm halo}(\gamma_1,i-i') \Omega_{\rm halo}(2\gamma_1,j-j')\\ 
&+\frac18 \sum_{\substack{i+j=N\\ 0\leq i'\leq i; k+l\leq j}}
 \kscom{|j-2i|  \gamma_{12}} \,
 \kscom{|l-k| \gamma_{12}} 
\bOm^+(\gamma_1+i'\gamma_2)\, \bOm^+(\gamma_1+k \gamma_2)\, 
 \bOm^+(\gamma_1+l \gamma_2)\, \\ &
\hspace*{3cm} \Omega_{\rm halo}(\gamma_1,i-i') \Omega_{\rm halo}(2\gamma_1,j-k-l)
\\ &
-\frac{1}{12} \sum_{\substack{2i+j=N\\ 0\leq i',i''\leq i,0\leq j'\leq j}} 
\left[ \kscom{(j-i) \gamma_{12}} \right]^2 \, 
\bOm^+(\gamma_1+i'\gamma_2)\,\bOm^+(\gamma_1+i''\gamma_2)\,\bOm^+(\gamma_1+j'\gamma_2)\, \\
&
\hspace*{3cm}
\Omega_{\rm halo}(\gamma_1,i-i')\, \Omega_{\rm halo}(\gamma_1,i-i'')\, 
\Omega_{\rm halo}(\gamma_1,j-j')
\\&
-  \sum_{\substack{
i+j+k=N;  i>j>k\geq 0\\
0\le i'\le i; 0\le j'\le j; 0\le k'\le k}}
\kappa(i,j,k)\, 
\bOm^+\argu{}{i'}\, \bOm^+\argu{}{j'} \, \bOm^+\argu{}{k'}\,\\
&
\hspace*{3cm}\,  \Omega_{\rm halo}(\gamma_1,i-i') \, \Omega_{\rm halo}(\gamma_1,j-j')
\, \Omega_{\rm halo}(\gamma_1,k-k')
\\&
+  \frac18 \!\!\!\!\sum_{\substack{i+j+k=N\\
 0\leq i'\leq i,0\leq j'\leq j,0\leq k'\leq k}}\!\!\!\!
\kscom{|j+k-2i| \gamma_{12}}\, 
\kscom{|j-k| \gamma_{12}}
\bOm^+\argu{}{i'}\, \bOm^+\argu{}{k'} \, \bOm^+\argu{}{j'}\,\\
&
\hspace*{3cm}\,  \Omega_{\rm halo}(\gamma_1,i-i') \, \Omega_{\rm halo}(\gamma_1,j-j')
\, \Omega_{\rm halo}(\gamma_1,k-k') .
\end{split}
\ee

As before, we may introduce 
the ``modified partition functions"
\be
\tilde Z_3^\pm(q) = \sum_{N=0}^{\infty} \tilde\Omega^\pm_3(N)\, q^N ,
\ee
in terms of which 
the wall-crossing relation takes the simple form
\be
\label{Z3pm}
\tilde Z_3^-(q) = \tilde Z_3^+(q)\, Z_{\rm halo}(3\gamma_1,q)\ ,
\ee
where $Z_{\rm halo}(3\gamma_1,q)$  is defined in \eqref{Z1pmh}.
The modified partition functions are related to the original partition functions
$\bar Z^\pm(3,q)$ by a relation of the form
\be
\tilde Z_3^\pm(q) =  \bar Z^\pm(3,q) + 
\bar Z^\pm(1,q) \star_\pm \bar Z^\pm(2,q)+ 
 \bar Z^\pm(1,q) \star_\pm \bar Z^\pm(1,q) \star_\pm
 \bar Z^\pm(1,q)
\ee
with the  definitions of the 'star' products following from
\eqref{defOm3p}.

\subsection{$\gamma_{12}>0$ case} \label{soppo}

We shall now briefly discuss what happens 
when $\gamma_{12}>0$. 
We can of course use all the formul\ae\  derived in this
section with $\gamma_1\leftrightarrow \gamma_2$, but
then {\it e.g.} eq, \eqref{Om3pm} will give the wall-crossing
formula in the charge sector $N\gamma_1+3\gamma_2$.
So if we want to find the wall-crossing formula for
$3\gamma_1+N\gamma_2$ we cannot get the
result by exchanging $\gamma_1$ and $\gamma_2$
in \eqref{Om3pm}. Instead we keep 
$\gamma_{12}>0$ and carefully examine how the
subsequent equations are affected. It is easy to see that
the only place where the sign of $\gamma_{12}$ enters
is in eq, \eqref{etest}; for $\gamma_{12}>0$ the
$<$ sign in \eqref{etest} is replaced
by a $>$ sign. So if we continue to define $\Omega^\pm$
as in \S\ref{spril} then $\Omega^+$ will
denote the index in the
chamber in which the multi-centered
bound states of black hole molecules exist. We would however
like to define $\Omega^+$ as the index associated with
single black hole molecules, and for this
we exchange the definitions of $\Omega^+$
and $\Omega^-$. 
Thus for example in eq, \eqref{Om3pm} we have to now
exchange $\Omega^+$ with $\Omega^-$ so that we have
an expression for $\Omega^+$ in terms of $\Omega^-$.
We can in principle solve these equations iteratively to find
$\Omega^-$ in terms of $\Omega^+$, but we shall now
suggest a simpler method. For this note that exchanging
$\Omega^+$ and $\Omega^-$ in 
the wall-crossing formula
\eqref{ewallfin} is equivalent to changing the order of the
products on both sides of \eqref{ewallfin}.
This in turn is equivalent to keeping the same order
as in \eqref{ewallfin} but 
changing the sign of all the structure
constants in the algebra \eqref{KSalg}. This
can be
achieved by changing $\gamma_{12}$ to
$-\gamma_{12}$.
Thus 
if we replace $\gamma_{12}$ by $-|\gamma_{12}|$
in the formul\ae\ we have derived ({\it e.g.}
\eqref{Om3pm}), we shall get the
correct wall-crossing formula for both signs of
$\gamma_{12}$.

\subsection{Refined wall-crossing and motivic invariants} \label{srefined}

We have already introduced the refined invariants in
\S\ref{smotivic}. In this subsection we shall review the
KS motivic wall-crossing formula that computes the
jump in the refined index across walls of marginal stability
and compare it with our wall-crossing formula based
on the analysis of supergravity bound states.

In order to describe the motivic generalization of the wall-crossing
formula \cite{ks}, we 
consider a set of generators $\{\hat e_\gamma, \gamma\in\Gamma\}$ 
satisfying the quantum torus relations
\be \label{edefhate}
\hat e_{\gamma} \, \hat e_{\gamma'} = (-y)^{ \langle \gamma,\gamma'\rangle}
\, \hat e_{\gamma+\gamma'}\ .
\ee
The associated Lie algebra is 
\be \label{eliealg}
[\hat e_{\gamma} , \hat e_{\gamma'}] = 
\left( (-y)^{\langle \gamma,\gamma'\rangle} 
- (-y)^{- \langle \gamma,\gamma'\rangle}\right)\, 
\, \hat e_{\gamma+\gamma'}\ .
\ee
Let us also introduce the quantum dilogarithm,
\be
\qli2(x) = \sum_{n=0}^{\infty} \frac{(x y)^n}{(1-y^2)\dots (1-y^{2n})} =
\prod_{n=0}^{\infty} (1+(-y)^{2n+1} x)^{-1} .
\ee
This satisfies the pentagon identity
\be
\label{penta}
\qli2(x_1) \qli2(x_2) = \qli2(x_2) \qli2(x_{12}) \qli2(x_1)
\ee
where  $x_1, x_2$ are two non-commutative variables satisfying 
$x_1 x_2/y = y x_2 x_1 \equiv - x_{12}$, 
and reduces to the ordinary dilogarithm for $\log y\equiv 
\nu \to 0$,
\be
\qli2(x) = \exp\left( -\frac{1}{2\nu} \Li_2(x) + \frac{x}{12(1-x)}
 \nu+ \frac{7x(1+x) }{720(x-1)^3}\nu^3+ \dots \right)\, .
\ee

We attach to the charge vector $\gamma$ the 
generator\footnote{In supersymmetric gauge theories,
for a vector multiplet 
with unit degeneracy, $\hat U_\gamma$ reduces to
$
\hat U_\gamma = \qli2\left( y \hat e_\gamma\right)\,
\qli2\left( y^{-1} \hat e_\gamma\right)
$
while for a hypermultiplet one has
$
\hat U_\gamma = \qli2\left( e_\gamma\right)^{-2}\ .
$
}
\be
\hat U_\gamma = \prod_{n\in\IZ} \qli2\left( y^n \hat e_\gamma\right)^{-(-1)^n  
\Omega_{{\rm ref},n}(\gamma)}\ .
\ee
The motivic  version of the KS wall-crossing formula \cite{ks,Dimofte:2009bv,Dimofte:2009tm} 
again states that the product
\be
\label{KSprodmot}
\hat A_{\gamma_1,\gamma_2} = \prod_{\substack{\gamma=M\gamma_1+N\gamma_2,\\
M\geq 0, N\geq 0}} \hat U_\gamma\ ,
\ee
ordered 
so that as we move from left to right the corresponding
$Z_\gamma$'s are ordered clockwise, 
stays
constant across the hyperplane of marginal stability $\cP(\gamma_1,\gamma_2)$.

As in the classical case, it is advantageous to combine the generators $\hat U_{k\gamma}$
for $k\geq 1$ in a single factor $\hat V_\gamma$. For this purpose, rewrite the 
operator $\hat U_\gamma$, using the identity for the quantum dilogarithm
\be
\qli2(x) = \exp\left( \sum_{k=1}^{\infty} \frac{(x y
    )^k}{k(1-y^{2k})}\right)\ ,
\ee
as
\be
\hat U_\gamma = \exp\left(  \sum_{k=1}^{\infty} 
\frac{\Omega_{\rm ref}(\gamma, y^k)}{k\, (y^k - y^{-k})} \hat e_{k\gamma} \right)\, .
\ee
Then the product of $\hat U_{\ell\gamma}$ over all $\ell\geq 1$, $\gamma$ being a primitive
vector, can be written in terms of $\bar \Omega_\mathrm{ref}(\gamma,y)$ defined in
\eqref{bOmref}:
\be
\label{Vquant}
\hat V_{\gamma} = \prod_{\ell\geq 1} \hat U_{\ell\gamma} = 
\exp\left(\sum_{N=1}^{\infty} \bar
\Omega_{\rm ref}(N\gamma, y)\, 
\tilde e_{N\gamma} \right), \quad 
\tilde e_\gamma \equiv \frac{ \hat e_\gamma}{y-y^{-1}}\, .
\ee
This is the motivic generalization of \eqref{Vclas}. 
The wall-crossing formula now takes the form
\be \label{ewallfinref}
\prod_{\substack{M\geq 0,N\geq 0>0, \\ \gcd(M,N)=1, M/N\downarrow}}
\hat V^+_{M\gamma_1+N\gamma_2}
=
\prod_{\substack{M\geq 0,N\geq 0>0, \\ \gcd(M,N)=1, M/N\uparrow}}
\hat V^-_{M\gamma_1+N\gamma_2}\, ,
\ee
where $\hat V^\pm$ are computed using the 
$\bOm_{\rm ref}(\gamma, y)$ in the chambers
$c^\pm$.
It follows from \eqref{eliealg}, \eqref{edefte}, \eqref{Vquant}
 and
\eqref{ewallfinref} that expressed in terms of
$\bar\Omega_{\rm ref}(\gamma, y)$ the wall-crossing
formula will satisfy manifest `charge conservation laws'.

To see how this reduces to the classical KS formula \eqref{ewallfin} in the limit $y\to 1$,
note that the generators
\be \label{edefte}
e_\gamma = \lim_{y\to 1} \tilde e_\gamma\, , \qquad 
\tilde e_\gamma \equiv \frac{ \hat e_\gamma}{y-y^{-1}}
\ee
satisfy the commutation relations \eqref{KSalg}. 
Moreover in this limit $\Omega_{\rm ref}(\gamma, y)$
approaches $\Omega(\gamma)$. Thus 
$\hat V^\pm$ defined in \eqref{Vquant}
reduce to $V^\pm$ and we
recover \eqref{ewallfin}.

\subsection{Semi-primitive refined wall-crossings
and its generalizations} \label{ssemiref}

The rescaled generators 
$\tilde e_\gamma=\hat e_\gamma/(y-y^{-1})$
satisfy the same Lie algebra as \eqref{KSalg}, 
\be
[\tilde e_{\gamma} , \tilde e_{\gamma'}] = 
\kappa(\langle\gamma,\gamma'\rangle,y)\, 
\, \tilde e_{\gamma+\gamma'}\ ,
\ee
provided $\kappa(x)$ is replaced by its quantum deformation 
\be
 \kappa(x,y) \equiv \displaystyle{\frac{ (-y)^{x} - (-y)^{-x}}{y-1/y}}
= (-1)^x \sinh(\nu\, x) / \sinh\nu, 
\quad \nu\equiv \ln y \ . 
\ee
Moreover, the generators $\hat V_\gamma$ in \eqref{Vquant} can be
obtained from their classical counterpart \eqref{Vclas} by replacing 
$\bOm(\gamma)\mapsto \bOm_{\rm ref}(\gamma,y), e_\gamma\mapsto 
\tilde e_\gamma$. Therefore, the wall-crossing formul\ae\  derived in 
Section \ref{sec_KS} and Appendix \ref{sswall} carry over to the motivic
case by just replacing 
\be \label{ereplace}
\Omega(\gamma)\mapsto
\Omega_{\rm ref}(\gamma,y)\ ,\qquad
\bOm(\gamma)\mapsto
\bOm_{\rm ref}(\gamma,y)\ , \qquad \kappa(x) \mapsto 
\kappa(x,y) \ .
\ee
In particular,  the primitive wall-crossing formula takes
the form\cite{Diaconescu:2007bf,Dimofte:2009bv}
\be
\label{wcref}
\Delta\Omega_{\rm ref}(\gamma\to \gamma_1+\gamma_2,y) 
=\frac{
(-y)^{\langle \gamma_1,\gamma_2\rangle}-
(-y)^{- \langle\gamma_1,\gamma_2\rangle}}{y-1/y}\,
\Omega_{\rm ref}(\gamma_1,y)\,
 \Omega_{\rm ref}(\gamma_2,y)
\ee
while the refined semi-primitive wall-crossing formula is given by 
\be
\label{Z1pmy}
Z^-(1,q,y) = Z^+(1,q,y)\,Z_{\rm halo}(\gamma_1,q,y) 
\ee
where
\be \label{ezhalo}
Z_{\rm halo}(\gamma_1,q,y) \equiv 
\exp\left(
\sum_{\ell=1}^{\infty} 
 \frac{ (-y)^{\langle \gamma_1, \ell \gamma_2 \rangle} - 
(-y)^{- \langle \gamma_1, \ell \gamma_2 \rangle}}{y-y^{-1}}
\,  \bOm_{\rm ref}(\ell\gamma_2,y)\, q^\ell \right)\ .
\ee
On the right hand side 
$\bOm_{\rm ref}(\ell\gamma_2,y)$ can be computed
in either chamber. 
\eqref{ezhalo} is in perfect agreement
with \eqref{eomp1ref}, showing that the Boltzmann gas
picture correctly reproduces the semi-primitive
motivic wall-crossing formula. To compare \eqref{ezhalo}
with known results, we note that
in terms of the ``integer" motivic invariants 
$\Omega_{{\rm ref},n}(\gamma)$, \eqref{ezhalo} becomes an infinite product 
\cite{Dimofte:2009bv}
\be
\label{Zhalomotprod}
Z_{\rm halo}(\gamma_1,q,y) = \prod_{k=1}^{\infty} 
\prod_{j=1}^{k |\gamma_{12}|} 
\prod_{n\in \IZ} 
\left( 1- (-1)^{2j-k |\gamma_{12}|} q^k y^{n+2j
-1-k |\gamma_{12}|}\right)^{ (-1)^n\, 
\Omega_{{\rm ref},n}(k\gamma_2)}\ .
\ee
To see this, note that the logarithm of the r.h.s. of \eqref{Zhalomotprod}
can be rewritten as 
\be
-
\sum_{d\geq 1} \sum_{k\geq 1} \sum_n \sum_{j=1}^{k
\left|\gamma_{12}\right|}
\frac{1}{d}\, (-1)^n\, 
\Omega_{{\rm ref},n}(k\gamma_2) \, \left[
(-1)^{2j-k |\gamma_{12}|} q^k y^{n+2j
-1-k |\gamma_{12}} |\right]^d\ .
\ee
The sum over $n$ leads to 
$\Omega_{{\rm ref},n}(k\gamma_2, y^d)$, while the
sum over $j$ is geometric, leading to 
\be \label{ekshalo}
\log Z_{\rm halo}(\gamma_1,q,y) =
\sum_{d\geq 1} \sum_{k\geq 1} 
\frac{1}{d}\, 
\Omega_{\rm ref}(k\gamma_2, y^d) \, 
\frac{(-y)^{k d \gamma_{12}}-(-y)^{-k d \gamma_{12}}}
{y^{d}-y^{-d}} \, q^{kd}\ ,
\ee
where we have used $\gamma_{12}<0$ to replace 
$|\gamma_{12}|$ by $-\gamma_{12}$.
Setting $\ell=k d$, the sum runs over divisors $d$ of $N$ and 
reproduces \eqref{ezhalo}.

The order two and three motivic wall-crossing formula can
be obtained by making the replacements \eqref{ereplace}
in eqs.\eqref{eordtwo}, \eqref{Om3pm}.

Finally, let us consider the problem
of determining a generic 3-body contribution
to the wall-crossing formula:
given three charge vectors $\alpha_1$, $\alpha_2$ and
$\alpha_3$ in $\tilde \Gamma$, what is the coefficient
of $\bOm_{\rm ref}^+(\alpha_1,y) \bOm_{\rm ref}^{+}(\alpha_2,y)
\bOm_{\rm ref}^{+}(\alpha_3,y)$ in
the expression for $\Delta\bOm_{\rm ref}(\alpha_1
+\alpha_2+\alpha_3,y)$? The analysis is a
straightforward generalization of that in \S\ref{s3decay}
and the final result is obtained by replacing 
$\kappa(\alpha_{ij})$ by $\kappa(\alpha_{ij},y)$
in \eqref{ereorder}. This gives:
\ben \label{efres2ref}
\Delta\bOm(\alpha_1+\alpha_2+\alpha_3, y) &=&
(-1)^{\alpha_{12}+\alpha_{23}
 +\alpha_{13}}\, (\sinh\nu)^{-2}\,
 \sinh(\nu\alpha_{12}) 
 \sinh(\nu(\alpha_{13}+\alpha_{23})) \nonumber \\
&& \bOm^+(\alpha_1, y)\, \bOm^+(\alpha_2, y)\, 
\bOm^+(\alpha_3, y) + \cdots \, .
\een
This is in perfect agreement with \eqref{erei5}
computed from the spectrum of bound states
of a 3-centered configuration in supergravity.
Similarly the 4-body contribution can be computed by replacing
$\kappa(\alpha)$ by $\kappa(\alpha,y)$ in \eqref{e4jump3}.
The result is in perfect agreement with \eqref{e4bodyfin}
and \eqref{elocal7}. 
We have also carried out a similar computation for 5-body
contribution and compared with the results obtained by
following the procedure
of \S\ref{squiver}, but we shall not give the details.

\subsection{KS vs. supergravity} \label{skssugra}

Eventually one would like to prove that the KS wall
crossing formula given in \eqref{ewallfinref}
is equivalent to the one obtained from quantization
of multi-black hole solutions as given in \eqref{efirst3},
\eqref{higgsf}. We have not yet reached this goal,
but would like to point out some common aspects of these
two formul\ae. The summand in \eqref{higgsf}
depends analytically on the $\alpha_{ij}$'s, but the
analyticity of the sum
is broken by the third condition described below
\eqref{higgsf}. In particular this constraint measures
whether $\sum_a\beta^{(a)}$, represented as a vector
in the two dimensional plane in the convention described
below \eqref{esa2}, lies to the left or the right of the
vector $\alpha_1+\cdots + \alpha_n$. 
Let us denote by $\cB$ the set of all vectors of the
form $\sum_i m_i \alpha_i$ with $m_i=0$ or 1, and arrange
them in an order following the convention described
below \eqref{esa2}. 
Let $\cB'$ denote the subset of elements
of $\cB$ which
lie
to the left of the central element
$\alpha_1+\cdots + \alpha_n$. 
In this case the expression
for $g_{\rm ref}$ given in \eqref{higgsf} depends
on the subset $\cB'$.
As we vary
the $\alpha_i$'s this subset may change and in that case
$g_{\rm ref}$ will be given by a different analytic 
function of the $\alpha_{ij}$'s. 
Note however
that $g_{\rm ref}$ does not depend on the relative
ordering of the vectors inside the subset $\cB'$.

Now this lack of analyticity is also manifest in the KS
wall-crossing formula. To see this,  recall the procedure
for manipulating the KS formula given in
\S\ref{s3decay}. Here we are supposed to begin with
the product $e_{\alpha_n}\cdots e_{\alpha_1}$ and
bring it to the order in which the vectors appear
in the set $\cB$.
Changing this order leads to 
a different final order of the $e_\beta$'s
and hence we expect $g_{\rm ref}(\alpha_1,\dots, \alpha_n,
y)$, given by the coefficient of 
$e_{\alpha_1+\cdots + \alpha_n}$
in the final state, to change. This in turn prevents
$g_{\rm ref}$ to be given by an analytic formula involving
the $\alpha_{ij}$'s for all choices of $\alpha_i$.

Now, the KS prescription for computing
$g_{\rm ref}$ {\it a priori}  seems to depend on
more information than \eqref{higgsf} since the KS formula
requires the detailed ordering of the 
vectors in $\cB$,  rather than just the unordered
list of the ones which
lie to the left of $\alpha_1+\cdots + \alpha_n$. We shall now
show that the KS formula in fact 
only depends on the unordered 
list of vectors which lie to the
left of $\alpha_1+\cdots + \alpha_n$. For this let us
consider a given order of all the $\alpha_i$'s and supppose
that we have brought $e_{\alpha_n}\cdots e_{\alpha_1}$
to the required order. Now consider the effect of switching
the relative order between two neighbouring
vectors $\beta_1$
and $\beta_2$ on the left of
$\alpha_1+\cdots + \alpha_n$. This will require  to switch
the corresponding $e_{\beta_1}$ and $e_{\beta_2}$
and will produce an extra factor of $e_{\beta_1+\beta_2}$.
But since $\beta_1$ and $\beta_2$ both lie to the
left of $\alpha_1+\cdots +\alpha_n$, $\beta_1+\beta_2$ will
also lie to the left of $\alpha_1+\cdots +\alpha_n$.
Thus such switchings can never produce a factor of
$e_{\alpha_1+\cdots + \alpha_n}$. The same argument
holds if we switch two vectors on the right of
$\alpha_1+\cdots +\alpha_n$. Thus a term proportional
to $e_{\alpha_1+\cdots +\alpha_n}$ can arise only if
we switch a vector from the left of 
$\alpha_1+\dots \alpha_n$ with a vector to the right of
$\alpha_1+\cdots + \alpha_n$. This shows that the
non-analyticity of 
\eqref{higgsf} and the KS formula 
are controlled by the same data.

\section{Wall-crossing from the Joyce-Song formula} \label{sjoyce}

In their work on Donaldson-Thomas invariants for coherent sheaves on a 
Calabi-Yau three-fold $\cX$ \cite{Joyce:2009xv} (which presumably count 
D6-D4-D2-D0 bound states in type IIA string theory compactified on $\cX$), 
Joyce and Song  give 
a fully explicit expression for the rational DT invariants $\bOm^-$ on
one side of the wall, in terms of the rational DT invariants $\bOm^+$
on the other side. Thus, the JS formula can be viewed as the solution
to the  implicit relation given by KS. In particular, it directly provides 
the functions $g(\{\alpha_i\})$ appearing in \eqref{esab2}, i.e. 
the solution to the black hole bound state problem.  It should be noted
however that the JS wall-crossing formula involves sums over many terms
with large denominators and large
cancellations, and is less computationally efficient that the KS formula
(compare for instance  table \eqref{et4body} with the analogous
computation for KS given in eq. (\ref{eD4body})).
In addition, the simple rule for dealing with identical particles
mentioned at the end of \S\ref{s3decay}  
is not at all obvious from the JS formula.

One way of understanding the large redundancy is that
JS work with Abelian categories, where constituents  
are either a subobject or a quotient object of the complete object. 
In physical terms this means that different terms in the
JS wall-crossing formula keep track of the order in which
the constituents (molecules) make the complete
object (bound molecular cluster). 
But in physics (and in the derived category
on which KS analysis is based) 
such a distinction is not
present.  For example
the existence and index of
a bound state of two primitive constituents 
carrying charges $\gamma_1$ and $\gamma_2$ is 
independent of the
order in which we choose $\gamma_1$ and $\gamma_2$.
For this reason, the JS formula contains many terms which
must combine at the end to ensure the independence
of the final answer of the order in which the constituents
are chosen. The KS formula is less redundant, at the cost of being
 implicit and perhaps 
less rigorously
established.

After stating the JS formula in \S\ref{sstJS} and its
implication for the black hole bound state problem
in \S\ref{sjsindex}, we verify 
the equivalence of the
JS and KS 
formul\ae\ for generic three and four body 
contribution, and semi-primitive
wall-crossing in \S\ref{sexampleJS} and \S\ref{ssemiJS}.

\subsection{Statement of the JS formula}
\label{sstJS}
In \cite{Joyce:2008pc,Joyce:2009xv}, the authors define rational-valued 
generalized Donaldson-Thomas  invariants $\DTb^\gamma(\tau)$ 
for any class $\gamma\in C(\cX)$,  where $\cX$ is a Calabi-Yau three-fold,
$C(\cX)$ is a positive cone\footnote{This is the
analog of the wedge $\tilde \Gamma$ introduced in 
eq. \eqref{ebpsset}.}
inside $K(\cX)$ and $\tau$ is a stability condition. They furthermore establish a general 
wall-crossing formula for the variation of $\DTb^\gamma(\tau)$ under a change
of stability condition. 
Conjecturally, the rational invariants $\DTb^\gamma(\tau)$
are related to integer invariants ${\rm DT}^\gamma(\tau)$ by a relation identical to 
\eqref{efirst}, and 
$\DTb^\gamma(\tau)$, ${\rm DT}^\gamma(\tau)$ 
are  particular instance of  the  invariants $\bOm(\gamma; t),
\Omega(\gamma;t)$ considered in \cite{ks}. The stability condition
$\tau$ is determined by a point in K\"ahler moduli space $t$.
In the convention we have chosen, 
\be 
\tau(\gamma,t)=-\arg Z(\gamma,t)\, .
\ee 
In this section we shall assume that 
the conjectured relation between the rational
and integer invariants holds, and denote the rational DT invariants of \cite{Joyce:2008pc,
Joyce:2009xv} as $\bOm(\gamma; t)$. The JS wall-crossing formula then
furnishes the solution to the KS formula \eqref{ewallfinU}, i.e. expresses
$\bar\Omega^-(\gamma)$ in terms of $\bar\Omega^+(\gamma')$.

Let $\alpha_1,\alpha_2,\dots,\alpha_n$ be $n$ charge vectors in the
positive cone $C(\cX)$ inside the charge lattice
described by  eq, \eqref{ebpsset} , and ($t$,$\tilde t$) be
a pair of points on the \kahler moduli space with associated
stability conditions $(\tau,\tilde \tau)$. 
To express the JS wall-crossing formula, we first 
need to introduce two functions
$S(\alpha_1,\dots, \alpha_n;t,\tilde t)$ and
$U(\alpha_1,\dots, \alpha_n;t,\tilde t)$, 
whose role is to capture the relevant information
about  the ordering of phases of
$\{Z_{\alpha_i}\}$.   
We define $S(\alpha_1,\ldots,\alpha_n;t,\tilde t)\in \{0,\pm 1\}$ as follows.
If $n=1$, set $S(\alpha_1;t,\tilde t)=1$. 
If $n>1$ and, for every $i=1,\dots,  n-1$, either 
\beq
(a) &&\qquad \tau(\alpha_i)\leq \tau(\alpha_{i+1})\quad \mbox{and}\quad
\tilde \tau(\alpha_1+\cdots+\alpha_i)>\tilde \tau(\alpha_{i+1}+\cdots+\alpha_n), \quad\mbox{or}\, 
\nn \\
(b) && \qquad \tau(\alpha_i)>\tau(\alpha_{i+1})\quad \mbox{and}\quad
\tilde \tau(\alpha_1+\cdots+\alpha_i)\le\tilde \tau(\alpha_{i+1}+\cdots+\alpha_n)\ ,
\label{abalt}
\eeq
let $S(\alpha_1,\ldots,\alpha_n;t,\tilde t)=(-1)^r$, where $r$ is the 
number of times option (a) is realized; otherwise, 
$S(\alpha_1,\ldots,\alpha_n;t,\tilde t)=0$. 

To define $U(\alpha_1,\ldots,\alpha_n;t,\tilde t)$, 
consider all ordered partitions of the $n$ vectors 
$\alpha_i$ into $1\leq m\leq n$ packets $\{\alpha_{a_{j-1}+1},\cdots,\alpha_{a_j}\}$, $j=1,\dots ,m$,
with $0=a_0<a_1<\cdots<a_m=n$, such that all vectors in each packet have the same phase $\tau(\alpha_i)$.
Let $\beta_j=\alpha_{a_{j-1}+1}+\cdots+\alpha_{a_j}, j=1,\dots, m$ be the sum of the charge vectors in each packet. We refer to the ordered set $\{\beta_j, j=1,\dots ,m\}$ as a contraction 
of $\{\alpha_i\}$. 
Next, consider all ordered partitions of the $m$ vectors 
$\beta_j$
into $1\leq l\leq m$ packets
$\{\beta_{b_{k-1}+1},\cdots,\beta_{b_k}\}$, with
$0=b_0<b_1<\cdots<b_l=m$, $k=1,\dots, l$, such that the total charge
vectors $\delta_k=\beta_{b_{k-1}+1}+\cdots+\beta_{b_k}, k=1,\dots, l$ in
each packets all have the same phase $\tilde\tau(\delta_k)$ (which is
then equal to $\tilde \tau(\sum\alpha_i)$).  
Now associate to each of the $l$ packets in the 
contraction a 
factor $S(\beta_{b_{k-1}+1},\beta_{b_{k-1}+2},\ldots,\beta_{b_k}; t,
\tilde t)$ as defined above, and
define the $U$-factor as the sum
\be
U(\alpha_1,\ldots,\alpha_n;t,\tilde t)\equiv \sum
\frac{(-1)^{l-1}}{l}\cdot\prod\nolimits_{k=1}^l S(\beta_{b_{k-1}+1},
\beta_{b_{k-1}+2},\ldots,\beta_{b_k}; t,\tilde t)\, 
\cdot\prod_{j=1}^m\frac{1}{(a_j-a_{j-1})!}\,
\ee
over all partitions of $\alpha_i$ and partitions of $\beta_j$
satisfying the conditions above. 

Finally, departing slightly from the presentation in \cite{Joyce:2008pc}, 
let us define the $\cL$ factor
\be
\cL(\alpha_1,\dots ,\alpha_n) \equiv 
 \sum_{{\rm connected\,oriented\, trees:} \atop {\rm vertices}\, \{1,\dots,n \},\, 
{\rm edge}\, i\to j,\,
{\rm implies}\, i<j }
 \prod\limits_{{\rm edges}\,i\to j}  \langle \alpha_i, 
 \alpha_j \rangle ,
\ee
where the sum runs over all connected  trees $g$ 
with $n$ vertices labelled from $i=1$ to 
$i=n$. We denote by $g^{(0)}=\{1,\dots ,n\}$ 
the set of vertices, and by $g^{(1)}$ 
the set of oriented edges $(i,j)$, with the 
orientation inherited from the standard order 
$i<j$ on $g^{(0)}$. In other words given any 
labelled tree, and an edge of this tree
connecting $i$ to $j$,
we associate to this edge an orientation from $i\to j$
if $i<j$.
In order to implement this formula on a computer, 
it is useful to note that there are 
$n^{(n-2)}$ labeled trees with $n$ vertices, which 
are in one-to-one correspondence 
with their Pr\"ufer code, an element in 
$\{1,\dots ,n\}^{n-2}$.\footnote{Note that these 
labeled trees differ qualitatively from
the attractor flow trees. One can view these labeled trees as connecting the
endpoints of the attractor flow trees.}

Having defined the $S$, $U$ and $\cL$ factors, we can now state  
the JS wall-crossing formula (eq. (5.13) in \cite{Joyce:2008pc}): 
\be
\label{jswcf0}
\begin{split}
\bOm(\gamma;\tilde t\,)=&\sum_{n\geq 1}
\sum_{\substack{ (\alpha_1,\dots ,\alpha_n)\in C(\cX)\\
\gamma= \alpha_1+\dots +\alpha_n}}\, 
\frac{(-1)^{n -1 +\sum_{i<j} | \langle \alpha_i, \alpha_j \rangle
    |}}{2^{n-1}}
    \, U(\alpha_1,\dots ,\alpha_n;t,\tilde t)\, \cL(\alpha_1,\dots ,\alpha_n) 
     \, \prod_{i=1}^n \, \bOm(\alpha_i,t) \ ,
\end{split} 
\ee
where the second sum runs over all {\it ordered} decompositions 
$\gamma=\alpha_1+\dots+\alpha_n$ with 
$\alpha_i\in C(\cX)$. 
Note that due to this constraint,  eq, \eqref{jswcf0} is automatically consistent with 
the charge conservation property.

The JS formula is valid for any pair of points
in the moduli space with
stability conditions $\tau,\tilde\tau$. Now we restrict to the vicinity
of a wall of marginal stability $\cP(\gamma_1,\gamma_2)$. 
The only states whose BPS
invariants can jump are those whose charges lie in 
$\tilde\Gamma\subset C(\cX)$.
Moreover, their discontinuities only depend on the BPS 
invariants of states in $\tilde\Gamma$.
We denote the states $\gamma=M\gamma_1+N\gamma_2$ by 
$\gamma=(M,N)$.
As in previous section, let us assume that $\gamma_{12}<0$, 
and  take $t$ in the 
chamber $c^+$ where multi-centered configurations are absent, 
and $\tilde t$  in the 
chamber $c^-$  where they are present. From \eqref{etest}, 
we see that, in the vicinity
of the wall,
\be
\pm \, \langle\gamma,\gamma'\rangle \, 
\left[\tau(\gamma, c_\pm) - \tau(\gamma'; c_\pm)\right] > 0\ .
\ee
Therefore, the conditions \eqref{abalt} become
\beq
\label{econditions}
(a) &&\qquad \langle \alpha_i,\alpha_{i+1}\rangle \leq 0 
\quad \mbox{and}\quad
\langle \alpha_1+\cdots+\alpha_i,
\alpha_{i+1}+\cdots+\alpha_n \rangle < 0 , \quad\mbox{or}\, 
\nn \\
(b) && \qquad  \langle \alpha_i,\alpha_{i+1} \rangle > 0 
\quad \mbox{and}\quad
 \langle \alpha_1+\cdots+\alpha_i, \alpha_{i+1}+\cdots+\alpha_n 
 \rangle \geq 0 \ ,
\label{abalt2}
\eeq
and \eqref{jswcf0} gives an expression for $\bOm^-$ in terms
of $\bOm^+$.

\begin{figure}
\centerline{\includegraphics[totalheight=9cm]{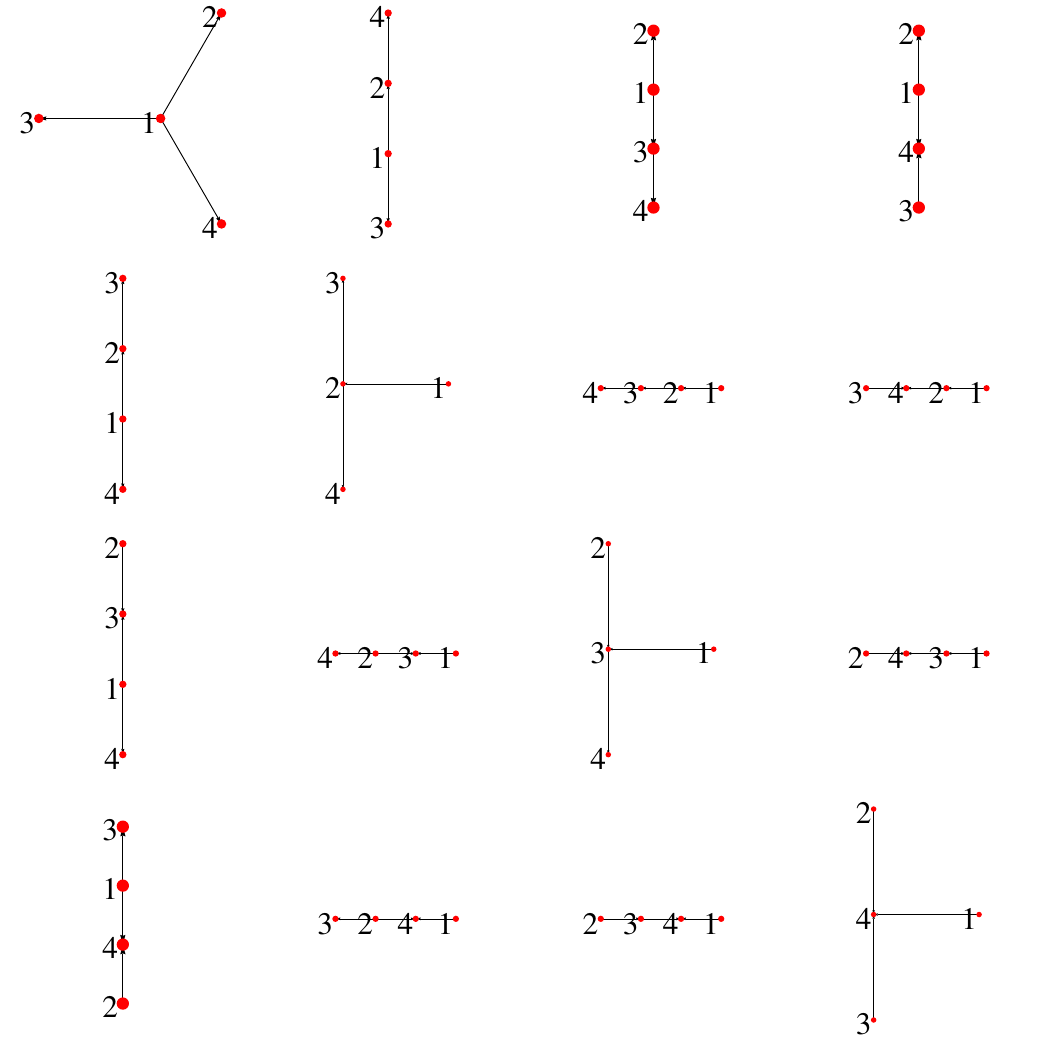}}
\caption{The 16 labelled trees contributing to 
$\cL(\alpha_1,\alpha_2,\alpha_3,\alpha_4)$.
\label{fig4body}}
\end{figure}

To illustrate the afore going prescription, we 
evaluate in detail the 
term proportional to $\bOm^+(\gamma_1)^2 
\bOm^+(\gamma_2)^2$ in the expressions for
$\bOm^-(2\gamma_1+2\gamma_2)$. 
We first  evaluate the $U$ factors, 
\beq \label{euexample}
 U(\gamma_2,\gamma_2,\gamma_1,\gamma_1) &=& \frac{1}{4}
   S(2 \gamma_2,2 \gamma_1)+\frac{1}{2}
   S(\gamma_2,\gamma_2,2 \gamma_1)+\frac{1}{2}
   S(2  \gamma_2,\gamma_1,\gamma_1)+S(\gamma_2,\gamma_2,
   \gamma_1,\gamma_1) = \frac{1}{4} \nn \\
U(\gamma_2,\gamma_1,\gamma_2,\gamma_1) &=&
   S(\gamma_2,\gamma_1,\gamma_2,\gamma_1)
   -\frac{1}{2} S(\gamma_2,\gamma_1)^2 = -\frac{1}{2} \nn \\
U(\gamma_2,\gamma_1,\gamma_1,\gamma_2) &=& -\frac{1}{2}
   S(\gamma_1,\gamma_2)
   S(\gamma_2,\gamma_1)+\frac{1}{2}
   S(\gamma_2,2\gamma_1,\gamma_2)
   +S(\gamma_2,\gamma_1,\gamma_1,\gamma_2) = 0  \\
 U(\gamma_1,\gamma_2,\gamma_2,\gamma_1)&=& -\frac{1}{2}
   S(\gamma_1,\gamma_2)
   S(\gamma_2,\gamma_1)+\frac{1}{2}
   S(\gamma_1,2
   \gamma_2,\gamma_1)+S(\gamma_1,\gamma_2,\gamma_2,\gamma_1) = 0 \nn \\
U( \gamma_1,\gamma_2,\gamma_1,\gamma_2)&=&
   S(\gamma_1,\gamma_2,\gamma_1,\gamma_2)
   -\frac{1}{2} S(\gamma_1,\gamma_2)^2 = \frac{1}{2} \nn \\
U( \gamma_1,\gamma_1,\gamma_2,\gamma_2 ) &=& \frac{1}{4}
   S(2 \gamma_1,2 \gamma_2)+\frac{1}{2}
   S(\gamma_1,\gamma_1,2 \gamma_2)+\frac{1}{2}
   S(2   \gamma_1,\gamma_2,\gamma_2)+S(\gamma_1,\gamma_1,
   \gamma_2,\gamma_2) = -\frac{1}{4}\nn
\eeq
For example in the first line the only non-vanishing
contribution comes from the $S(2\gamma_2,2\gamma_1)/4$
term. For the other terms the condition 
\eqref{econditions} fails for at least one $i$.
To compute the $\cL$ factors we observe that there
are 16 trees with 4 labelled nodes (see Figure \ref{fig4body}).
Out of those, twelve are obtained
by various inequivalent 
permutations of the tree connecting
nodes 1 to 2, 2 to 3 and 3 to 4, and four are
obtained by inequivalent permutations of a tree that
connects node 1 to each of the nodes 2, 3 and 4.
The $\cL$-factors are computed by adding the
contributions from each of these 16 trees, leading to 
\be \label{egenL}
\begin{split}
\cL(\alpha_1,\alpha_2,\alpha_3,\alpha_4) =& 
\alpha_{12}\, \alpha_{13}\, \alpha_{14}\,+\alpha_{12}\, \alpha_{23}\,
   \alpha_{14}\,+\alpha_{13}\, \alpha_{23}\, \alpha_{14}\,+\alpha_{13}\, \alpha_{24}\, 
 \alpha_{14} \\ &+\alpha_{23}\, \alpha_{24}\, \alpha_{14}\,+\alpha_{12}\, \alpha_{34}\, 
 \alpha_{14} +\alpha_{23}\, \alpha_{34}\, \alpha_{14}\,+\alpha_{24}\, \alpha_{34}\, 
 \alpha_{14} \\ &+\alpha_{12}\, \alpha_{13}\, \alpha_{24}\,+\alpha_{12}\, \alpha_{23}\, 
 \alpha_{24} +\alpha_{13}\, \alpha_{23}\, \alpha_{24}\,+\alpha_{12}\, \alpha_{13}\, 
 \alpha_{34} \\ &+\alpha_{12}\, \alpha_{23}\, \alpha_{34}\,+\alpha_{13}\, \alpha_{23}\, 
 \alpha_{34} +\alpha_{12}\, \alpha_{24}\, \alpha_{34}\,+\alpha_{13}\, \alpha_{24}\, \alpha_{34}\,.
 \end{split}
\ee
This gives
\beq
\cL(\gamma_2,\gamma_2,\gamma_1,\gamma_1) =  -4 \gamma_{12}^3 &\ ,& 
\cL(\gamma_2,\gamma_1,\gamma_2,\gamma_1) =  2 \gamma_{12}^3 \nn \\
\cL(\gamma_2,\gamma_1,\gamma_1,\gamma_2) =  0 &\ ,& 
\cL(\gamma_1,\gamma_2,\gamma_2,\gamma_1)= 0 \\\cL( \gamma_1,\gamma_2,\gamma_1,\gamma_2)=  -2 \gamma_{12}^3 &\ ,& 
\cL( \gamma_1,\gamma_1,\gamma_2,\gamma_2 ) =  4 \gamma_{12}^3\nn
\eeq
The total contribution from these terms to the right hand
side of \eqref{jswcf0} is thus given by
\be
\bOm^-(2\gamma_1+2\gamma_2) = \frac12  
\gamma_{12}^3\, [\bOm^+(\gamma_1)]^2\, 
[\bOm^+(\gamma_2)]^2+\dots ,
\ee
in agreement with the formul\ae\  \eqref{eordtwo} and \eqref{DOm22}.

\subsection{Index of supersymmetric 
bound states from the JS formula} \label{sjsindex}

It is useful to rewrite the JS formula \eqref{jswcf0} as a
sum  over unordered decompositions $\{\alpha_i\}$ of the 
charge vector $\gamma$,
\be
\label{jswcf2}
\Delta\bOm(\gamma) =
\sum_{n\geq 2 }\, 
  \sum_{\substack{ \{\alpha_1,\dots ,\alpha_n\} \in C(\cX)\\
\gamma= \alpha_1+\dots +\alpha_n}}\, 
\frac{g(\{\alpha_i\})}{|{\rm Aut}(\{\alpha_i\})|}
 \prod\nolimits_{i=1}^n \bOm^+(\alpha_i) \ .
 \ee
 where $|{\rm Aut}(\{\alpha_i\})|$ is 
 defined as follows. We choose any specific order of the
 $\alpha_i$'s and identify $|{\rm Aut}(\{\alpha_i\})|$ as
 the order of 
 the subgroup of 
 the permutation group $\Sigma_n$ on $n$ elements 
 which leaves invariant this particular ordering
 of the $\{\alpha_i\}$'s. 
 In particular if all the $\alpha_i$'s are different then the
 automorphism subgroup is trivial and
 $|{\rm Aut}(\{\alpha_i\})|=1$. On the other hand if $m$
 of the $\alpha_i$'s are the same then
 $|{\rm Aut}(\{\alpha_i\})|=m!$.
 The coefficient $g(\{\alpha_i\})$ 
 is then given by 
\be {
\label{egfromjs}
\begin{split}
g(\{\alpha_i\})\,  = &
\frac{(-1)^{n-1+\sum_{i<j} \langle \alpha_i,\alpha_j\rangle}
}{2^{n-1}}
\sum_{\sigma\in\Sigma_n} 
U\left(\alpha_{\sigma(1)},\dots , \alpha_{\sigma(n)}; t_+, t_-\right)\,
\cL\left(\alpha_{\sigma(1)},\dots ,\alpha_{\sigma(n)}\right)\ ,
\end{split}
}
\ee
where the sum runs over all permutations of $n$ elements.
Relabelling the set $\{\alpha_i\}$ into a partition $\{m_{r,s}\}$ such that 
$\gamma=(M,N)=\sum_{r,s} (r,s) m_{r,s}$, 
the coefficient $g(\{\alpha_i\})$ is the same as the one 
appearing in \eqref{esab2}.
Thus it 
should be identified with the index of the supersymmetric
quantum mechanics of $n$ distinguishable charged 
dyons in $\IR^{1,3}$,
along the lines of  \cite{Denef:2002ru}.

\subsection{Generic 2-body, 3-body and 4-body contributions}
\label{sexampleJS}

Let us first derive the primitive wall-crossing formula
$\gamma\to\gamma_1+\gamma_2$ from the JS wall
crossing formula. In this case there is only one
tree -- connecting the nodes 1 and 2 with the arrow
directed from 1 to 2. If we choose the first node to be
$\gamma_1$ and the second node to be $\gamma_2$
then since $\gamma_{12}<0$ possibility (a) in 
\eqref{econditions} is realized, and we have $
U(\gamma_1,\gamma_2)=S(\gamma_1,
\gamma_2)=-1$. Since $n=2$, \eqref{egfromjs} now
gives a contribution of ${1\over 2} (-1)^{1+\gamma_{12}}
\gamma_{12} (-1)$. An identical contribution comes from
the term where we put $\gamma_2$ in the first
node and $\gamma_1$ in the second node
since now we have $U=S=1$ and  the
$\langle \alpha_{\sigma(i)}, \alpha_{\sigma(j)} \rangle$
factor in \eqref{egfromjs} is now equal to
$\gamma_{21}$. Adding the two contributions we
recover the primitive wall-crossing formula 
\eqref{eprimks}. We summarize the $S,U$ and $\cL$
factors for the two permutations in the table below.
\be
\label{tg12}
\begin{array}{|c|c|c|c|}
\hline
\sigma(12) & S & U & \cL \\
\hline
  12 & a & -1 & \gamma_{12} \\
  21 & b & 1 & -\gamma_{12}\\
\hline
\end{array}
\ee

Next we 
reproduce the result of Sec. \ref{s3decay} for the generic three body
contribution to the wall-crossing from centers
carrying charges  $(\alpha_1,\alpha_2,\alpha_3)$. 
The order of the $\alpha_i$'s is given by (\ref{eorder}).
With the definition of the ordering explained below
\eqref{esa2} 
we see that if $(\alpha,\beta)$
follow a clockwise order then $\langle\alpha, \beta\rangle
> 0$.
Since the phases are assumed to be generic,
the $U$ and $S$ factors coincide, \i.e.\
$U(\alpha_i,\alpha_j,\alpha_k; t,\tilde t)=S(\alpha_i,\alpha_j,\alpha_k; t,\tilde t)$, for any permutation $\{i,j,k\}$
of $\{1,2,3\} $.  
In this case, there are three trees contributing. 
The $S,U,\cL$-factors are summarized in the table below. 
Substitution of these data into eq. (\ref{jswcf0})
reproduces directly Eqs. \eqref{eyto2}, (\ref{efres}).
\be
\label{tg123}
\begin{array}{|c|c|c|c|}
\hline
\sigma(123) & S & U & \cL \\
\hline
123 & \text{bb} & 1 & \alpha_{12}\alpha_{13}+\alpha_{13}\alpha_{23}+\alpha_{12}\alpha_{23}\\
132 & \text{b-} & 0 & \alpha_{12}\alpha_{13}-\alpha_{13}\alpha_{23}-\alpha_{12}\alpha_{23}\\
213 &\text{ab} & -1 & -\alpha_{12}\alpha_{23}+\alpha_{13}\alpha_{23}-\alpha_{12}\alpha_{13}\\
231 &  \text{-a} & 0 & \alpha_{12}\alpha_{13}-\alpha_{13}\alpha_{23}-\alpha_{12}\alpha_{23}\\
312 & \text{ab} & -1 & \alpha_{13}\alpha_{23}-\alpha_{12}\alpha_{23}-\alpha_{13}\alpha_{12} \\
321 & \text{aa} & 1 & \alpha_{13}\alpha_{23}+\alpha_{12}\alpha_{13}+\alpha_{12}\alpha_{23} \\
\hline
\end{array}
\ee

Next, we turn to the generic 4-body case. We use the same phase
ordering as eq. (\ref{e4body}). The non-vanishing 
contributions are given in the following table
\be
\label{et4body}
\begin{array}{|c|c|c|c|}
\hline
\sigma(1234) & S & U & \cL \\ \hline 
 1234 & \text{bbb} & 1 & \alpha_{12}\, \alpha_{13}\, \alpha_{14}\,+\alpha_{12}\, \alpha_{23}\,
   \alpha_{14}\,+\alpha_{13}\, \alpha_{23}\, \alpha_{14}\,+\alpha_{13}\, \alpha_{24}\, 
 \alpha_{14} \\ &&&+\alpha_{23}\, \alpha_{24}\, \alpha_{14}\,+\alpha_{12}\, \alpha_{34}\, 
 \alpha_{14} +\alpha_{23}\, \alpha_{34}\, \alpha_{14}\,+\alpha_{24}\, \alpha_{34}\, 
 \alpha_{14} \\ &&&+\alpha_{12}\, \alpha_{13}\, \alpha_{24}\,+\alpha_{12}\, \alpha_{23}\, 
 \alpha_{24} +\alpha_{13}\, \alpha_{23}\, \alpha_{24}\,+\alpha_{12}\, \alpha_{13}\, 
 \alpha_{34} \\ &&&+\alpha_{12}\, \alpha_{23}\, \alpha_{34}\,+\alpha_{13}\, \alpha_{23}\, 
 \alpha_{34} +\alpha_{12}\, \alpha_{24}\, \alpha_{34}\,+\alpha_{13}\, \alpha_{24}\, \alpha_{34}\, \\
 1342 & \text{bba} & -1 & \alpha_{12}\, \alpha_{13}\, \alpha_{14}\,-\alpha_{12}\, \alpha_{23}\,
   \alpha_{14}\,-\alpha_{13}\, \alpha_{23}\, \alpha_{14}\,-\alpha_{13}\, \alpha_{24}\, 
 \alpha_{14} \\ &&& +\alpha_{23}\, \alpha_{24}\, \alpha_{14}\,+\alpha_{12}\, \alpha_{34}\, 
 \alpha_{14} -\alpha_{23}\, \alpha_{34}\, \alpha_{14}\,-\alpha_{24}\, \alpha_{34}\, 
 \alpha_{14} \\ &&& -\alpha_{12}\, \alpha_{13}\, \alpha_{24}\,+\alpha_{12}\, \alpha_{23}\, 
 \alpha_{24} +\alpha_{13}\, \alpha_{23}\, \alpha_{24}\,+\alpha_{12}\, \alpha_{13}\, 
 \alpha_{34}  \\ &&&-\alpha_{12}\, \alpha_{23}\, \alpha_{34}\,-\alpha_{13}\, \alpha_{23}\, 
 \alpha_{34} -\alpha_{12}\, \alpha_{24}\, \alpha_{34}\,-\alpha_{13}\, \alpha_{24}\, \alpha_{34}\, \\
 1423 & \text{bab} & -1 & \alpha_{12}\, \alpha_{13}\, \alpha_{14}\,+\alpha_{12}\, \alpha_{23}\,
   \alpha_{14}\,+\alpha_{13}\, \alpha_{23}\, \alpha_{14}\,-\alpha_{13}\, \alpha_{24}\, 
 \alpha_{14} \\ &&& -\alpha_{23}\, \alpha_{24}\, \alpha_{14}\,-\alpha_{12}\, \alpha_{34}\, 
 \alpha_{14} -\alpha_{23}\, \alpha_{34}\, \alpha_{14}\,+\alpha_{24}\, \alpha_{34}\, 
 \alpha_{14} \\ &&& -\alpha_{12}\, \alpha_{13}\, \alpha_{24}\,-\alpha_{12}\, \alpha_{23}\, 
 \alpha_{24} -\alpha_{13}\, \alpha_{23}\, \alpha_{24}\,-\alpha_{12}\, \alpha_{13}\, 
 \alpha_{34} \\ &&& -\alpha_{12}\, \alpha_{23}\, \alpha_{34}\,-\alpha_{13}\, \alpha_{23}\, 
 \alpha_{34}+\alpha_{12}\, \alpha_{24}\, \alpha_{34}\,+\alpha_{13}\, \alpha_{24}\, \alpha_{34}\, \\
 1432 & \text{baa} & 1 & \alpha_{12}\, \alpha_{13}\, \alpha_{14}\,-\alpha_{12}\, \alpha_{23}\,
   \alpha_{14}\,-\alpha_{13}\, \alpha_{23}\, \alpha_{14}\,-\alpha_{13}\, \alpha_{24}\, 
 \alpha_{14} \\ &&& +\alpha_{23}\, \alpha_{24}\, \alpha_{14}\,-\alpha_{12}\, \alpha_{34}\, 
 \alpha_{14}+\alpha_{23}\, \alpha_{34}\, \alpha_{14}\,+\alpha_{24}\, \alpha_{34}\, 
 \alpha_{14} \\ &&& -\alpha_{12}\, \alpha_{13}\, \alpha_{24}\,+\alpha_{12}\, \alpha_{23}\, 
 \alpha_{24} +\alpha_{13}\, \alpha_{23}\, \alpha_{24}\,-\alpha_{12}\, \alpha_{13}\, 
 \alpha_{34} \\ &&& +\alpha_{12}\, \alpha_{23}\, \alpha_{34}\,+\alpha_{13}\, \alpha_{23}\, 
 \alpha_{34} +\alpha_{12}\, \alpha_{24}\, \alpha_{34}\,+\alpha_{13}\, \alpha_{24}\, \alpha_{34}\, \\
 2341 & \text{bba} & -1 & -\alpha_{12}\, \alpha_{13}\, \alpha_{14}\,+\alpha_{12}\, \alpha_{23}\,
   \alpha_{14}\,+\alpha_{13}\, \alpha_{23}\, \alpha_{14}\,+\alpha_{13}\, \alpha_{24}\, 
 \alpha_{14} \\ &&& -\alpha_{23}\, \alpha_{24}\, \alpha_{14}\,+\alpha_{12}\, \alpha_{34}\, 
 \alpha_{14} -\alpha_{23}\, \alpha_{34}\, \alpha_{14}\,-\alpha_{24}\, \alpha_{34}\, 
 \alpha_{14} \\ &&& +\alpha_{12}\, \alpha_{13}\, \alpha_{24}\,-\alpha_{12}\, \alpha_{23}\, 
 \alpha_{24} -\alpha_{13}\, \alpha_{23}\, \alpha_{24}\,+\alpha_{12}\, \alpha_{13}\, 
 \alpha_{34} \\ &&& -\alpha_{12}\, \alpha_{23}\, \alpha_{34}\,-\alpha_{13}\, \alpha_{23}\, 
 \alpha_{34}\, -\alpha_{12}\, \alpha_{24}\, \alpha_{34}\,-\alpha_{13}\, \alpha_{24}\, \alpha_{34}\, \\
2413 & \text{bab} & -1 & \alpha_{12}\, \alpha_{13}\, \alpha_{14}\,+\alpha_{12}\, \alpha_{23}\,
   \alpha_{14}\,-\alpha_{13}\, \alpha_{23}\, \alpha_{14}\,-\alpha_{13}\, \alpha_{24}\, 
 \alpha_{14} \\ &&& -\alpha_{23}\, \alpha_{24}\, \alpha_{14}\,-\alpha_{12}\, \alpha_{34}\, 
 \alpha_{14} +\alpha_{23}\, \alpha_{34}\, \alpha_{14}\,+\alpha_{24}\, \alpha_{34}\, 
 \alpha_{14} \\ &&& -\alpha_{12}\, \alpha_{13}\, \alpha_{24}\,-\alpha_{12}\, \alpha_{23}\, 
 \alpha_{24} +\alpha_{13}\, \alpha_{23}\, \alpha_{24}\,+\alpha_{12}\, \alpha_{13}\, 
 \alpha_{34} \\ &&& +\alpha_{12}\, \alpha_{23}\, \alpha_{34}\,-\alpha_{13}\, \alpha_{23}\, 
 \alpha_{34} +\alpha_{12}\, \alpha_{24}\, \alpha_{34}\,-\alpha_{13}\, \alpha_{24}\, \alpha_{34}\, \\
  2431 & \text{baa} & 1 & -\alpha_{12}\, \alpha_{13}\, \alpha_{14}\,+\alpha_{12}\, \alpha_{23}\,
   \alpha_{14}\,+\alpha_{13}\, \alpha_{23}\, \alpha_{14}\,+\alpha_{13}\, \alpha_{24}\, 
 \alpha_{14} \\ &&& -\alpha_{23}\, \alpha_{24}\, \alpha_{14}\,-\alpha_{12}\, \alpha_{34}\, 
 \alpha_{14} +\alpha_{23}\, \alpha_{34}\, \alpha_{14}\,+\alpha_{24}\, \alpha_{34}\, 
 \alpha_{14} \\ &&& +\alpha_{12}\, \alpha_{13}\, \alpha_{24}\,-\alpha_{12}\, \alpha_{23}\, 
 \alpha_{24} -\alpha_{13}\, \alpha_{23}\, \alpha_{24}\,-\alpha_{12}\, \alpha_{13}\, 
 \alpha_{34} \\ &&& +\alpha_{12}\, \alpha_{23}\, \alpha_{34}\,+\alpha_{13}\, \alpha_{23}\, 
 \alpha_{34} +\alpha_{12}\, \alpha_{24}\, \alpha_{34}\,+\alpha_{13}\, \alpha_{24}\, \alpha_{34}\, \\
 3124 & \text{abb} & -1 & -\alpha_{12}\, \alpha_{13}\, \alpha_{14}\,-\alpha_{12}\, \alpha_{23}\,
   \alpha_{14}\,+\alpha_{13}\, \alpha_{23}\, \alpha_{14}\,-\alpha_{13}\, \alpha_{24}\, 
 \alpha_{14} \\ &&& -\alpha_{23}\, \alpha_{24}\, \alpha_{14}\,+\alpha_{12}\, \alpha_{34}\, 
 \alpha_{14} -\alpha_{23}\, \alpha_{34}\, \alpha_{14}\,+\alpha_{24}\, \alpha_{34}\, 
 \alpha_{14} \\ &&& -\alpha_{12}\, \alpha_{13}\, \alpha_{24}\,-\alpha_{12}\, \alpha_{23}\, 
 \alpha_{24} +\alpha_{13}\, \alpha_{23}\, \alpha_{24}\,-\alpha_{12}\, \alpha_{13}\, 
 \alpha_{34} \\ &&& -\alpha_{12}\, \alpha_{23}\, \alpha_{34}\,+\alpha_{13}\, \alpha_{23}\, 
 \alpha_{34} +\alpha_{12}\, \alpha_{24}\, \alpha_{34}\,-\alpha_{13}\, \alpha_{24}\, \alpha_{34}\, \\
 3142 & \text{aba} & 1 & -\alpha_{12}\, \alpha_{13}\, \alpha_{14}\,-\alpha_{12}\, \alpha_{23}\,
   \alpha_{14}\,+\alpha_{13}\, \alpha_{23}\, \alpha_{14}\,+\alpha_{13}\, \alpha_{24}\, 
 \alpha_{14} \\ &&& +\alpha_{23}\, \alpha_{24}\, \alpha_{14}\,+\alpha_{12}\, \alpha_{34}\, 
 \alpha_{14} -\alpha_{23}\, \alpha_{34}\, \alpha_{14}\,-\alpha_{24}\, \alpha_{34}\, 
 \alpha_{14} \\ &&& +\alpha_{12}\, \alpha_{13}\, \alpha_{24}\,+\alpha_{12}\, \alpha_{23}\, 
 \alpha_{24} -\alpha_{13}\, \alpha_{23}\, \alpha_{24}\,-\alpha_{12}\, \alpha_{13}\, 
 \alpha_{34} \\ &&& -\alpha_{12}\, \alpha_{23}\, \alpha_{34}\,+\alpha_{13}\, \alpha_{23}\, 
 \alpha_{34} -\alpha_{12}\, \alpha_{24}\, \alpha_{34}\,+\alpha_{13}\, \alpha_{24}\, \alpha_{34}\, \\
 3241 & \text{aba} & 1 & -\alpha_{12}\, \alpha_{13}\, \alpha_{14}\,-\alpha_{12}\, \alpha_{23}\,
   \alpha_{14}\,-\alpha_{13}\, \alpha_{23}\, \alpha_{14}\,+\alpha_{13}\, \alpha_{24}\, 
 \alpha_{14} \\ &&& +\alpha_{23}\, \alpha_{24}\, \alpha_{14}\,+\alpha_{12}\, \alpha_{34}\, 
 \alpha_{14} +\alpha_{23}\, \alpha_{34}\, \alpha_{14}\,-\alpha_{24}\, \alpha_{34}\, 
 \alpha_{14} \\ &&& +\alpha_{12}\, \alpha_{13}\, \alpha_{24}\,+\alpha_{12}\, \alpha_{23}\, 
 \alpha_{24} +\alpha_{13}\, \alpha_{23}\, \alpha_{24}\,+\alpha_{12}\, \alpha_{13}\, 
 \alpha_{34} \\ &&& +\alpha_{12}\, \alpha_{23}\, \alpha_{34}\,+\alpha_{13}\, \alpha_{23}\, 
 \alpha_{34} -\alpha_{12}\, \alpha_{24}\, \alpha_{34}\,-\alpha_{13}\, \alpha_{24}\, \alpha_{34}\, \\
 \hline
\end{array}
\ee
\be
\begin{array}{|c|c|c|c|}
\hline
\sigma(1234) & S & U & \cL \\ \hline 
4213 & \text{aab} & 1 & 
 \alpha_{12}\, \alpha_{13}\, \alpha_{14}\,
 +\alpha_{12}\, \alpha_{23}\,  \alpha_{14}\,
 -\alpha_{13}\, \alpha_{23}\, \alpha_{14}\,
 +\alpha_{13}\, \alpha_{24}\, \alpha_{14}\\ &&&
 +\alpha_{23}\, \alpha_{24}\, \alpha_{14}\,
 -\alpha_{12}\, \alpha_{34}\, \alpha_{14} 
 +\alpha_{23}\, \alpha_{34}\, \alpha_{14}\,
 -\alpha_{24}\, \alpha_{34}\, \alpha_{14}\, \\ &&&
 +\alpha_{12}\, \alpha_{13}\, \alpha_{24}\,
 +\alpha_{12}\, \alpha_{23}\,  \alpha_{24}
  -\alpha_{13}\, \alpha_{23}\, \alpha_{24}\,
  +\alpha_{12}\, \alpha_{13}\, \alpha_{34} \\ &&&
  +\alpha_{12}\, \alpha_{23}\, \alpha_{34}\,
  -\alpha_{13}\, \alpha_{23}\, \alpha_{34}\, 
  -\alpha_{12}\, \alpha_{24}\, \alpha_{34}\,
  +\alpha_{13}\, \alpha_{24}\, \alpha_{34}\, \\
 4321 & \text{aaa} & -1 & -\alpha_{12}\, \alpha_{13}\, \alpha_{14}\,-\alpha_{12}\, \alpha_{23}\,
   \alpha_{14}\,-\alpha_{13}\, \alpha_{23}\, \alpha_{14}\,-\alpha_{13}\, \alpha_{24}\, 
 \alpha_{14} \\ &&& -\alpha_{23}\, \alpha_{24}\, \alpha_{14}\,-\alpha_{12}\, \alpha_{34}\, 
 \alpha_{14} -\alpha_{23}\, \alpha_{34}\, \alpha_{14}\,-\alpha_{24}\, \alpha_{34}\, 
 \alpha_{14} \\ &&& -\alpha_{12}\, \alpha_{13}\, \alpha_{24}\,-\alpha_{12}\, \alpha_{23}\, 
 \alpha_{24} -\alpha_{13}\, \alpha_{23}\, \alpha_{24}\,-\alpha_{12}\, \alpha_{13}\, 
 \alpha_{34} \\ &&& -\alpha_{12}\, \alpha_{23}\, \alpha_{34}\,-\alpha_{13}\, \alpha_{23}\, 
 \alpha_{34}\, -\alpha_{12}\, \alpha_{24}\, \alpha_{34}\,-\alpha_{13}\, \alpha_{24}\, \alpha_{34}\,
\\\hline
\end{array}\nn
\ee
Adding up all these contributions, one finds again the result of Eqs. \eqref{enonmot4}, (\ref{eD4body}).

\subsection{Semi-primitive wall-crossing formula from JS}
\label{ssemiJS}
Let us now derive the semi-primitive wall-crossing formula from \eqref{jswcf0},  i.e. compute $\bOm(\gamma;c_-)$ in terms
of $\bOm(\gamma;c_+)$ for $\gamma=(1,N)\in \tilde\Gamma$.  We mostly follow
the discussion in \cite{Stoppa:2009}, Section 3, suitably generalized.

At given order $n$, the most general
ordered decomposition $\gamma=\sum_{i=1,\dots ,n}\alpha_i$ is 
\be
\label{spjs1}
\gamma = (0,N_1) + \dots + (0,N_{i_*-1}) + (1,N_{i_*}) 
+ (0,N_{i_*+1}) + \dots + (0,N_n)
\ee
where $\{N_i\}$ is a partition of $N$ of length $n$, $\sum_{1\leq i\leq n} N_i = N$, 
and $1\leq i_* \leq N$ labels the position of the charge $(1,0)$ in this decomposition. 
For such a pair $(\{N_i\}, i_*)$, one may check that the $S$ factor is given by 
\be
\label{spjs2}
S(\{N_i\}, i_*;c_+,c_-) = (-1)^{n-i_*}
\ee
if $i_*=1$ or $i_*=2$, and vanishes otherwise. To compute the $U$ factor, note that due to the phase constraints, 
the ordered partition of $\{\alpha_i\}$ labelled by
$\{a_j\}$ must decompose into  
ordered partitions of the `head' set $\{N_i\}, i=1,\dots ,i_*-1$ and `tail' set $\{N_i\}, i=i_*+1,\dots ,n$, 
while the vector $(1,N_{i_*})$ must lie in its own packet. Moreover, since the phases of all the vectors $\beta_j$,
except the vector $(1, N_{i_*})$, are the same
in the chamber $c_-$, all the $\beta_j$'s must be
be grouped into a single packet at the second step in order
to satisfy the phase constraint on the $\delta_k$'s.
Otherwise the packet containing $(1, N_{i_*})$ will have
different phase from the others.
The $U$-factor therefore reduces to 
\be
\label{spjs3}
U( \{N_i\}, i_* ;c_+,c_-) = \sum_{\substack{1\leq m\leq n\\
0=a_0< a_1<\dots a_m =n}} \, S(\beta_{1},
\beta_{2},\ldots,\beta_{m}; c_+,c_-)\, 
\cdot\prod_{j=1}^m\frac{1}{(a_j-a_{j-1})!}.
\ee
Now, let $j_*$ be the packet in which the vector $(1,N_{i_*})$ lies. In view of \eqref{spjs2},
the factor $S(\beta_{1},\beta_{2},\ldots,\beta_{m}; c_+,c_-)$ vanishes unless $j_*=1$ (which happens if $i_*=1$) or $j_*=2$ (which happens whenever $i_*>1$ and we group
all the vectors $(0,N_1),\dots, (0,N_{i_*-1})$ in a
single packet). In either case, the contraction 
of the head set is trivial, and we are left with 
\be
\label{spjs4old}
U(\{N_i\}, i_* ;c_+,c_-) = \frac{1}{(i_*-1)!} \sum_{\{p_r\}} 
\frac{(-1)^{{\rm len}(p)}}{\prod_r p_r!},
\ee
where the sum runs over ordered partitions  of the $n-i_*$ elements in the tail set, 
i.e. integer sequences $\{p_r, 1\leq r\leq {\rm len}(p) \}, p_r\geq 1$ such that 
$\sum_r p_r=n-i_\star$. This evaluates, for all $i_*\geq 1$,  to the binomial coefficient
\be
U(\{N_i\}, i_* ;c_+,c_-) = \frac{(-1)^{n-i_*}}{(i_*-1)!\, (n-i_*)!}\ ,
\ee
a result which is in particular  independent of the partition $\{N_i\}$.

Now, we turn to the sum over graphs $g$. Due to the Landau factors 
$\langle \alpha_i, \alpha_j \rangle$ in \eqref{jswcf0}, the only contributing 
graph is a tree rooted at $i_*$, with leaves $1, \dots, i_*-1, i_*+1,
\dots ,n$. The Landau factor 
is then given by 
\be
\cL= \prod_{i=1}^{i_*-1} \, \langle(0,N_i),(1,N_{i_*})\rangle
\,
\prod_{i=i_*+1}^n  \, \langle(1,N_{i_*}),(0,N_i)\rangle .
\ee
Taking into account the additional factor $1/2^{n-1}$, 
we thus arrive at  
\be
\label{spjs4}
\begin{split}
\bOm^-(1,N)=\sum_{n\geq 1} \sum_{\substack{1\leq i_*\leq n\\ \sum_{1\leq i\leq n} N_i = N}}
& \frac{ (-1)^{n-1}\,  (-1)^{i_\star -1+\sum_{i\neq i_*} N_i}\,  (-1)^{n-i_*} \gamma_{12}^{n-1}}
 {2^{n-1}\, (i_*-1)!\, (n-i_*)!}\\
& \times \bOm^+(1,N_{i_*})\, \prod\nolimits_{i\neq i_*} N_i\, \bOm^+(0,N_i) \ .
\end{split}
\ee
Plugging this expression in the partition function \eqref{defZNq}, we can easily 
carry out the sum over $N_i$ and obtain
\be \label{jsjs1}
\bar Z^-(1,q) = \bar Z^+(1,q) \, \sum_{n\geq 1} \sum_{1\leq i_*\leq n}
\frac{ \left[ \log Z_{\rm halo}(\gamma_1, q) \right]^{n-1}}{2^{n-1} (i_*-1)!\, (n-i_*)!}
\ee
where $Z_{\rm halo}(\gamma_1, q)$ is the same function introduced
in \eqref{Z1pmh}.
The sum over $i_*$ leads to 
\be \label{jsjs2}
\bar Z^-(1,q) = \bar Z^+(1,q) \, \sum_{n\geq 1} \frac{\left[ \log Z_{\rm halo}(\gamma_1, q) \right]^{n-1} }{ (n-1)!}
\ee
and the sum over $n$ finally leads to 
\be \label{jsjs3}
\bar Z^-(1,q) = \bar Z^+(1,q) \, Z_{\rm halo}(\gamma_1, q)\ .
\ee

Finally we note that the derivation given above can be
simplified using the Boltzmann gas picture in which we
analyze identical particle contribution to wall-crossing
as a limit of non-identical particle contribution and then
include an extra symmetry factor $1/N!$ for $N$
identical particles. To see how this works, we consider
the case where we have $(N+1)$ different $\alpha_i$'s
satisfying $\alpha_{ij}>0$ for $i<j$, with the
understanding that we shall eventually take the limit
in which the first $N$ $\alpha_i$'s approach $\gamma_2$
or its multiple
and $\alpha_{N+1}$ approaches $\gamma_1$.
Now the $S$ and the $U$ factors coincide as in
\S\ref{sexampleJS}. Furthermore since eventually
we shall take the limit in which the first $N$ $\alpha_i$'s
coincide, the requirement of a non-vanishing $\cL$
tells us that only trees which contribute are
those in which $\alpha_{N+1}$ is connected to
all the other $\alpha_i$'s. We can still choose the
direction of the arrows arbitrarily. Let us
consider a configuration in which $m$ of the arrows
are directed towards $\alpha_{N+1}$ and $n=N-m$ are
directed away from it. 
In this case the arrows go from  the nodes 1 to $m$
towards the central node $(m+1)$ and from the central node
to the nodes $(m+2)$ to $(N+1)$. 
The Landau factor associated with these nodes is
$(-1)^{m} \prod_{i=1}^N \langle  \alpha_{N+1}, \alpha_i
\rangle$.
Next we need to assign the $\alpha_i$'s to the different
nodes. First of all there are ${N\choose m}$ ways of
deciding which of the $\alpha_i$'s will be
assigned to the first $m$ nodes. Once this is done
there is no further freedom of rearranging the $\alpha_i$'s
among the first $m$ nodes or the last $(N-m)$ nodes;
in order that $U=S$ does not vanish,  the $\alpha_i$'s
must be arranged in a clockwise order among the first
$m$ nodes and anti-clockwise order among the
last $(N-m)$ nodes. In this arrangement
the possibility $(a)$ is realized $(N-m)$ times and the
possibility $(b)$ is realized $m$ times.
Hence the corresponding $U$ is given by 
$(-1)^{N-m}$. Substituting these into eq, \eqref{egfromjs}
we now get
\be \label{ealtgjs}
g(\{\alpha_i\}) = 2^{-N} (-1)^{\sum_{i<j}
\langle\alpha_i,\alpha_j\rangle} \sum_{m=0}^N 
\, {N\choose m} \, \prod_{i=1}^{N}\, 
\langle \alpha_{N+1},  \alpha_i\rangle
= (-1)^{\sum_{i<j}
\langle\alpha_i,\alpha_j\rangle} \prod_{i=1}^N
\langle\alpha_{N+1},  \alpha_i\rangle\, .
\ee
If we now take the set $(\alpha_1,\dots, \alpha_N)$
to contain $m_l$ copies of $l\gamma_2$ with
$\sum lm_l=N$ then,
including the symmetry factor $\prod_l {1/ m_l!}$,
we get the coefficient of the 
$\bOm^+(\gamma_1) \prod_l \bOm^+(l\gamma_2)^{m_l}$
term to be
\be \label{efincoeff}
\, \prod_l  {1\over m_l!} \left[ (-1)^{
 l\gamma_{12}}(l\gamma_{12}) \right]^{m_l}\, ,
\ee
which is the desired result.

\bigskip

{\bf Acknowledgement:}    
We are grateful to D.~Joyce and J.~Stoppa for useful correspondence.
J.M. thanks the LPTHE for hospitality during part of this work. 
The work of J.M. is partially supported by ANR grant
BLAN06-3-137168.
A.S. acknowledges the support of 
Chaires Internationales de Recherche Blaise Pascal
during his stay at LPTHE where the work began,
of the J. C.  Bose fellowship of the Department of
Science and Technology, India and of the
project 11-R\& D-HRI-5.02-0304.

\appendix

\section{Wall crossing formul\ae\  in special cases} 
\label{sswall}

In this appendix we give explicit wall-crossing
formul\ae\ in some special cases. These cases 
illustrate the general results of \S\ref{sec_KS} 
and \S\ref{sjoyce}, and serve as tests
of the equivalence of the JS and KS wall-crossing formul\ae\ .
For brevity we state the results for the classical rational
invariants $\bar\Omega$, but the same formulae also hold
for the motivic rational invariants 
with the replacement \eqref{ereplace}.
For notational convenience we shall denote
$\bOm^\pm(M\gamma_1+N\gamma_2)$
by $\bOm^\pm(M,N)$. The  result for $(M,N)=(2,2)$ below
is in agreement with \cite{Manschot:2010xp}, Eq (2.13).

\be
\label{DOm12}
\begin{split}
\Delta\bOm(1,2)=
& \kscom{2 \gamma_{12}} \, \bOm^+(0,2)\, \bOm^+(1,0) \\&
+\frac{1}{2} [\kscom{\gamma_{12}]}^2\,  [\bOm^+(0,1)]^2 \, \bOm^+(1,0)
+\kscom{\gamma_{12}}\, 
   \bOm^+(0,1) \, \bOm^+(1,1)
\end{split}
\ee
\be
\label{DOm22}
\begin{split}
\Delta\bOm&(2,2) =  
 \kappa (4  \gamma_{12})
\,   \bOm^+(0,2) \, \bOm^+(2,0)   + \kappa (2 \gamma_{12})
   \left[ \bOm^+(1,0) \bOm^+(1,2)+\bOm^+(0,1) \bOm^+(2,1) \right]\\
   & + \kappa  (\gamma_{12}) \, \kappa (2 \gamma_{12}) \, 
   \bOm^+(0,1) \bOm^+(1,0) \bOm^+(1,1) 
   +\frac{1}{4}  [\kappa (\gamma_{12})]^2
    \, \kappa (2\gamma_{12}) \, 
   \bOm^+(0,1)^2 \bOm^+(1,0)^2\\
   & 
   +\frac{1}{2} [\kappa (2 \gamma_{12})]^2 
   \left[\bOm^+(2,0) \bOm^+(0,1)^2+\bOm^+(0,2)
   \bOm^+(1,0)^2\right] 
   \end{split}
\ee
\be
\label{DOm13}
\begin{split}
\Delta\bOm&(1,3)=
 \kappa (3   \gamma_{12})\, \bOm^+(0,3) \, \bOm^+(1,0) 
+\kappa (2  \gamma_{12})\, \bOm^+(0,2)\, \bOm^+(1,1) 
+\kappa (\gamma_{12})\, \bOm^+(0,1)\, \bOm^+(1,2) \\
   & +\kappa (\gamma_{12}) \, \kappa (2 \gamma_{12})\, 
    \bOm^+(0,1) \, \bOm^+(0,2)\,   \bOm^+(1,0) 
    +\frac12 [\kappa (\gamma_{12})]^2
    \bOm^+(0,1)^2 \,  \bOm^+(1,1)   \\
   & +\frac{1}{6}[\kappa
   (\gamma_{12})]^3\,  \bOm^+(0,1)^3\, \bOm^+(1,0) 
   \end{split}
\ee
 \be
\label{DOm23}
\begin{split}
\Delta\bOm&(2,3) = 
   \kappa   (\gamma_{12})\, \bOm^+(1,1)\,  \bOm^+(1,2) 
   +\kappa (3 \gamma_{12})\, \bOm^+(1,0) \, \bOm^+(1,3) 
   +\kappa (6  \gamma_{12}) \bOm^+(0,3) \, \bOm^+(2,0) \\
   &+ \kappa (4 \gamma_{12}) \bOm^+(0,2) \, \bOm^+(2,1)
   +\kappa (2 \gamma_{12})\bOm^+(0,1) \, \bOm^+(2,2) \\
   & +\frac{1}{2}[\kappa (\gamma_{12})]^2     \bOm^+(0,1)\, \bOm^+(1,1)^2
   +\frac12 [\kappa (3 \gamma_{12})]^2 \bOm^+(0,3) \bOm^+(1,0)^2\\
   &
    +\frac12 
    \left[ [\kappa (\gamma_{12})]^2 + [\kappa (2\gamma_{12})]^2
    + \kappa ( \gamma_{12})\, \kappa (3 \gamma_{12}) \right]
    \bOm^+(0,1)\, \bOm^+(1,0) \, \bOm^+(1,2)
    \\
    &+\frac{1}{2} \left[
     \kappa(\gamma_{12}) \, \kappa (2 \gamma_{12})
     + \kappa(\gamma_{12}) \, \kappa (4 \gamma_{12})+ 
      \kappa(2\gamma_{12}) \, \kappa (3 \gamma_{12}) \right]
    \bOm^+(0,2) \bOm^+(1,0) \, \bOm^+(1,1)
  \\
   &
    + \kappa (2 \gamma_{12})\,  \kappa (4 \gamma_{12}) \, 
   \bOm^+(0,1)\,  \bOm^+(0,2)\,  \bOm^+(2,0)
    +\frac12  [\kappa (2   \gamma_{12})]^2 \, 
   \bOm^+(0,1)^2 \bOm^+(2,1) \\
   & +\frac{1}{4} \left( 
   3 [\kappa (\gamma_{12})]^3 + \kappa (3 \gamma_{12})\,  [\kappa(  \gamma_{12})]^2
   + \kappa ( \gamma_{12})\,  [\kappa (2 \gamma_{12})]^2 \right)\, 
   [\bOm^+(0,1)]^2 \, \bOm^+(1,0) \, \bOm^+(1,1) \\
   & +\frac12  \kappa (\gamma_{12})\, \kappa (2\gamma_{12})\,
   [\kappa (\gamma_{12}) +\kappa (3\gamma_{12})]\, 
    [\bOm^+(1,0)]^2\,  \bOm^+(0,1)\, \bOm^+(0,2) \\
   & +\frac16 [\kappa (2\gamma_{12})]^3\, [\bOm^+(0,1)]^3\, \bOm^+(2,0)  
    +\frac{1}{12}     [\kappa (\gamma_{12})]^3 \, [3 \kappa (\gamma_{12})+\kappa (3
   \gamma_{12})]\, 
   \bOm^+(0,1)^3 \, \bOm^+(1,0)^2
\end{split}
 \ee
\be
\label{DOm24} 
\begin{split}
\Delta\bOm(2,4) = & \left[ \frac{1}{12} \kappa(2 \gamma_{12}) \kappa(\gamma_{12})^4
+\frac{1}{48} \kappa(4 \gamma_{12}) \kappa(\gamma_{12})^4 \right]
\bOm^+(0,1)^4 \bOm^+(1,0)^2 
   \\
   + & \frac{1}{12} \left[ 4 \kappa(2 \gamma_{12})
   \kappa(\gamma_{12})^3
     + \kappa(4 \gamma_{12}) \kappa(\gamma_{12})^3
   +  \kappa(2
  \gamma_{12})^3 \kappa(\gamma_{12})
   \right]
   \bOm^+(0,1)^3 \bOm^+(1,0) \bOm^+(1,1) \\ 
 +& \left[ \frac{1}{2}\kappa(2 \gamma_{12})^2 \kappa(\gamma_{12})^2
      +\frac{1}{4}   \kappa(2 \gamma_{12}) \kappa(4 \gamma_{12})
   \kappa(\gamma_{12})^2 \right]
    \bOm^+(0,1)^2 \bOm^+(0,2) \bOm^+(1,0)^2
 \\ 
   +& \frac{1}{4} \bOm^+(0,1)^2 \bOm^+(1,1)^2
   \kappa(2 \gamma_{12}) \kappa(\gamma_{12})^2\\ 
   +& \left[ \frac{1}{2} \kappa(2 \gamma_{12})   \kappa(\gamma_{12})^2
     +\frac{1}{4}  \kappa(4 \gamma_{12}) \kappa(\gamma_{12})^2
     +\frac{1}{4}  \kappa(2 \gamma_{12})^3
   \right]
   \bOm^+(0,1)^2 \bOm^+(1,0)
   \bOm^+(1,2) 
 \\ 
  +& \left[ \kappa(2 \gamma_{12})^2 \kappa(\gamma_{12})
  + \kappa(2 \gamma_{12}) \kappa(4 \gamma_{12}) \kappa(\gamma_{12})\right]
  \bOm^+(0,1) \bOm^+(0,2) \bOm^+(1,0) \bOm^+(1,1)
   \\ 
   + & \kappa(2\gamma_{12}) \kappa(\gamma_{12})
   \bOm^+(0,1) \bOm^+(1,1) \bOm^+(1,2)
 +\frac{1}{2} \kappa(4 \gamma_{12})^2 \bOm^+(0,4) \bOm^+(1,0)^2 
   \\ 
   +& \frac{1}{2}  \left[\kappa(2 \gamma_{12})   \kappa(\gamma_{12}) 
   +\kappa(4 \gamma_{12}) \kappa(\gamma_{12}) 
    +  \kappa(2 \gamma_{12}) \kappa(3 \gamma_{12}) \right]
    \bOm^+(0,1) \bOm^+(1,0) \bOm^+(1,3) 
        \\ 
   +& \frac{1}{2} \left[  \kappa(2 \gamma_{12})   \kappa(3 \gamma_{12}) \kappa(\gamma_{12})
+ \kappa(3 \gamma_{12}) \kappa(4 \gamma_{12}) \kappa(\gamma_{12}) \right]
    \bOm^+(0,1) \bOm^+(0,3) \bOm^+(1,0)^2
       \\   
      + & \frac{1}{2} \left[
      \kappa(6 \gamma_{12}) \kappa(\gamma_{12})
      +   \kappa(2 \gamma_{12}) \kappa(3 \gamma_{12}) 
      +\kappa(3 \gamma_{12}) \kappa(4 \gamma_{12})
      \right]
      \bOm^+(0,3)  \bOm^+(1,0) \bOm^+(1,1) 
      \\ 
   + & \frac{1}{24}\kappa(2 \gamma_{12})^4  \bOm^+(0,1)^4 \bOm^+(2,0) 
   +\frac{1}{6} \kappa(2\gamma_{12})^3
   \bOm^+(0,1)^3 \bOm^+(2,1)  \\ 
   +&\frac{1}{2} \kappa(2 \gamma_{12})^2 \bOm^+(0,2) \bOm^+(1,1)^2 
   +\frac{1}{2}  \kappa(2 \gamma_{12})^2 \bOm^+(0,1)^2 \bOm^+(2,2)
     \\ 
   +&\frac{1}{2} \kappa(4 \gamma_{12})^2 \bOm^+(0,2)^2 \bOm^+(2,0)
   +\kappa(2 \gamma_{12}) \bOm^+(1,1) \bOm^+(1,3) 
   +\kappa(2 \gamma_{12}) \bOm^+(0,1)\bOm^+(2,3) 
      \\ 
   +&\frac{1}{4} \kappa(2\gamma_{12})^2 \kappa(4 \gamma_{12}) \bOm^+(0,2)^2 \bOm^+(1,0)^2 
   +\frac{1}{2}  \kappa(2 \gamma_{12})^2 \kappa(4 \gamma_{12}) 
   \bOm^+(0,1)^2 \bOm^+(0,2) \bOm^+(2,0)
   \\ 
   +&\kappa(4\gamma_{12}) \bOm^+(1,0) \bOm^+(1,4) 
   +\kappa(4 \gamma_{12})  \bOm^+(0,2) \bOm^+(2,2) 
   +\kappa(6  \gamma_{12})  \bOm^+(0,3) \bOm^+(2,1) \\ 
   +& \kappa(2 \gamma_{12}) \kappa(4 \gamma_{12})
 \bOm^+(0,2) \bOm^+(1,0)   \bOm^+(1,2)    
 +\kappa(2 \gamma_{12}) \kappa(4 \gamma_{12}) \bOm^+(0,1) \bOm^+(0,2)
   \bOm^+(2,1) \\ 
   +& \kappa(2 \gamma_{12}) \kappa(6 \gamma_{12})
\bOm^+(0,1) \bOm^+(0,3) \bOm^+(2,0)    
+ \kappa(8 \gamma_{12}) \bOm^+(0,4) \bOm^+(2,0)\, . 
   \end{split}  
   \ee 

To derive these results from the JS wall-crossing formula described in
\S\ref{sjoyce}, we use results for the $U$ and $\cL$ factors tabulated
below.  In these tables, the first
column describes the total charge, the second column
lists the allowed ordered decompositions  and the third and 
the fourth columns
give the $U$ and $\cL$ factors introduced in 
\S\ref{sstJS}. Note that the $U$ factor typically 
involves a sum of multiple S-factors representing possible
partitioning of the constituents as in \eqref{euexample},
while the $\cL$ factor comes from a sum of several trees
as in \eqref{egenL}.

 \be\nn
\begin{array}{|c|c|c|c|c|}
\hline
\gamma & (\alpha_i)  & U & \cL \\
\hline
(1, 2) & \text{$\{(0,1)$, $(0,1)$, $(1,0)\}$}  &
   \frac{1}{2} & \gamma_{12}^2 \\
 & \text{$\{(0,1)$, $(1,0)$, $(0,1)\}$}  & -1 &
   -\gamma_{12}^2 \\
 & \text{$\{(1,0)$, $(0,1)$, $(0,1)\}$}  &
   \frac{1}{2} & \gamma_{12}^2
\\ \hline 
(1, 3) &  \text{$\{(0,1)$, $(0,2)$, $(1,0)\}$}  & \frac{1}{2} & 2 \gamma_{12}^2 \\
& \text{$\{(0,1)$, $(1,0)$, $(0,2)\}$} & -1 & -2 \gamma_{12}^2 \\
 &\text{$\{(0,2)$, $(0,1)$, $(1,0)\}$}  & \frac{1}{2} & 2 \gamma_{12}^2 \\
 &\text{$\{(0,2)$, $(1,0)$, $(0,1)\}$}  & -1 & -2 \gamma_{12}^2 \\
 &\text{$\{(1,0)$, $(0,1)$, $(0,2)\}$}  & \frac{1}{2} & 2 \gamma_{12}^2 \\
 & \text{$\{(1,0)$, $(0,2)$, $(0,1)\}$}  & \frac{1}{2} & 2 \gamma_{12}^2 \\
 & \text{$\{(0,1)$, $(0,1)$, $(1,1)\}$}  &
   \frac{1}{2} & \gamma_{12}^2 \\
 &\text{$\{(0,1)$, $(1,1)$, $(0,1)\}$}  & -1 &
   -\gamma_{12}^2 \\
 &\text{$\{(1,1)$, $(0,1)$, $(0,1)\}$}  &
   \frac{1}{2} & \gamma_{12}^2
\\ \hline
 (2, 2) & \text{$\{(0,2)$, $(1,0)$, $(1,0)\}$}  & \frac{1}{2} & 4 \gamma_{12}^2 \\
 &\text{$\{(1,0)$, $(0,2)$, $(1,0)\}$} & -1 & -4 \gamma_{12}^2 \\
 & \text{$\{(1,0)$, $(1,0)$, $(0,2)\}$}  & \frac{1}{2} & 4 \gamma_{12}^2\\
 & \text{$\{(0,1)$, $(1,0)$, $(1,1)\}$}  &
   -\frac{1}{2} & -\gamma_{12}^2 \\
 &\text{$\{(0,1)$, $(1,1)$, $(1,0)\}$}  & 1 &
   3 \gamma_{12}^2 \\
 &\text{$\{(1,0)$, $(0,1)$, $(1,1)\}$}  &
   -\frac{1}{2} & -\gamma_{12}^2 \\
 &\text{$\{(1,0)$, $(1,1)$, $(0,1)\}$}  & 1 &
   3 \gamma_{12}^2 \\
 &\text{$\{(1,1)$, $(0,1)$, $(1,0)\}$}  &
   -\frac{1}{2} & -\gamma_{12}^2 \\
 &\text{$\{(1,1)$, $(1,0)$, $(0,1)\}$}  &
   -\frac{1}{2} & -\gamma_{12}^2\\
  &  \text{$\{(0,1)$, $(0,1)$, $(2,0)\}$}  &
   \frac{1}{2} & 4 \gamma_{12}^2 \\
 &\text{$\{(0,1)$, $(2,0)$, $(0,1)\}$}  & -1 &
   -4 \gamma_{12}^2 \\
 &\text{$\{(2,0)$, $(0,1)$, $(0,1)\}$}  &
   \frac{1}{2} & 4 \gamma_{12}^2
\\ \hline
\end{array}
\qquad
\begin{array}{|c|c|c|c|c|}
\hline
\gamma & (\alpha_i)  & U & \cL \\
\hline
 (2, 3)  &
 \text{$\{(0,3)$, $(1,0)$, $(1,0)\}$}  &
   \frac{1}{2} & 9 \gamma_{12}^2 \\
 &\text{$\{(1,0)$, $(0,3)$, $(1,0)\}$}  & -1 &
   -9 \gamma_{12}^2 \\
 &\text{$\{(1,0)$, $(1,0)$, $(0,3)\}$}  &
   \frac{1}{2} & 9 \gamma_{12}^2 \\
& \text{$\{(0,2)$, $(1,0)$, $(1,1)\}$} & 0 &
   0 \\
& \text{$\{(0,2)$, $(1,1)$, $(1,0)\}$}  & 1 &
   8 \gamma_{12}^2 \\
& \text{$\{(1,0)$, $(0,2)$, $(1,1)\}$}  & -1 &
   -4 \gamma_{12}^2 \\
 &\text{$\{(1,0)$, $(1,1)$, $(0,2)\}$}  & 1 &
   8 \gamma_{12}^2 \\
& \text{$\{(1,1)$, $(0,2)$, $(1,0)\}$}  & -1 &
   -4 \gamma_{12}^2 \\
 &\text{$\{(1,1)$, $(1,0)$, $(0,2)\}$} & 0 &
   0 \\
 &\text{$\{(0,1)$, $(1,1)$, $(1,1)\}$}  &
   \frac{1}{2} & \gamma_{12}^2 \\
& \text{$\{(1,1)$, $(0,1)$, $(1,1)\}$}  & -1 &
   -\gamma_{12}^2 \\
& \text{$\{(1,1)$, $(1,1)$, $(0,1)\}$}  &
   \frac{1}{2} & \gamma_{12}^2 \\
& \text{$\{(0,1)$, $(1,0)$, $(1,2)\}$}  & -1 &
   -3 \gamma_{12}^2 \\
& \text{$\{(0,1)$, $(1,2)$, $(1,0)\}$} & 1 &
   5 \gamma_{12}^2 \\
& \text{$\{(1,0)$, $(0,1)$, $(1,2)\}$}  & 0 &
   -\gamma_{12}^2 \\
& \text{$\{(1,0)$, $(1,2)$, $(0,1)\}$}  & 1 &
   5 \gamma_{12}^2 \\
& \text{$\{(1,2)$, $(0,1)$, $(1,0)\}$} & 0 &
   -\gamma_{12}^2 \\
& \text{$\{(1,2)$, $(1,0)$, $(0,1)\}$}  & -1 &
   -3 \gamma_{12}^2 \\
&\text{$\{(0,1)$, $(0,2)$, $(2,0)\}$}  &
   \frac{1}{2} & 8 \gamma_{12}^2 \\
& \text{$\{(0,1)$, $(2,0)$, $(0,2)\}$}  & -1 &
   -8 \gamma_{12}^2 \\
& \text{$\{(0,2)$, $(0,1)$, $(2,0)\}$}  &
   \frac{1}{2} & 8 \gamma_{12}^2 \\
& \text{$\{(0,2)$, $(2,0)$, $(0,1)\}$}  & -1 &
   -8 \gamma_{12}^2 \\
& \text{$\{(2,0)$, $(0,1)$, $(0,2)\}$}  &
   \frac{1}{2} & 8 \gamma_{12}^2 \\
& \text{$\{(2,0)$, $(0,2)$, $(0,1)\}$} &
   \frac{1}{2} & 8 \gamma_{12}^2 \\
& \text{$\{(0,1)$, $(0,1)$, $(2,1)\}$}  &
   \frac{1}{2} & 4 \gamma_{12}^2 \\
& \text{$\{(0,1)$, $(2,1)$, $(0,1)\}$}  & -1 &
   -4 \gamma_{12}^2 \\
& \text{$\{(2,1)$, $(0,1)$, $(0,1)\}$}  &
   \frac{1}{2} & 4 \gamma_{12}^2\\
\hline
\end{array}
\ee

\be
\begin{array}{|c|c|c|c|c|}
\hline
\gamma & (\alpha_i)  & U & \cL \\
\hline
(2,3) 
   & \text{$\{(0,1)$, $(0,1)$, $(0,1)$, $(1,0)$,
   $(1,0)\}$}  & \frac{1}{12} & -\gamma_{12}^3 \\
   & \text{$\{(0,1)$, $(0,1)$, $(1,0)$, $(0,1)$,
   $(1,0)\}$}  & 0 & \gamma_{12}^3 \\
   & \text{$\{(0,1)$, $(0,1)$, $(1,0)$, $(1,0)$,
   $(0,1)\}$}  & -\frac{1}{4} & -4 \gamma_{12}^3 \\
   & \text{$\{(0,1)$, $(1,0)$, $(0,1)$, $(0,1)$,
   $(1,0)\}$}  & -\frac{1}{2} & -\gamma_{12}^3 \\
   & \text{$\{(0,1)$, $(1,0)$, $(0,1)$, $(1,0)$,
   $(0,1)\}$}  & 1 & 2 \gamma_{12}^3 \\
   & \text{$\{(0,1)$, $(1,0)$, $(1,0)$, $(0,1)$,
   $(0,1)\}$}  & -\frac{1}{4} & 0 \\
   & \text{$\{(1,0)$, $(0,1)$, $(0,1)$, $(0,1)$,
   $(1,0)\}$}  & \frac{1}{3} & \gamma_{12}^3 \\
   & \text{$\{(1,0)$, $(0,1)$, $(0,1)$, $(1,0)$,
   $(0,1)\}$}  & -\frac{1}{2} & 0 \\
   & \text{$\{(1,0)$, $(0,1)$, $(1,0)$, $(0,1)$,
   $(0,1)\}$}  & 0 & -2 \gamma_{12}^3 \\
   & \text{$\{(1,0)$, $(1,0)$, $(0,1)$, $(0,1)$,
   $(0,1)\}$} & \frac{1}{12} & 4 \gamma_{12}^3 \\
   \hline   
\end{array}
\ee

\be
 \begin{array}{|c|c|c|c|c|}
\hline
\gamma & (\alpha_i)  & U & \cL \\
\hline
(1,3) 
   & \text{$\{(0,1)$, $(0,1)$, $(0,1)$, $(1,0)\}$}
   & \frac{1}{6} & -\gamma_{12}^3 \\
   & \text{$\{(0,1)$, $(0,1)$, $(1,0)$, $(0,1)\}$}
    & -\frac{1}{2} & \gamma_{12}^3 \\
   & \text{$\{(0,1)$, $(1,0)$, $(0,1)$, $(0,1)\}$}
    & \frac{1}{2} & -\gamma_{12}^3 \\
   & \text{$\{(1,0)$, $(0,1)$, $(0,1)$, $(0,1)\}$}
   & -\frac{1}{6} & \gamma_{12}^3
\\ \hline
(2,2) 
   & \text{$\{(0,1)$, $(0,1)$, $(1,0)$, $(1,0)\}$}
    & \frac{1}{4} & -4 \gamma_{12}^3 \\
   & \text{$\{(0,1)$, $(1,0)$, $(0,1)$, $(1,0)\}$}
    & -\frac{1}{2} & 2 \gamma_{12}^3 \\
   & \text{$\{(0,1)$, $(1,0)$, $(1,0)$, $(0,1)\}$}
    & 0 & 0 \\
   & \text{$\{(1,0)$, $(0,1)$, $(0,1)$, $(1,0)\}$}
   & 0 & 0 \\
   & \text{$\{(1,0)$, $(0,1)$, $(1,0)$, $(0,1)\}$}
   & \frac{1}{2} & -2 \gamma_{12}^3 \\
   & \text{$\{(1,0)$, $(1,0)$, $(0,1)$, $(0,1)\}$}
    & -\frac{1}{4} & 4 \gamma_{12}^3
\\ \hline
\{2,3\} 
   & \text{$\{(0,1)$, $(0,2)$, $(1,0)$, $(1,0)\}$}
    & \frac{1}{4} & -12 \gamma_{12}^3 \\
   & \text{$\{(0,1)$, $(1,0)$, $(0,2)$, $(1,0)\}$}
    & -1 & 8 \gamma_{12}^3 \\
   & \text{$\{(0,1)$, $(1,0)$, $(1,0)$, $(0,2)\}$}
    & \frac{1}{2} & -4 \gamma_{12}^3 \\
   & \text{$\{(0,2)$, $(0,1)$, $(1,0)$, $(1,0)\}$}
    & \frac{1}{4} & -12 \gamma_{12}^3 \\
   & \text{$\{(0,2)$, $(1,0)$, $(0,1)$, $(1,0)\}$}
    & 0 & 4 \gamma_{12}^3 \\
   & \text{$\{(0,2)$, $(1,0)$, $(1,0)$, $(0,1)\}$}
    & -\frac{1}{2} & 4 \gamma_{12}^3 \\
   & \text{$\{(1,0)$, $(0,1)$, $(0,2)$, $(1,0)\}$}
    & \frac{1}{2} & 0 \\
   & \text{$\{(1,0)$, $(0,1)$, $(1,0)$, $(0,2)\}$}
    & 0 & -4 \gamma_{12}^3 \\
   & \text{$\{(1,0)$, $(0,2)$, $(0,1)$, $(1,0)\}$}
    & -\frac{1}{2} & 0 \\
   & \text{$\{(1,0)$, $(0,2)$, $(1,0)$, $(0,1)\}$}
   & 1 & -8 \gamma_{12}^3 \\
   & \text{$\{(1,0)$, $(1,0)$, $(0,1)$, $(0,2)\}$}
    & -\frac{1}{4} & 12 \gamma_{12}^3 \\
   & \text{$\{(1,0)$, $(1,0)$, $(0,2)$, $(0,1)\}$}
    & -\frac{1}{4} & 12 \gamma_{12}^3 \\
   & \text{$\{(0,1)$, $(0,1)$, $(1,0)$, $(1,1)\}$}
   & 0 & 0 \\
   & \text{$\{(0,1)$, $(0,1)$, $(1,1)$, $(1,0)\}$}
   & \frac{1}{2} & -8 \gamma_{12}^3 \\
   & \text{$\{(0,1)$, $(1,0)$, $(0,1)$, $(1,1)\}$}
    & -1 & 2 \gamma_{12}^3 \\
   & \text{$\{(0,1)$, $(1,0)$, $(1,1)$, $(0,1)\}$}
    & 1 & -4 \gamma_{12}^3 \\
   & \text{$\{(0,1)$, $(1,1)$, $(0,1)$, $(1,0)\}$}
    & 0 & 2 \gamma_{12}^3 \\
   & \text{$\{(0,1)$, $(1,1)$, $(1,0)$, $(0,1)\}$}
  & -1 & 4 \gamma_{12}^3 \\
   & \text{$\{(1,0)$, $(0,1)$, $(0,1)$, $(1,1)\}$}
    & \frac{1}{2} & 0 \\
   & \text{$\{(1,0)$, $(0,1)$, $(1,1)$, $(0,1)\}$}
   & 0 & -2 \gamma_{12}^3 \\
   & \text{$\{(1,0)$, $(1,1)$, $(0,1)$, $(0,1)\}$}
   & -\frac{1}{2} & 8 \gamma_{12}^3 \\
   & \text{$\{(1,1)$, $(0,1)$, $(0,1)$, $(1,0)\}$}
   & -\frac{1}{2} & 0 \\
   & \text{$\{(1,1)$, $(0,1)$, $(1,0)$, $(0,1)\}$}
    & 1 & -2 \gamma_{12}^3 \\
   & \text{$\{(1,1)$, $(1,0)$, $(0,1)$, $(0,1)\}$}
   & 0 & 0 \\
   & \text{$\{(0,1)$, $(0,1)$, $(0,1)$, $(2,0)\}$}
    & \frac{1}{6} & -8 \gamma_{12}^3 \\
   & \text{$\{(0,1)$, $(0,1)$, $(2,0)$, $(0,1)\}$}
    & -\frac{1}{2} & 8 \gamma_{12}^3 \\
   & \text{$\{(0,1)$, $(2,0)$, $(0,1)$, $(0,1)\}$}
    & \frac{1}{2} & -8 \gamma_{12}^3 \\
   & \text{$\{(2,0)$, $(0,1)$, $(0,1)$, $(0,1)\}$}
   & -\frac{1}{6} & 8 \gamma_{12}^3
\\ \hline
\end{array}
\ee

\section{D6-D0 bound states\label{secD0D6}}

In this subsection, we test and apply the wall-crossing  
formula on generalized DT-invariants
for dimension zero sheaves on a Calabi-Yau threefold $\cX$, for which many
results are already known in the literature
\cite{ks,Toda:2009,Stoppa:2009,Nagao:2010}. 

The stability conditions for coherent sheaves on $\cX$ depend on the  complexified
K\"ahler moduli  $t^a=B^a+\I J^a$. The holomorphic central
charge $Z_\gamma$ is given, in the large $J$ limit, by $Z_\gamma=-\int_\cX
e^{-t}\sqrt{\mathrm{Td}(\cX)}\wedge \gamma$. Let us denote by $\Omega(r,n;t)$ the
generalized DT-invariant for a sheaf of rank $r$, vanishing first and second
Chern class, and third Chern class $n$. Physically, $\Omega(r,n)$ counts the number of bound states of $r$ D6-branes and $n$ D0-branes,
with charge vector $\gamma= r \gamma_1 + n \gamma_2$,
$\gamma_{12}=-1$. 

It is known that   for infinite volume and small $B$-field,  a configuration of 
$r\geq 1$ D6-branes and $n\geq 1$ D0 branes do not form any 
bound state \cite{Witten:2000mf}. Moreover, there are no bound states of 
$r>1$ D6-branes, while $n\geq 1$ D0 bind into 
 precisely $|\chi|$ bosonic (fermionic)
bound states 
for negative (positive) $\chi$,
where  $\chi$ is the Euler number of $\cX$. 
Thus the only non-vanishing
DT invariants in this chamber are \cite{ks, Joyce:2008pc} 
\be
\label{D0D6Omp}
\Omega^+(1,0) = 1\ ,\qquad \Omega^+(0,n)=-\chi\quad (n>0)\ ,
\ee
where we have used the notation $\Omega^\pm(m,n)
=\Omega^\pm(m\gamma_1+n\gamma_2)$. Similarly, 
the motivic invariants are given by 
\be
\label{D0D6Ompmot}
\Omega^+_{\rm ref}(1,0,y) = 1\ ,\qquad \Omega^+_{\rm ref}(0,n,y)
=-P(y)/y^3\quad (n>0)\ ,
\ee
where $P(y)=1+b_2 y^2-b_3 y^3+b_2 y^4+y^6$ is the Poincar\'e polynomial
of $\cX$, such that \eqref{D0D6Ompmot} reduces to \eqref{D0D6Omp} in the 
classical limit $y\to 1$. 

By increasing the magnitude of the $B$-field, one reaches the wall 
of marginal stability $\cP(\gamma_1,\gamma_2)$. We refer to the
chamber across this wall as the  `DT' chamber. We shall obtain the 
motivic DT invariants in this chamber for $r\leq 3$ by applying the formul\ae\ 
derived in Section \ref{sec_KS}, suitably generalized to the motivic case
according to the discussion in Section \ref{srefined}.

To apply the semi-primitive wall-crossing formul\ae\  and
its higher order generalizations,
it is useful to introduce partition functions
\be
\label{defZNqmot}
Z^\pm_{\rm ref}(M,q,y) = \sum_{N=0}^{\infty} \Omega^\pm_{\rm ref}\argy{M}{N}{y}\,q^{N}\ ,\qquad
\bar Z^\pm_{\rm ref}(M,q,y) = \sum_{N=0}^{\infty} \bOm^\pm_{\rm ref}\argy{M}{N}{y}\,q^{N}\ ,
\ee
for fixed value of $M$. These two objects are related by 
\be
\begin{split}
\bar Z^\pm_{\rm ref}(M,q,y) 
=& \sum_{d|M} \frac{y-y^{-1}}{d(y^{d}-y^{-d})} Z^\pm_{\rm ref}(M/d,q^d,y^d)\ ,\qquad\\
Z^\pm_{\rm ref}(M,q,y) =& \sum_{d|M} \mu(d) \frac{(y-y^{-1}}{d(y^{d}-y^{-d})}  
\bar Z^\pm_{\rm ref}(M/d,q^d,y^d)\ .
\end{split}
\ee
In the chamber $c_+$, the modified
partition functions \eqref{defZNqmot} are simply
\be
\bar Z_{\rm ref}^+(1,q,y)=1\ ,\qquad
\bar Z_{\rm ref}^+(2,q,y)=\frac{y}{2(1+y^2)}\ ,\qquad
\bar Z_{\rm ref}^+(3,q,y)=\frac{y^2}{3(1+y^2+y^4)}\ .
\ee
Moreover, the partition function of the halo degeneracies 
follows from \eqref{Zhalomotprod},
\be
\begin{split}
Z_{\rm halo}(\gamma_1,q,y) = \prod_{k=1}^{\infty} 
\prod_{j=1}^{k} &
\Big\{
\left( 1- (-q)^k y^{2j-k+2} \right)^{-1}\, 
\left( 1- (-q)^k y^{2j-k} \right)^{-b_2}\, 
\left( 1- (-q)^k y^{2j-1-k} \right)^{b_3}\\ &
\left( 1- (-q)^k y^{k-2j} \right)^{-b_2}\,
\left( 1- (-q)^k y^{k-2j-2} \right)^{-1}\Big\}
\end{split}
\ee
where we have made a change of variable
$j\to k+1-j$ in the last two terms to make the
$y\to y^{-1}$ symmetry manifest.
Applying \eqref{Z1pmy}, we immediately obtain the partition function of
the motivic invariants with $r=1$,
\be
Z_{\rm ref}^-(1,q,y) = \bar Z_{\rm ref}^-(1,q,y) = Z_{\rm halo}(\gamma_1,q,y) \ .
\ee
In the classical limit $y\to 1$, 
this reduces to 
\be
Z^-(1,q) = \bar Z^-(1,q) = [M(-q)]^\chi\ ,\qquad
\ee
where $M(q)=\prod_{k=1}^\infty (1-q^k)^{-k}$ is the Mac-Mahon 
function 
\cite{gw-dt,MR2407118,MR2485880, MR2284053}.
For comparison with the higher rank formul\ae\  of \cite{Toda:2009,Stoppa:2009} below,
it is useful to note that the expansion of this formula in powers of $\chi$ reads
\be
e^{-\chi \sum_{k=1}^{\infty} k \log(1-(-q)^k)} = 
1+ \sum_{p=1}^{\infty}
\frac{\chi^p}{p!} \prod_{i=1}^p \sum_{k_i=1}^{\infty} 
\sum_{n_i=1}^{\infty} \frac{k_i}{n_i} (-q)^{k_i n_i}\ .
\ee

For $r=2$, we have
\be
Z_{\rm ref}^-(2,q,y) = \bar Z_{\rm ref}^-(2,q,y) - \frac{y}{2(1+y^2)} \bar Z_{\rm ref}^-(1,q^2,y^2) \, .
\ee
We now use
\eqref{defOm2p}, \eqref{Z2pm} together with 
the replacement 
\eqref{ereplace} to obtain
\be
\begin{split}
Z_{\rm ref}^-(2,q,y)= &\frac{y}{2(1+y^2}\left[ 
Z_{\rm halo}(2\gamma_1,q,y)-Z_{\rm halo}(\gamma_1,q^2,y^2)\right]\\
&-\frac14 \sum_{n_1,n_2} \kscom{ |n_1-n_2|,y}\, \Omega_{\rm ref}^-(1,n_1,y) \, 
\Omega_{\rm ref}^-(1,n_2,y)\, q^{n_1+n_2}\ .
\end{split}
\ee
In the classical limit, this reduces to \be
Z^-(2,q)= \frac14 [M(q)]^{2\chi} - \frac14 [M(-q^2)]^\chi
-\frac14 \sum_{n_1,n_2} (-1)^{n_1-n_2} 
 |n_1-n_2|\, \Omega^-(1,n_1) \, \Omega^-(1,n_2)\, q^{n_1+n_2}\ .
\ee
This agrees with  \cite{Toda:2009},Thm 1.2 and \cite{Stoppa:2009},(2.9), 
who obtain
\be
\begin{split}
\Omega^-(2,n)=&{1\over 4}\delta_{n,0}+
\sum_{p=1}^{\infty} 2^{p-2}
\frac{\chi^p}{p!}  \sum_{\substack{k_i>0,n_i>0\\\sum_i
k_i n_i=n}} \prod_{i=1}^p \frac{k_i}{n_i}  
-\frac14 \Omega^-(1,n/2) \\
&
-\frac14\sum_{\substack{
n_1\geq 0,n_2\geq 0\\ n_1+n_2=n}} (-1)^{n_1-n_2} 
 |n_1-n_2|\, \Omega^-(1,n_1) \, \Omega^-(1,n_2)
\end{split}
\ee
where it is understood that the third term is zero if $n$ is odd.

For $r=3$, we use similarly
\be
Z_{\rm ref}^-(3,q,y) = \bar Z_{\rm ref}^-(3,q,y) - \frac{y^2}{3(1+y^2+y^4)} 
\bar Z_{\rm ref}^-(1,q^3,y^3) 
\ee
and \eqref{defOm3p}, \eqref{Z3pm} to obtain
\be
\begin{split}
Z_{\rm ref}^-(3,q,y)= &\frac{y^2}{3(1+y^2+y^4}\left[ 
Z_{\rm halo}(3\gamma_1,q,y)-Z_{\rm halo}(\gamma_1,q^3,y^3)\right]\\
 &-\frac12 \sum_{n_1,n_2} \kscom{|n_2-2 n_1|,y}
 \,  \Omega_{\rm ref}^-(1,n_1,y)\, \Omega_{\rm ref}^-(2,n_2,y)  \, q^{n_1+n_2}\\
 &-\frac{y}{4(1+y^2)} \sum_{n_1,n_2}
 \kappa(2|n_2-n_1|,y)\,  \Omega_{\rm ref}^-(1,n_1,y)\, \Omega_{\rm ref}^-(1,n_2,y^2)  \, q^{n_1+2n_2}\\
& -\frac{1}{12} \sum_{n_1,n_2} 
\kscom{n_1-n_2,y}^2 \, [\Omega_{\rm ref}^-(1,n_1,y)]^2 \, 
\Omega_{\rm ref}^-(1,n_2,y) \, q^{2n_1+n_2}\\
&- \sum_{n_1>n_2>n_3}
\kappa(n_1,n_2,n_3,y)\, \Omega_{\rm ref}^-(1,n_1,y)\,\Omega_{\rm ref}^-(1,n_2,y)\,
\Omega_{\rm ref}^-(1,n_3,y)\, q^{n_1+n_2+n_3}\ .
\end{split}
\ee
In the classical limit, this reduces to 
\be
\begin{split}
Z^-(3,q) = &\frac19 [M(-q)]^{3\chi} - \frac19 [M(-q^3)]^{\chi}\\
 &-\frac12 \sum_{n_1,n_2}
(-1)^{n_2-2n_1} |n_2-2 n_1|\,  \Omega^-(1,n_1)\, \Omega^-(2,n_2)  \, q^{n_1+n_2}\\
 &-\frac14 \sum_{n_1,n_2}
 |n_2-n_1|\,  \Omega^-(1,n_1)\, \Omega^-(1,n_2)  \, q^{n_1+2n_2}\\
& -\frac{1}{12} \sum_{n_1,n_2} 
(n_1-n_2)^2  [\Omega^-(1,n_1)]^2 \, \Omega^-(1,n_2) \, q^{2n_1+n_2}\\
&- \sum_{n_1>n_2>n_3}
\kappa(n_1,n_2,n_3)\, \Omega^-(1,n_1)\,\Omega^-(1,n_2)\,\Omega^-(1,n_3)\, q^{n_1+n_2+n_3}\ .
\end{split}
\ee
This can be 
rewritten as in \cite{Stoppa:2009}, eq. (2.12), after correcting the coefficient of the 3-body term
from 1/4 into 1/6 in that equation:
\be
\label{stoppa3}
\begin{split}
\Omega^-(3,n) & = {1\over 9}\delta_{n,0}+
\sum_{p=1}^{\infty} 3^{p-2}
\frac{\chi^p}{p!}  \sum_{\substack{k_i>0,n_i>0\\k_i n_i=n}} \prod_{i=1}^p \frac{k_i}{n_i} 
-\frac19 \Omega^-(1,n/3) \\
& -\frac12 \sum_{\substack{n_1\geq 0,n_2\geq 0\\ n_1+n_2=n}}
(-1)^{n_2-2n_1} |n_2-2 n_1| \Omega^-(1,n_1)\Omega^-(2,n_2) 
\\
& -\frac14 \sum_{\substack{n_1\geq 0,n_2\geq 0\\ 2n_1+n_2=n}}
|n_1-n_2| \Omega^-(1,n_1)\Omega^-(1,n_2) \\ &
-\frac1{12} 
\sum_{\substack{n_1\geq 0,n_2\geq 0\\ 2n_1+n_2=n}}
(n_1-n_2)^2  [\Omega^-(1,n_1)]^2 \Omega^-(1,n_2) \\
&-\frac16 \sum_{\substack{0\leq n_1 <n_2 < n_3\\ n_1+n_2+n_3=n}}
\left[ (n_1-n_2)(n_1+n_2-2 n_3) + (n_2-n_3)(2n_1-n_2-n_3) \right]\\
&\qquad \qquad \Omega^-(1,n_1)\Omega^-(1,n_2)\Omega^-(1,n_3)
\end{split}
\ee
where it is understood that the third term vanishes
if $n$ is not a multiple of 3.

The classical invariants  $\Omega^-(r,n)$ for low values of $(r,n)$ are summarized in the table 
below:
\be
\begin{array}{|c|c|c|c|c|c|}
\hline
r \backslash n & \ 0\  & \ 1\  & 2 & 3 & 4  \\\hline
0 &  \cdot  & -\chi & -\chi & -\chi & -\chi \\
1 &  1  & -\chi & \frac12(\chi^2+5\chi) & -\frac16(\chi^3+15\chi^2+20\chi)&
\frac1{24}(\chi^4+30\chi^3+155\chi^2+126\chi)  \\
2 &  0  & 0      & -\chi &  -\frac16(\chi^3+15\chi^2+20\chi) & 
-\frac1{12}(\chi^4+30\chi^3+119 \chi^2 +102\chi)\\
3 &  0 & 0 & 0 & -\chi & \frac1{24}(\chi^4+30\chi^3+155\chi^2+126\chi) \\
\hline
\end{array}
\ee
\be
\begin{array}{|c|c|}
\hline
r \backslash n & 5  \\
\hline
0 & -\chi \\
1 & -\frac{1}{120} \chi  \left(\chi ^4+50 \chi ^3+575 \chi ^2+1630 \chi +624\right)  \\
2 & -\frac{1}{120} \chi  \left(7 \chi ^4+250 \chi ^3+1925 \chi ^2+3890 \chi +1248\right) \\
3 &   -\frac{1}{120} \chi  \left(7 \chi ^4+250 \chi ^3+1925 \chi ^2+3890 \chi +1248\right) \\
\hline
\end{array}
\ee
For low values of $(r,n)$, the motivic invariants can also be expressed 
directly in terms of the Poincar\'e polynomial of the CY threefold $\cX$,
\be
\begin{split}
(-y)^6\, \Omega^-_{\rm ref}(1,2,y) =& \frac{1}{2}
\left[ P(y^2)+[P(y)]^2\right] +
 \left(y^4+y^2\right) P(y)\\
=& 1+(b_2+1) y^2-b_3 y^3+\frac{1}{2} \left(b_2^2+5 b_2+2\right)
   y^4-(b_2+1) b_3 y^5\\
   &+\frac{1}{2} y^6 \left(2 b_2^2+4
   b_2+b_3^2-b_3+2\right)-(b_2+1) b_3 y^7+\frac{1}{2}
   \left(b_2^2+5 b_2+2\right) y^8\\
   &-b_3 y^9+(b_2+1)
   y^{10}+y^{12}
   \end{split}
\ee
\be
\begin{split}
(-y)^9\, \Omega^-_{\rm ref}(1,3,y) =& 
\frac{1}{6}  [P(y)]^3+\frac{1}{2}  P(y)\, P(y^2) 
   + \left(y^4+y^2\right) [P(y)]^2 \\
   &+
   \left(\frac{1}{3} P(y^3)+\left(y^8+y^6+y^4\right)
   P(y)\right) 
   \end{split}
\ee

\be 
\begin{split}
(-y)^{13}\, \Omega^-_{\rm ref}(2,4,y) 
= & 
\frac{1}{24} \bigg[ 
12 \left(y^3+y\right)^2 [P(y)]^3+\left(y^2+1\right)
   [P(y)]^4   \\ & 
   +3 \left(y^2+1\right) \left(2
   P\left(y^4\right)+[P\left(y^2\right)]^2+4 \left(y^8+y^4\right)
   P\left(y^2\right)\right)  \\ & 
   +6 \left(y^2+1\right)
   \left(P\left(y^2\right)+6 y^8+4 y^6+6 y^4\right) [P(y)]^2
    \\ &  
    +4 \left(3
   \left(y^6+y^2\right) P\left(y^2\right)+2 \left(\left(y^2+1\right)
   P\left(y^3\right) 
   \right. \right.   \\ & \left. \left. 
   +3 \left(y^8+y^6+2 y^4+y^2+1\right) y^6\right)\right)
   P(y) \bigg]
   \end{split}  
\ee
\be 
\begin{split} 
 (-y)^{17}\, &\Omega^-_{\rm ref}(3,5,y) 
=  \frac{1}{30}  \left(y^4+1\right) \left(6 P\left(y^5\right)+5 P\left(y^2\right)
   P\left(y^3\right)\right) \\
   & +\frac{1}{8} \left(y^4+y^2+1\right) P(y) \left(2
   P\left(y^4\right)+P\left(y^2\right)^2\right)\\
   & +\frac{1}{6}  \left(y^2+1\right) P(y)
   \left(\left(y^2+1\right) P(y)+2 \left(y^6+y^2\right)\right)
   P\left(y^3\right)\\
   & +\frac{1}{12} P(y)\, P\left(y^2\right)\, \left[ \left(y^4+3 y^2+1\right) P(y)^2+6
   \left(y^8+2 y^6+2 y^4+y^2\right) P(y)\right.\\
   & \left. \qquad +12 \left(y^8+y^6+2 y^4+y^2+1\right) y^4\right]
   \\
 & +\frac{1}{120} P(y) \left[ 120 \left(y^2+1\right)^2
   \left(y^4+y^2+1\right) y^4 P(y)^2+\left(y^4+5 y^2+1\right) P(y)^4\right. \\
   & \left. \qquad +20 \left(y^8+4 y^6+4
   y^4+y^2\right) P(y)^3+240 \left(y^{10}+2 y^8+3 y^6+3 y^4+2 y^2+1\right) y^6 P(y)\right.\\
   & \left. \qquad +120
   \left(y^{12}+y^{10}+2 y^8+2 y^6+2 y^4+y^2+1\right)
   y^8\right] 
   \end{split}  
   \ee

The motivic 
invariants $\Omega_{\rm ref}^-(r,n,-y)$ 
have an expansion in positive and negative powers of
$y$ of the form:
\be \label{epolmot}
\sum_{n=-K}^K c_n y^n\, ,
\ee
for some integer $K$
with 
\be \label{econd}
c_{n}=c_{-n}, \quad (-1)^n c_n \ge 0,
\quad c_n\in\mathbf{Z}\, , \quad (-1)^n \left( c_n - c_{n+2}
\right)
\ge 0 \quad \hbox{for}\quad n\ge 0\, .
\ee
These are required for consistency of the interpretation
\eqref{refdeg} for the refined index. 
In particular $(-1)^n \left( c_n - c_{n+2}\right)$ measures
the total number of states with angular momentum $n/2$
contributing to the index (without counting the angular
momentum factor $(n+1)$).
Thus our results
provide strong support to the motivic wall-crossing formula.
Furthermore in all examples the refined index
seems to satisfy the  symmetry property 
\be
\label{symprop}
\Omega_{\rm ref}^-(r,n,y) = \Omega_{\rm ref}^-(n-r,n,y)
\ee
when $0\leq r\leq n$. 
We have tested this for all $r\le 3$, $n-r\le 3$.
In the classical case, this 
was established in  
\cite{Nagao:2010}.

\section{Generalizations of Seiberg-Witten spectra}
\label{egensw}

Assume that in the region $c_+$, the spectrum consists only of two states 
$\gamma_1,\gamma_2$ with BPS degeneracies $\Omega^+(\gamma_1)=q$,
$\Omega^+(\gamma_2)=p$, respectively. Using the formul\ae\
derived in the text, 
we find that the BPS invariants 
$\Omega^-\argu{M}{N}$ in the region $c_-$
are given, for $\min(M,N)\leq 3$  by 
\be \label{eswstudy}
{\tiny \left(
\begin{array}{cccccc}
  0 & \vbox{\hbox{$\frac{1}{6} (-1)^{\gamma} \gamma p q$}
 \hbox{$(\gamma q+1) (\gamma q+2)$}} &
  \vbox{\hbox{$\frac{1}{6} \gamma p q (\gamma q+1)$}
  \hbox{$(3 \gamma p+\gamma (3 \gamma p+2)
   q+1)$}} & \vbox{\hbox{$
   \frac{1}{2} (-1)^{\gamma} \gamma^2 p (\gamma p+1) q $}
  \hbox{$(\gamma q+1)
   (\gamma q p+p+q) $}}& 
   \vbox{\hbox{$\frac{1}{24} \gamma p (\gamma p+1) q 
   $}
  \hbox{$(3 \gamma p+4
   \gamma (\gamma p+1) q+1) $}
  \hbox{$(3 \gamma p+4 \gamma (\gamma p+1) q+2)$}} 
  &
  \vbox{\hbox{$ \frac{1}{120} (-1)^\gamma\,
  \gamma p q (\gamma p+1)$}
  \hbox{$\Big(60 \gamma^5 p^3 q^2+80 
  \gamma^4 p^3 q+220 \gamma^4 p^2
   q^2$}\hbox{$+27 \gamma^3 p^3+240 
   \gamma^3 p^2 q+260 \gamma^3 p q^2
   $}\hbox{$+63 \gamma^2 p^2
   +220 \gamma^2 p q+100
   \gamma^2 q^2$}\hbox{$+42 \gamma p
   +60 \gamma q+8\Big)$}}
    \\ \\
 0 & \vbox{\hbox{$ \frac{1}{2} \gamma p q$}
 \hbox{$ (\gamma q+1)$}}  & \vbox{\hbox{
 $\frac{1}{4} \gamma p q $} \hbox{$
 (2
   \gamma (\gamma q p+p+q)$} \hbox{$
   -(-1)^{\gamma}+1)$}}
   & \vbox{\hbox{$ \frac{1}{6} \gamma p
   (\gamma p+1) q $} \hbox{$
(2 \gamma p+3 \gamma (\gamma p+1) q+1)$}} 
& \vbox{\hbox{$\frac{1}{12}
   \gamma^2 p (\gamma p+1) (\gamma p+2) $}
   \hbox{$q (3 q+p (3 \gamma q+2))$}} &
   \vbox{\hbox{$ \frac{1}{120} \gamma p (\gamma p+1) 
   (\gamma p+2) q $} \hbox{$(6 + 16 \gamma p + 8
    \gamma^2 p^2 
   + 30 \gamma q$}\hbox{$ + 45 \gamma^2 p q + 15 
   \gamma^3 p^2 q)$}} \\ \\
 q & (-1)^{\gamma} \gamma p q & 
 \frac{1}{2} \gamma p (\gamma p+1) q &
 \vbox{\hbox{$\frac{1}{6} (-1)^{\gamma} \gamma p 
$} \hbox{$ (\gamma p+1) (\gamma p+2) q $}}&
   \vbox{\hbox{$
   \frac{1}{24} \gamma p (\gamma p+1)$} \hbox{$
    (\gamma p+2) (\gamma p+3) q $}}&
\vbox{\hbox{$   \frac{1}{120} (-1)^{\gamma} \gamma p (\gamma p+1) (\gamma p+2)$} \hbox{$
   (\gamma p+3) (\gamma p+4) q $}}\\ \\
 0 & p & 0 & 0 & 0 & 0
\end{array}
\right)}
\vspace*{1cm}
\ee
where $\gamma\equiv \gamma_{12}$ (In this table, $M$ increases from bottom to top, 
while $N$ increases from left to right).
This table has been given for $\gamma<0$; as discussed
in \S\ref{soppo}, for a general sign of $\gamma$ the result
is given by replacing $\gamma\to -|\gamma|$ in this table.
With this replacement the table is symmetric under
$p\leftrightarrow q$, $M\leftrightarrow N$, and we can
recover some of the results for $M=4$ and $5$ from the
corresponding results for $N=4$ and 5.
For $(p,q,\gamma)=(1,1,-1)$
\eqref{eswstudy} agrees with the pentagonal 
identity \eqref{penta}. On the other hand for
$(p,q,\gamma)=
 (1,1,-2), (2,2,-1),(1,4,-1)$,
 $\Omega^+(\gamma)$ 
 gives the strong
 coupling spectrum of the $SU(2)$ 
 Seiberg-Witten theories 
with $N_f=0,2,3$ flavors.
 For example (in a suitable basis)
 the strong coupling spectrum 
of the Seiberg-Witten theory~\cite{Seiberg:1994aj,Bilal:1996sk} has 
a single state of charge $\gamma_1=(2,1)$ 
and a single state of charge
$\gamma_2=(0,-1)$ for $N_f=0$,\footnote{Here
$(n_e,n_m)$ denotes a state carrying electric charge
$n_e$ and magnetic charge $n_m$.}  
two states of charge $\gamma_1=(1,1)$ 
and two states of charge
$\gamma_2= (0,-1)$ for $N_f=2$ transforming respectively
in the (2,1) and (1,2) representations of the $SO(4)$ flavour
group, and four states of charge
$\gamma_1= (0,1)$ and 
a single state of charge $\gamma_2=(1,-2)$ 
for $N_f=3$ transforming respectively
in the {\bf 4} and {\bf 1} representation of the 
$SO(6)$ flavour
group.
Thus \eqref{eswstudy}
 reproduces 
 the weak coupling spectrum of the $SU(2)$ 
 Seiberg-Witten theories 
with $N_f=0,2,3$ flavors, as demonstrated in \cite{Gaiotto:2008cd,Dimofte:2009tm}.

If instead the spectrum in $c_+$ consists of three states $\gamma_1,\gamma_2,\gamma_1+\gamma_2,$ with respective BPS degeneracies $p,q,r$, the spectrum in $c_-$ is given by
\be{\tiny
\left(
\begin{array}{cccccc}
  0 & \vbox{\hbox{$ \frac{1}{6} \gamma q (\gamma q+1) 
 $}\hbox{$ ((-1)^{\gamma} p (\gamma q+2)+3
   r ) $}}& 
  \vbox{\hbox{$ 
  \frac{1}{6} \gamma q  (3 \gamma (\gamma q p+p)^2
  $}\hbox{$ +3 r
   (\gamma r+1)$}\hbox{$
   +p  (\gamma q (2 \gamma q+3)
   $}\hbox{$ 3 (-1)^{\gamma} \gamma
   (5 \gamma q+3) r+1 ) ) $}}
   &  \vbox{\hbox{$ \frac{1}{4} \gamma^2 p q  (3  (5 p q
   \gamma^2$}\hbox{$ +3 (p+q) \gamma+1 ) r
   $}\hbox{$ +2 (-1)^{\gamma}  (3 \gamma r^2+3
   r
   $}\hbox{$
   +(\gamma p+1) (\gamma q+1)$}\hbox{$
    (\gamma q p+p+q) ) )$}}
   &  \vbox{\hbox{$ \frac{1}{24}
   \gamma p  (q  (12 \gamma^2 (9 \gamma p+4) r^2
   $}\hbox{$ +12 \gamma (5
   \gamma p+2) r
   $}\hbox{$ +(\gamma p+1) (3 \gamma p
   $}\hbox{$ +4 \gamma (\gamma p+1) q+1) $}\hbox{$
   (3
   \gamma p$}\hbox{$ +4 \gamma (\gamma p+1) q+2) )
   $}\hbox{$ +2 (-1)^{\gamma} r
    (\gamma  (2 \gamma r^2+6 r $}\hbox{$
    +q (\gamma p (31 \gamma
   p+33)$}\hbox{$
   +\gamma (\gamma p (67 \gamma p+93)+32) q
   $}\hbox{$ +8) )+4 ) ) $}}&
   \vbox{\hbox{$  \frac{1}{720} \gamma p  (-20 (-1)^{\gamma} p^2 q  (2 (\gamma p+1) $}\hbox{$
   (\gamma p+2) q^2+3 r^2 ) \gamma^4
   $}\hbox{$ +30 p r  (4 \gamma r^2+6
   r+(\gamma p+1) q$}\hbox{$ (11 \gamma p+\gamma (37 \gamma p+32) q+16) )$}\hbox{$
   \gamma^2+2 (-1)^{\gamma} q  (30 \gamma^2 (\gamma p
   $}\hbox{$  (67 \gamma
   p+90)+30) r^2
   $}\hbox{$ +60 \gamma (2 \gamma p+1) (8 \gamma p+5) r
   $}\hbox{$ +(\gamma p+1)
    (\gamma  (20 \gamma (\gamma p (5 \gamma p 
    $}\hbox{$ (2 \gamma
   p+7)+39)+15) q^2
   $}\hbox{$ 
   +60 (\gamma p+1) (2 \gamma p+1) (2 \gamma p+3) q
   $}\hbox{$ +9 p (3
   \gamma p (3 \gamma p+7)+14) )+24 ) )
   $}\hbox{$ +15 r  (\gamma
    (\gamma^3 q (163 \gamma q+75) p^3
    $}\hbox{$ 
    +12 \gamma^2 q (40 \gamma q+13)
   p^2 $}\hbox{$
   + (\gamma  (467 \gamma q^2+111 q
   $}\hbox{$ +4 r (4 \gamma
   r+9) )+24 ) p
   $}\hbox{$ +150 \gamma q^2+30 q
   $}\hbox{$ +8 r (2 \gamma
   r+3) )+8 ) )   $}}\\ \\ \\
0 &  \vbox{\hbox{$ 
 \frac{1}{2} (-1)^{\gamma} \gamma q  $}\hbox{$
 ((-1)^{\gamma} p (\gamma q+1)$}\hbox{$ 
 +2
   r ) $}}&  \vbox{\hbox{$ 
   \frac{1}{4} \gamma p q  $}\hbox{$
   (2 \gamma  (\gamma q p+p+q$}\hbox{$ +4
   (-1)^{\gamma} r )-(-1)^{\gamma}+1 ) $}}
   &  \vbox{\hbox{$ \frac{1}{6} (-1)^{\gamma}
   \gamma p  $}\hbox{$
   (3 \gamma (5 \gamma p+3) q r+(-1)^{\gamma}  
   $}\hbox{$ (3
   \gamma r^2+3 r+(\gamma p+1) 
   $}\hbox{$ 
   q (2 \gamma p+3 \gamma (\gamma p+1)
   q+1) ) ) $}}& 
    \vbox{\hbox{$ 
    \frac{1}{12} \gamma p  (\gamma  (\gamma^2 q (3
   \gamma q+2) p^3
   $}\hbox{$ 
   +2 \gamma q  (6 \gamma q+10 (-1)^{\gamma} \gamma
   r$}\hbox{$ +3 ) p^2
   $}\hbox{$ 
   + (15 \gamma q^2+4  (9 (-1)^{\gamma} \gamma r+1 )
   q$}\hbox{$ +6 r (\gamma r+1) ) p
   $}\hbox{$ 
   +16 (-1)^{\gamma} q r
   $}\hbox{$ +6  (q^2+r^2 ) )$}\hbox{$ 
   -3
    (-1+(-1)^{\gamma} ) r )$}} & 
     \vbox{\hbox{$ \frac{1}{120} (-1)^{\gamma} \gamma p
   (\gamma p+1)  $}\hbox{$
   (5 \gamma (\gamma p (23 \gamma p+55)+30)$}\hbox{$
    q
   r+(-1)^{\gamma}  (60 \gamma (\gamma p+1) r^2
   $}\hbox{$ +20 (2 \gamma p+1)
   r+(\gamma p+2) q 
   $}\hbox{$ (\gamma (\gamma p+2) (8 p
   $}\hbox{$ +15 (\gamma p+1)
   q)+6) ) )$}} \\ \\ \\
 q & (-1)^{\gamma} \gamma p q+r & \frac{1}{2} \gamma p  (\gamma p q+q+2
    (-1)^{\gamma} r ) & \vbox{\hbox{$
    \frac{1}{6} \gamma p (\gamma p+1)
  $}\hbox{$
   ((-1)^{\gamma} (\gamma p+2) q+3 r )$}} & 
    \vbox{\hbox{$ \frac{1}{24} \gamma p
   (\gamma p+1) (\gamma p+2) $}\hbox{$
    ((\gamma p+3) q+4 (-1)^{\gamma} r ) $}} &
 \vbox{\hbox{$  \frac{1}{120} \gamma p (\gamma p+1)
    (\gamma p+2) $}\hbox{$ (\gamma p+3)
    ((-1)^{\gamma} (\gamma p+4) q
    $}\hbox{$ +5 r ) $}}\\ \\ \\
 0 & p & 0 & 0 & 0 & 0
\end{array}
\right)}
\ee
For $(p,q,r,\gamma)=(1,1,1,1)$ this reproduces  again the pentagonal 
identity \eqref{penta} (read backwards). 
For $(p,q,r,\gamma)=(1,1,1,-1)$ it reproduces 
the weak coupling spectrum of the $SU(2)$ Seiberg-Witten  theory with $N_f=1$ 
flavor \cite{Dimofte:2009tm}, since the strong coupling
spectrum of the latter has three states of unit degeneracy
each, carrying charges $\gamma_1=(0,1)$,
$\gamma_2=(1,-1)$ and $\gamma_1+\gamma_2=(1,0)$.

\section{U(N) quiver quantum mechanics from
Boltzmann black hole halos} \label{sUNquiver}

In this appendix we shall show that our Boltzmann gas
picture of multi-centered black hole system also makes
a specific prediction on quiver quantum mechanics
without oriented loops.
Consider
a multi-black hole configuration
with $(n+N)$ centers, with the first $n$ centers carrying
charges $\alpha_1,\dots, \alpha_n$ and each of
the last $N$
centers carrying charge $\beta_0$.
We take all the $\alpha_i$'s and $\beta_0$ to lie
in a two dimensional plane, and assume, without any
loss of generality, that
$\alpha_{ij}
\equiv \langle\alpha_i,\alpha_j\rangle>0$ for $i<j$.
Suppose further that
each center
carries unit intrinsic index, \i.e.\ $\Omega^+(\alpha_i)=
\Omega^+(\beta_0)=1$, and that
$\Omega^+(k\beta_0)=0$ for $k>1$. 
In this case the
system is described by a $U(1)^n \times U(N)$ quiver
quantum mechanics~\cite{Denef:2002ru}, 
and the refined index 
$\Tr(-y)^{2J_3}$ of this quantum mechanics
is given by
\be \label{ec1}
(-y)^{-C_{U(1)^n \times U(N)} (\alpha_1,
\dots, \alpha_n, \beta_0)}\, 
P_{U(1)^n \times U(N)} (\alpha_1,
\dots, \alpha_n, \beta_0, y)\, .
\ee
Here  $P_{U(1)^n \times U(N)} 
(\alpha_1,
\dots, \alpha_n, \beta_0, y)$ is the Poincar\'e polynomial
associated with a quiver quantum mechanics with
$n+1$ nodes, with the first $n$-nodes carrying $U(1)$
factors and the last node carrying a $U(N)$ factor.
Furthermore there are $\alpha_{ij}$ 
arrows from the
$i$'th to the $j$'th node  for $i<j$, 
$|\langle\alpha_i,\beta_0\rangle|$ arrows from the 
$i$th node to the last node if 
$\langle\alpha_i,\beta_0\rangle>0$ and 
$|\langle\alpha_i,\beta_0\rangle|$ arrows from the 
last node to the $i$th node if 
$\langle\alpha_i,\beta_0\rangle<0$. 
$C_{U(1)^n \times U(N)} (\alpha_1,
\dots, \alpha_n, \beta_0)$ denotes the maximum power
of $y^2$ in $P_{U(1)^n \times U(N)} 
(\alpha_1,
\dots, \alpha_n, \beta_0, y)$, signifying the
maximum value of $2J_3$ carried by the
system.
An explicit expression for the Poincar\'e polynomial of
this quiver will be given in eq. \eqref{ec5} below. 
On the
other hand we can also calculate the index using
\eqref{esecond}-\eqref{higgsf}.
First note that we have, from \eqref{esecond},
\be \label{ec2}
\bOm^+(l\beta_0) =  {y-y^{-1}\over l(y^l
- y^{-l})} \, .
\ee
Using \eqref{efirst3} we find that this configuration 
contributes 
\be \label{ec3aa}
\sum_{\{k_l\}\atop \sum lk_l=N} 
g_{\rm ref}(\alpha_1,\dots, \alpha_n, \beta_0\times k_1,
2\beta_0\times k_2,\dots, y) \prod_l \Big\{
 {1\over k_l!} \bOm^+ (l\beta_0)^{k_l}\Big\} 
\ee
to $\Tr(-y)^{2J_3}$ , where $g_{\rm ref}$ is given by 
\eqref{higgsf} and the symbol $l\beta_0\times k_l$
means that the vector $l\beta_0$ is repeated $k_l$
times. To study the consequences of equating 
\eqref{ec3aa} to
\eqref{ec1}, note
that the expression for
$g_{\rm ref}$ given in \eqref{higgsf}, after stripping
off the first factor is nothing but the Poincar\'e polynomial
of a quiver theory with only $U(1)$ factors. This gives 
\be \label{egpoinc}
\begin{split}
& g_{\rm ref}(\alpha_1,\dots, \alpha_n, \beta_0\times k_1,
2\beta_0\times k_2,\cdots, y) 
=  (-y)^{n+\sum_l {k_l} -1-\sum_{i<j} \alpha_{ij}
- N\sum_i |\langle \alpha_i,\beta_0\rangle|} \cr & \qquad \times
P_{U(1)^n \times U(1)^{k_1}\times U(1)^{k_2}\times
\cdots}(\alpha_1,\dots, \alpha_n, \beta_0\times k_1,
2\beta_0\times k_2,\cdots, y)\, .
\end{split}
\ee
Using eqs.\eqref{ec1}, \eqref{ec2} 
and \eqref{egpoinc} we get
a direct relation
between the Poincar\'e polynomials of a quiver
carrying a $U(N)$ factor and those carrying only $U(1)$
factors:
\be \label{ec3}
\begin{split}
P_{U(1)^n \times U(N)} (\alpha_1,
\dots, \alpha_n, \beta_0, y)
= &\sum_{\{k_l\}\atop \sum lk_l=N} 
P_{U(1)^n \times \prod_l U(1)^{k_l}} (\alpha_1,
\dots, \alpha_n, \beta_0\times k_1, 2\beta_0\times k_2,
\cdots, y) \\
&  (-y)^{C_{U(1)^n \times U(N)} (\alpha_1,
\dots, \alpha_n,\beta_0) 
+
n+\sum_l {k_l} -1-\sum_{i<j} \alpha_{ij}
- N\sum_i |\langle \alpha_i,\beta_0\rangle|}\, \\
& \left\{
\prod_l {1\over k_l!} \left( {y-y^{-1}\over l(y^l
- y^{-l})}
\right)^{k_l}\right\}
\end{split}
\ee
We could also derive a similar relation where the first
$U(1)^n$ factor is replaced by a
product of $U(M)$ factors
by choosing among the set $(\alpha_1,\dots, \alpha_n)$
identical charges. 
Let us use a shorthand notation
where $P_{N_1,\dots, N_p}(\beta_1,\dots, 
\beta_p, y)$ denotes the Poincar\'e polynomial of
a quiver with the $i$-th 
node having a $U(N_i)$ factor and charge $\beta_i$
and furthermore arrange the $\beta_i$'s such that
$\beta_{ij}>0$ for $i>j$.
Similarly $C_{N_1,\dots, N_p}(\beta_1,\dots, 
\beta_p)$ will denote the maximum power of
$y^2$ in this polynomial.
Then the  generalization of \eqref{ec3} will take
the form
\be \label{ec3.5} 
\begin{split}
& y^{- C_{N_1,\dots, N_p}(\beta_1,\dots, 
\beta_p)} P_{N_1 \cdots N_p} 
(\beta_1, \dots, \beta_p,y)\\
=&  \sum_{\{k_l\}\atop \sum lk_l=N_r} 
P_{N_1,\dots ,N_{r-1}, 1\times k_1, 1\times k_2,\cdots,
N_{r+1},\dots, N_p}  (\beta_1,
\dots, \beta_{r-1},  \beta_r\times k_1, 2\beta_r\times k_2,
\cdots, \beta_{r+1}, \dots, \beta_p, y) \\
& y^{- C_{N_1,\dots ,N_{r-1}, 1\times k_1, 1\times k_2,\cdots,
N_{r+1},\dots, N_p}  (\beta_1,
\dots, \beta_{r-1},  \beta_r\times k_1, 2\beta_r\times k_2,
\cdots, \beta_{r+1}, \dots, \beta_p)} \left\{
\prod_l {1\over k_l!} \left( {y-y^{-1}\over l(y^l
- y^{-l})} (-1)^{l-1}
\right)^{k_l}\right\}
\end{split}
\ee
where we have substituted $y\to -y$, using the fact that the Poincar\'e
polynomial of a quiver is an even function of $y$. By
repeated use of \eqref{ec3.5} we can express the
Poincar\'e polynomial of a quiver with $U(N)$ factors
in terms of the Poincar\'e polynomial of quivers
with $U(1)$ factors only.

We now test e.q. \eqref{ec3.5} against the 
Poincar\'e polynomial of a quiver
quantum mechanics with $U(N)$ factors but no
oriented loops computed in~\cite{MR1974891} (see also \cite{Denef:2002ru}):
\begin{equation} 
\label{ec5}
\begin{split}
P_{N_1,\cdots, N_p} 
(\beta_1, \dots, \beta_p,y)
=& (y^2-1)^{1-\sum_i N_i} \, y^{-\sum_i N_i(N_i-1)} \,
\\ & \sum_{\rm partitions}(-1)^{s-1} y^{2\sum_{a\leq b}\sum_{i 
 <
j}\beta_{ij} \, 
N^b_i N^a_j} \prod_{a,i} ([N^a_i,y]!)^{-1}\, , \\
C_{N_1,\cdots, N_p} 
(\beta_1, \dots, \beta_p)
=& \sum_{i<j} N_i N_j \beta_{ij} - \sum_i N_i^2
+ 1 
\end{split}
\end{equation}
where 
\be \label{ec5.5}
[N,y] \equiv \frac{y^{2N}-1}{y^2-1}, \qquad
[N,y]! \equiv
[1,y][2,y]\ldots[N,y]\, ,
\ee
and
the sum over partitions in \eqref{ec5}
runs over all ordered partitions of 
the vector $\sum_i N_i \beta_i$ into 
non-zero vectors $\{ \sum_i 
N^a_i\beta_i, \quad a=1,\dots, s\}$
for $s=1,\dots, \sum_i {N_i}$,
satisfying $\sum_a N^i_a=N_i$ and\footnote{Normally
this relation is written as a condition on the phase of 
$Z_{\sum_{a=1}^b\sum_i N^a_i\beta_i}$, 
but we have used \eqref{enew1}
to expres this as a condition involving 
the symplectic products.}
\be \label{ec6}
\left\langle \sum_{a=1}^b\sum_{i=1}^p  N^a_i \beta_i , 
\sum_{j=1}^p N_j \beta_j \right\rangle > 0\, ,
\ee
for $b=1,\ldots ,s-1$. If $C_{N_1,\dots, N_p} 
(\beta_1, \dots, \beta_p)$ given in \eqref{ec5} becomes
negative then the Poincar\'e polynomial vanishes.

Based on \eqref{ec5}, eq. \eqref{ec3.5} can be proven by collecting
the terms multiplying a given partition on both sides of
this equation
and showing that they cancel. 
Consider for example a partition with $s$ vectors of the
form $\sum_{i=1}^p N^a_i \beta_i$ 
for $a=1,\dots, s$ 
and let $N^a_r=M_a$. 
Thus the $s$ vectors have the form $A_1+M_1\beta_r$,
$A_2+M_2\beta_r$, $\dots$ $A_s+M_s\beta_r$
where $A_1,\dots, A_s$ are linear combinations of
the $\beta_i$'s other than $\beta_r$. If we just focus
on the contribution involving the $U(N_r)$ factor then the
contribution to this particular partition from the
left hand side of \eqref{ec3.5} contains a factor of
\be \label{eleftfactor}
(y^2 -1)^{-N_r} y^{N_r} \prod_a ([M_a, y]!)^{-1}\, ,
\ee
where we have made use of \eqref{ec5} and ignored
the $y^{-\sum_{i<j} N_i N_j \beta_{ij}}$ term since it
is a common factor on both sides of \eqref{ec3.5}.
On the right hand side this particular partition can
come from many different terms. In particular
$M_a\beta_r$ can arise as a sum of $k^{(a)}_1$ copies
of $\beta_r$, $k^{(a)}_2$ copies
of $2\beta_r$ etc. with $M_a =\sum_l k^{(a)}_l l$. This
in turn can arise from the 
$P_{N_1,\dots ,N_{r-1}, 1\times k_1, 1\times k_2,\cdots,
N_{r+1},\dots, N_p}$ term if the $k^{(a)}_l$'s satisfy
$\sum_a k^{(a)}_l = k_l$. There is a combinatoric
factor of $k_l! / \prod_a k^{(a)}_l!$ associated with
different ways of choosing the $k^{(a)}_l$ copies of
$l\beta_r$ out of $k_l$ copies of $l\beta_r$.
Using \eqref{ec3.5} and \eqref{ec5} we see that the
relevant factor on the right hand side is
\be \label{erhs}
\sum_{\{k_l\}}\sum_{\{k^{(a)}_l\}\atop \sum_l k^{(a)}_l l= M_a,
\sum_a k^{(a)}_l = k_l}
{1\over \prod_{a,l} k^{(a)}_l!} \, (y^2-1)^{-\sum_l k_l}
y^{\sum_l k_l} 
\left\{
\prod_l  \left( {y-y^{-1}\over l(y^l
- y^{-l}) }(-1)^{l-1}
\right)^{k_l}\right\}\, .
\ee
We can simplify this as follows. First of all 
using the relation $N_r =
\sum_l k_l l$, we can express $\sum_l k_l$ in the
exponents of $(y^2-1)$ and $y$ as
$N_r - \sum_l k_l (l-1)= N_r -  \sum_{l,a} k^{(a)}_l (l-1)$.
Next the sum over
$\{k_l\}$ and the restriction $\sum_a k^{(a)}_l = k_l$ can
be removed if we replace all factors of $k_l$ by
$\sum_a k^{(a)}_l$. 
Using \eqref{ec5.5} 
this allows us to express \eqref{erhs} as
\be \label{rhs2}
(y^2-1)^{-N_r} y^{N_r} \prod_a \left\{
\sum_{k^{(a)}_l \atop \sum_l k^{(a)}_l l=M_a}
(y^2-1)^{\sum_l k^{(a)}_l (l-1)} \prod_l 
\left({(-1)^{l-1} \over [l,y]\, l} 
\right)^{k^{(a)}_l}
\right\}\, .
\ee
The equality of \eqref{eleftfactor} and \eqref{rhs2} 
is now reduced to the following identity:
\be \label{eidentity}
{1\over [N,y]!} \stackrel{?}{=} \sum_{\{k_l\} \atop \sum_l l k_l=N}
\prod_l {1\over k_l!} (-1)^{k_l (l-1)} (y^2-1)^{k_l(l-1)}\left(
{1\over [l,y]\, l}\right)^{k_l}\, .
\ee
To prove \eqref{eidentity}, consider the equivalent identity
for partition functions
\be
\sum_{N=0}^{\infty} \frac{z^N}{[N,y]!} \stackrel{?}{=} 
\exp\left( \sum_{l=1}^{\infty} \frac{(1-y^2)^l }{1-y^{2l}} \frac{z^l}{l}\right)\ .
\ee
Rewriting the r.h.s. as 
\be
\exp\left( \sum_{k=0}^{\infty} \sum_{l=1}^{\infty} \frac{[z(1-y^2)]^l\, y^{2lk}}{l}\right)=
\prod_{k=0}^{\infty}
\left( 1- z(1-y^2) y^{2k} \right)^{-1}\ ,
\ee
we see that \eqref{eidentity} follows from the known relation between the $q$-deformed 
exponential and the $q$-deformed Pochhammer symbol with $q=y^2$ (see e.g.
\cite{qexp}),
\be
\sum_{N=0}^{\infty} \frac{z^N}{[N,y]!}  
= \left( z(1-q) ; q\right)_{\infty} = \prod_{k=0}^{\infty}
\left( 1- z(1-q) q^{k} \right)^{-1}\ ,\qquad |q|<1\ .
\ee

\providecommand{\href}[2]{#2}\begingroup\raggedright\endgroup

\end{document}